\documentclass[twoside,twocolumn,9pt]{article}
\usepackage{extsizes}
\usepackage[super,sort&compress,comma]{natbib} 
\usepackage[version=3]{mhchem}
\usepackage[left=1.5cm, right=1.5cm, top=1.785cm, bottom=2.0cm]{geometry}
\usepackage{balance}
\usepackage{multirow}
\usepackage{mathptmx}
\usepackage{mathtools}
\usepackage{tabularx}
\usepackage{mdwlist}
\usepackage{amsmath,systeme}
\usepackage{sectsty}
\usepackage{graphicx} 
\usepackage{lastpage}
\usepackage[format=plain,justification=justified,singlelinecheck=false,font={stretch=1.125,small,sf},labelfont=bf,labelsep=space]{caption}
\usepackage{float}
\usepackage{fancyhdr}
\usepackage{fnpos}
\usepackage[english]{babel}
\addto{\captionsenglish}{%
  \renewcommand{\refname}{Notes and references}
}
\usepackage{array}
\usepackage{droidsans}
\usepackage{charter}
\usepackage[T1]{fontenc}
\usepackage[usenames,dvipsnames]{xcolor}
\usepackage{setspace}
\usepackage[compact]{titlesec}
\usepackage{hyperref}
\usepackage{array, makecell} 
\usepackage{ragged2e}
\newcolumntype{L}{>{\RaggedRight\arraybackslash}X}

\graphicspath{ {./figures/} }

\definecolor{cream}{RGB}{222,217,201}

\begin{document}

\pagestyle{fancy}
\thispagestyle{plain}
\fancypagestyle{plain}{
\renewcommand{\headrulewidth}{0pt}
}

\makeFNbottom
\makeatletter
\renewcommand\LARGE{\@setfontsize\LARGE{15pt}{17}}
\renewcommand\Large{\@setfontsize\Large{12pt}{14}}
\renewcommand\large{\@setfontsize\large{10pt}{12}}
\renewcommand\footnotesize{\@setfontsize\footnotesize{7pt}{10}}
\makeatother

\renewcommand{\thefootnote}{\fnsymbol{footnote}}
\renewcommand\footnoterule{\vspace*{1pt}%
\color{cream}\hrule width 3.5in height 0.4pt \color{black}\vspace*{5pt}} 
\setcounter{secnumdepth}{5}

\makeatletter 
\renewcommand\@biblabel[1]{#1}            
\renewcommand\@makefntext[1]%
{\noindent\makebox[0pt][r]{\@thefnmark\,}#1}
\makeatother 
\renewcommand{\figurename}{\small{Fig.}~}
\sectionfont{\sffamily\Large}
\subsectionfont{\normalsize}
\subsubsectionfont{\bf}
\setstretch{1.125} 
\setlength{\skip\footins}{0.8cm}
\setlength{\footnotesep}{0.25cm}
\setlength{\jot}{10pt}
\titlespacing*{\section}{0pt}{4pt}{4pt}
\titlespacing*{\subsection}{0pt}{15pt}{1pt}

\fancyfoot{}
\fancyfoot[RO]{\footnotesize{\sffamily{1--\pageref{LastPage} ~\textbar  \hspace{2pt}\thepage}}}
\fancyfoot[LE]{\footnotesize{\sffamily{\thepage~\textbar\hspace{2pt} 1--\pageref{LastPage}}}}
\fancyhead{}
\renewcommand{\headrulewidth}{0pt} 
\renewcommand{\footrulewidth}{0pt}
\setlength{\arrayrulewidth}{1pt}
\setlength{\columnsep}{6.5mm}
\setlength\bibsep{1pt}

\makeatletter 
\newlength{\figrulesep} 
\setlength{\figrulesep}{0.5\textfloatsep} 

\newcommand{\topfigrule}{\vspace*{-1pt}%
\noindent{\color{cream}\rule[-\figrulesep]{\columnwidth}{1.5pt}} }

\newcommand{\botfigrule}{\vspace*{-2pt}%
\noindent{\color{cream}\rule[\figrulesep]{\columnwidth}{1.5pt}} }

\newcommand{\dblfigrule}{\vspace*{-1pt}%
\noindent{\color{cream}\rule[-\figrulesep]{\textwidth}{1.5pt}} }

\makeatother

\newcommand{\SC}[1]{\color{red}{#1}\normalcolor}
\newcommand{\NMD}[1]{\color{blue}{#1}\normalcolor}
\newcommand{\MF}[1]{\color{green}{#1}\normalcolor}

\twocolumn[
  \begin{@twocolumnfalse}
\begin{tabular}{m{1cm} p{15cm}}

 & \noindent\LARGE{\textbf{CO$_2$ removal and 1.5$^{\circ}$C: what, when, where, and how?}}\\
\vspace{0.3cm} & \vspace{0.3cm} \\

 & \noindent\large{Solene Chiquier,\textit{$^{ab}$} Mathilde Fajardy,\textit{$^{c}$} and Niall MacDowell$^{\ast}$\textit{$^{ab}$}} \\
 \vspace{0.3cm} & \vspace{0.3cm} \\
 
 & \noindent\normalsize{
The international community aims to limit global warming to 1.5$^{\circ}$C, but little progress has been made towards a global, cost-efficient, and fair climate mitigation plan to deploy carbon dioxide removal (CDR) at the Paris Agreement’s scale.
Here, we investigate how different CDR options --- afforestation/reforestation (AR), bioenergy with carbon capture and storage (BECCS), and direct air carbon capture and storage (DACCS) --- might be deployed to meet the Paris Agreement's CDR objectives.
We find that international cooperation in climate mitigation policy is key for deploying the most cost-efficient CDR pathway --- comprised of BECCS, mainly (74\%), and AR (26\%) ---, allowing to take the most advantage of regional bio-geophysical resources and socio-economic factors, and time variations, and therefore minimising costs. 
Importantly, with international cooperation, the spatio-temporal evolution of the CDR pathway differs greatly from the regional allocation of the Paris Agreement's CDR objectives --- based on responsibility for climate change, here used as a proxy for their socio-economically fair distribution.
With limited, or no international cooperation, we find that the likelihood of delivering these CDR objectives decreases, as deploying CDR pathways becomes significantly more challenging and costly.
Key domestic bio-geophysical resources include geological CO$_2$ sinks, of which the absence or the current lack of identification undermines the feasibility of the Paris Agreement's CDR objectives, and land and biomass supply, of which the limited availability makes them more costly --- particularly when leading to the deployment of DACCS.
Moreover, we show that developing international/inter-regional cooperation policy instruments --- such as an international market for negative emissions trading --- can deliver, simultaneously, cost-efficient and equitable CDR at the Paris Agreement's scale, by incentivising participating nations to meet their share of the Paris Agreement's CDR objectives, whilst making up for the uneven distribution of CDR potentials across the world.
Crucially, we conclude that international cooperation --- cooperation policy instruments, but also robust institutions to monitor, verify and accredit their efficiency and equity --- is imperative, as soon as possible, to preserve the feasibility and sustainability of future CDR pathways, and ensure that future generations do not bear the burden, increasingly costlier, of climate mitigation inaction.
} 

\end{tabular}
 \end{@twocolumnfalse} \vspace{0.6cm}
]


\renewcommand*\rmdefault{bch}\normalfont\upshape
\rmfamily
\section*{}
\vspace{-1cm}


\footnotetext{\textit{$^{a}$Centre for Environmental Policy, Imperial College London, Exhibition Road, London, SW7 1NA, UK.}}
\footnotetext{\textit{$^{b}$Centre for Process Systems Engineering, Imperial College London, Exhibition Road, London, SW7 2AZ, UK.}}
\footnotetext{\textit{$^{c}$Cambridge Judge Business School, University of Cambridge, Trumpington Street, Cambridge CB2 1AG, UK.}}
\footnotetext{$^{\ast}$Corresponding author. Email: niall@imperial.ac.uk; Tel: +44 (0)20 7594 9298}



\section{Introduction}

\subsection{Carbon dioxide removal (CDR) and the Paris Agreement}

Through the 2015 Paris Agreement, Parties to the UNFCCC agreed to hold global warming to "well below" 2$^{\circ}$C and pursue efforts to limit it to 1.5$^{\circ}$C by reducing global CO$_2$ emissions as soon as possible and reaching net-zero by mid-century \cite{UNFCCC2015}. 
Because of the near-linear relationship between cumulative anthropogenic CO$_2$ emissions and temperature increase \cite{MacDougall2015,IPCC2014,Matthews2009,Zickfeld2009}, halting global warming to 1.5$^{\circ}$C requires CO$_2$ emissions to stay within a remaining carbon budget of about 420 Gt CO$_2$ \cite{IPCC2018,Huppmann2018,Matthews2017,Millar2017}. \newline
\indent If future anthropogenic CO$_2$ emissions are not reduced promptly enough and “overshoot” this remaining carbon budget, then negative CO$_2$ emissions will be required to return to it, \textit{i.e.} the CO$_2$ emissions level (and the temperature increase target of 1.5$^{\circ}$C) is first exceeded and then return to by deploying carbon dioxide removal (CDR).
However, delaying short-term climate mitigation will ultimately result into a more aggressive mid-term transformation of energy systems, higher long-term costs, and stronger transitional economic and societal impacts \cite{Luderer2013,Luderer2016,Sanderson2020,Victoria2020}. Particularly, the increased reliance on CDR might render the feasibility of the 1.5$^{\circ}$C objective of the Paris Agreement questionable \cite{McLaren2020,Hansen2017,Hanna2021}.
\newline
\indent Most Parties have committed to legally-binding net-zero targets by the second half of this century --- mostly 2050 but also, for instance, 2060 in China or 2070 in India --- since the close of COP26 \cite{ClimateAnalytics2021,UNFCCC2021}.
However, almost none are on track with their nationally determined contributions (NDCs) \cite{ClimateAnalytics2022}, which themselves, moreover and anyway, still fall short of the Paris Agreement's 1.5$^{\circ}$C ambition \cite{ClimateAnalytics2021,Geiges2020,Rogelj2016}.
Therefore, large-scale deployment of CDR is critical not only 1) to achieve net-zero by compensating for on-going CO$_2$ emissions, particularly residual ones from hard-to-abate sectors such as transport or agriculture, but also 2) to provide net negative emissions to return from any overshoots of the remaining carbon budget \cite{Luderer2018}.
\newline
\indent In Integrated Assessment Models (IAMs), most 1.5$^{\circ}$C-consistent scenarios require CO$_2$ emissions to decrease from 2030, reach net zero by 2050, and become net negative afterwards in order to return from overshoots \cite{IPCC2018,Rogelj2018}. “No or limited overshoot” scenarios (categorised as P1, P2 and P3 in the Special Report on Global Warming of 1.5$^{\circ}$C (SR15) published by the IPCC) rely on cumulative CO$_2$ removal of 246--689 GtCO$_2$ by 2100\footnote{These numbers account for negative emissions arising from both "CCS/Biomass" and "CO2/AFOLU", as categorised by the IPCC.}, and “higher overshoot” scenarios (categorised as P4) on as much as 1,186 GtCO$_2$ \cite{IPCC2018,Huppmann2018}. Such deployments of CDR are estimated to start immediately (\textit{i.e.,} between 2020--2030) and reach up to 4--24 GtCO$_2$/yr in 2100 \cite{IPCC2018,Huppmann2018}.
For these reasons, this study focuses on the 1.5$^{\circ}$C-consistent CDR scenarios of the IPCC SR15, rather than on the mid-century net-zero objectives set out by the Parties’ NDCs.

\subsection{The techno-economic challenges of CDR}

Various CDR options has been suggested --- including afforestation/reforestation (AR) \cite{Canadell2008}, bioenergy with carbon capture and storage (BECCS) \cite{Obersteiner2001}, direct air capture with carbon capture and storage (DACCS) \cite{Socolow2011, Keith2018}, ocean fertilisation \cite{Renforth2013}, enhanced weathering (EW) of minerals \cite{Lenton2006,Schuiling2006}, biochar \cite{Woolf2010,Smith2016b} or soil carbon sequestration \cite{Minasny2017,Smith2016b} --- but have scarcely been taken up in IAMs. 
To that date, they have included mainly AR and BECCS, DACCS as well (yet only recently), and seldom EW \cite{Strefler2021,Fuhrman2020,Realmonte2019,Strefler2018,Marcucci2017}, and that mainly because other CDR options are still highly speculative \cite{Fuhrman2019}. 
Particularly, only BECCS \cite{Drax2020} and DACCS \cite{CE2020,CW2020} have been deployed at the demonstration scale, yet nowhere near the scales required to deliver the Paris Agreement's 1.5$^{\circ}$C ambition, and whilst AR is a well-established and mature practice, projects with the aim of removing CO$_2$ from the atmosphere have only recently emerged, mostly in China \cite{ARChina_GreenWall,ARChina_Shandong}. 
\newline
\indent The nascent nature of most CDR options has raised heightened concerns about the feasibility and sustainability of the large-scale deployments of CDR in 1.5$^{\circ}$C-consistent scenarios, especially if achieved \textit{via} such limited portfolios of CDR options (\textit{e.g.,} only BECCS and AR) \cite{Fuss2018,Heck2018,Nemet2018,Smith2016}.
Particularly, Fuss \textit{et al.} \cite{Fuss2018} reduced the CDR potential of BECCS in 2050 from 8 to 0.5--5 GtCO$_2$/yr for sustainability safeguard, and suggested therefore that BECCS alone would be insufficient to deliver the Paris Agreement's most stringent CDR targets, such as in the P4 scenario of the SR15.
Despite the increasing focus on CDR in the academic literature, emphasising on CDR potential, cost, and up-scaling, as well as interactions with the sustainable development goals (SDGs), the CDR efficiency and permanence of most CDR options are still uncertain, and remain major challenges to their deployment \cite{Fuss2018,Nemet2018,NASEM2019,Honegger2021}.

\subsection{International cooperation and CDR policy}

With the principle of "common but differentiated responsibilities and respective capabilities" lying at the heart of the Paris Agreement, there has been recently an increasing reflection on the role and value of CDR at the national scale and the need for equity in sharing its global burden \cite{Mohan2021,Morrow2020}. 
Different burden-sharing principles \cite{Ringius2002}, based on equity, climate change responsibility, or financial capacity for instance, have been investigated in the context of CDR \cite{Pozo2020,Mohan2021}.
\newline
\indent Importantly, the amounts of CDR deployed in global and cost-optimal IAMs scenarios fail to reflect the responsibility of each nation for climate change, or any other socio-economically fair establishment of its share of the global CDR burden. 
As the CDR potentials (as well as feasibilities) of each nation vary due to bio-geophysical and socio-political factors --- including the availability of bioenergy resources, geological and/or biogenic CO$_2$ sinks and low-carbon and affordable energy, and the acceptability of the various CDR options---, they don’t necessarily match with national CDR targets, and that, regardless of how the global CDR burden is shared.
\newline
\indent As promoted by the Paris Agreement (in the general context of climate mitigation), international cooperation would certainly allow to deploy most-cost efficiently, sustainably, and feasibly CDR in line with the Paris Agreement 1.5$^{\circ}$C ambition.
For instance, Fajardy \textit{et al.} \cite{Fajardy2020a} emphasised the value of collaboration in delivering CDR at large-scales, \textit{via} BECCS, in a most cost-effective manner. 
Bauer \textit{et al.} \cite{Bauer2020} investigated the trade-off between cost-efficiency and national sovereignty --- the nation's ability to maintain governing control of economic resources by limiting international transfer payment, while contributing to climate mitigation actions --- in delivering the Paris Agreement, and showed the value of cooperation \textit{via} an hybrid combination of financial transfers and differentiated carbon prices. 
Finally, Strefler \textit{et al.} \cite{Strefler2021} also showed that large international financial transfer and strong international institutions would be required for delivering CDR at the Paris Agreement's 1.5$^{\circ}$C scale, while meeting fairness and sustainability criteria. 
\newline
\indent Deploying CDR with international cooperation will certainly involve the adoption of international/inter-regional policy instruments, such as markets for internationally transferred mitigation outcomes (ITMOs) (\textit{i.e.,} transfers between domestic ETS) or voluntary emission reductions (VERs) (\textit{i.e.,} international market). Such market-based approaches have been introduced in the Article 6 of the Paris Agreement \cite{UNFCCC2015}, the rulebook of which was recently completed at COP26. 
Importantly, these instruments should be combined with a transparent assessment of sustainable development implications, \textit{e.g.} the SDGs of the Paris Agreement, as advocated by Honegger and Reiner \cite{Honegger2018}. 
For example, at the EU-scale, Rickels \textit{et al.} \cite{Rickels2021} considered the integration of BECCS into the EU emissions trading system (ETS) and its potential implications for the EU ETS.  
It is still unclear, however, how CDR options might be integrated within such international/inter-regional market-based approaches, notably do to many challenges around the permanence, additivity, measurability, and verifiability of their CDR potentials \cite{Brander2021,Mace2021}. 

\subsection{Contribution of this study}

This study investigates the spatio-temporal potential, composition, and evolution of a portfolio of CDR options (AR, BECCS and DACCS) by exploring different climate policy options, while delivering CDR targets that are consistent with the Paris Agreement's 1.5$^{\circ}$C objective in the context of 5 regions (Brazil, China, the EU-28, India and the USA). 
\newline
\indent By doing so, and considering a range of feasibility and sustainability criteria, we aim to keep within reach the Paris Agreement's 1.5$^{\circ}$C objective by helping policymakers to understand 1) the real-world potential, implications and challenges of the different nascent CDR options, and 2) the benefits of international cooperation policy, at high spatio-temporal resolution. Particularly, we aim to bridge the gap between the IAMs top-down approach and the CDR assessments bottom-up approach.
Note that this study doesn’t contribute to define how much CDR should be required to meet the Paris Agreement's 1.5$^{\circ}$C objective, nor how it should be shared, \textit{i.e.} allocated regionally.
It doesn't either aim to determine policy design for the integration of negative emissions within carbon markets. 
\newline
\indent Firstly, Section \ref{sec: Methods} describes the Modelling and Optimisation of Negative Emissions Technologies (MONET) framework used in this study. 
In Section \ref{sec: Opt CDR}, cost-optimal CDR pathways (\textit{i.e.,} portfolios of CDR options), subject to alternative climate policies are outlined.
Section \ref{sec: Neg ETS} discusses the role and value of international cooperation in climate policy, \textit{via} an international market for negative emissions trading, and Section \ref{sec: Switch Policy} emphasises the urgency of shifting towards international cooperation policy, and discusses the impacts of delaying it.
Lastly, we present some conclusions in Section \ref{sec: Conclusions}.

\section{Methods}
\label{sec: Methods}

\begin{figure}
\centering
  \includegraphics[width=8.3cm]{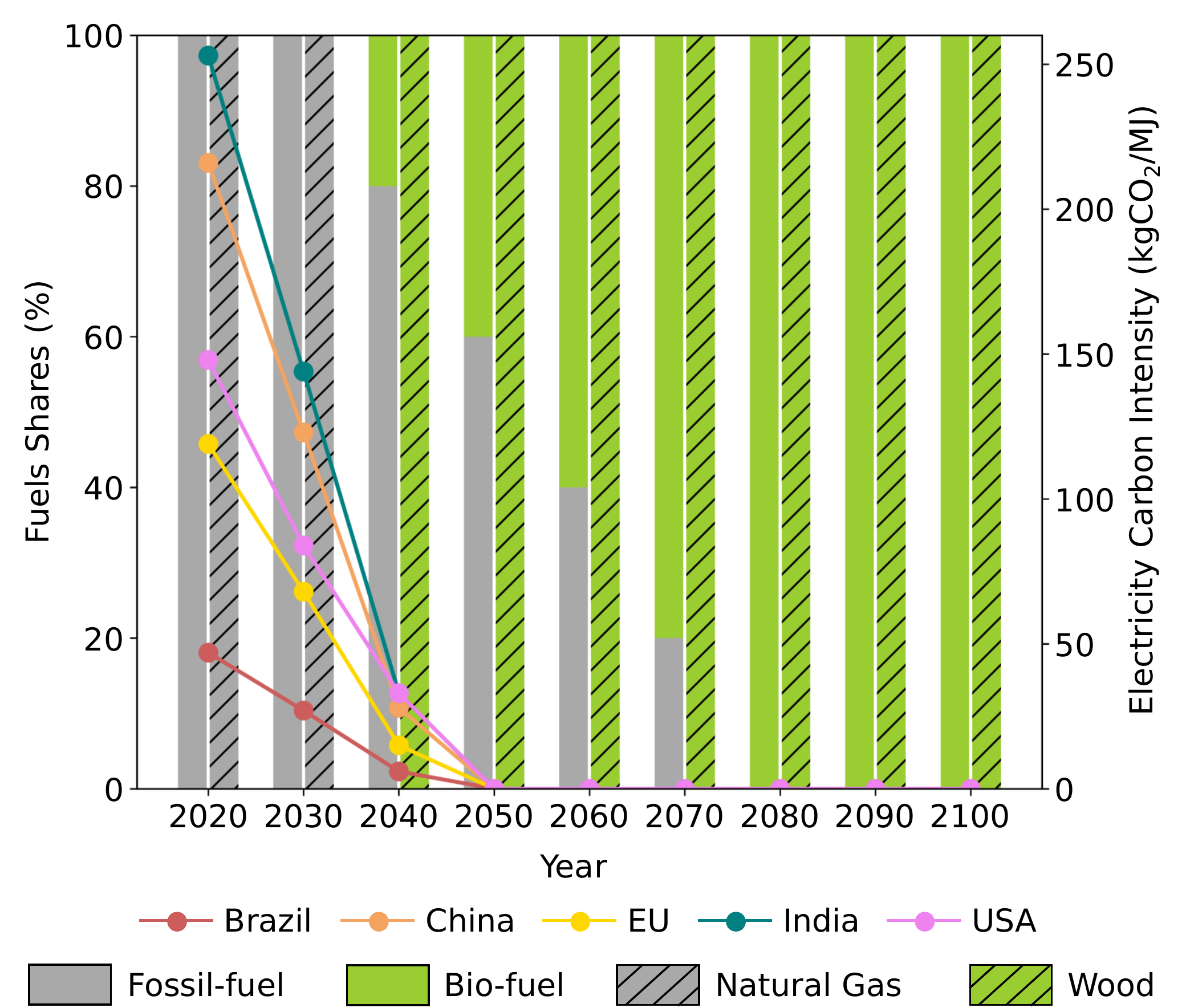}
  \caption{Transition of the worldwide economy towards net-zero. Following a decreasing carbon intensity (as projected by the IPCC P2 scenario \cite{IPCC2018, Huppmann2018}, the electricity system becomes carbon neutral in 2050. Fossil-fuels (\textit{i.e.,} diesel and petrol) are progressively replaced by 100\% bio-fuels (\textit{i.e.,} bio-diesel and bio-ethanol) in 2080. Natural gas is switched to 100\% wood as early as 2040 for biomass drying for BECCS.}
  \label{fig: ESD}
\end{figure}

In this study, we use the Modelling and Optimisation of Negative Emissions Technologies (MONET) framework to provide insights on the composition (\textit{i.e.,} what is deployed?) and spatio-temporal evolution (\textit{i.e.,} when and where is it deployed?) of cost-optimal CDR pathways deployed to deliver the Paris Agreement's 1.5$^{\circ}$C objective.
The MONET framework is spatio-temporally explicit, and it 1) provides whole-system analyses (\textit{e.g.,} CDR potential, cost, land use) for different CDR options, and 2) determines cost-optimal deployment of a portfolio of such CDR options between 2020--2100, subject to long-term CDR targets, CDR deployment conditions (\textit{i.e.,} build/expansion rates and operating lifetimes), and bio-geophysical constraints (\textit{i.e.,} land and geological CO$_2$ storage availabilities, maximum water stress).
All together, these constraints aim to encompass criteria of feasibility and sustainability.
\newline
\indent The current implementation of MONET describes the deployment of 3 CDR options --- AR, BECCS, and DACCS --- across 5 regions --- Brazil, China, the EU (EU-27 + UK), India and the USA. 
The spatial resolution is at the state/province scale (national scale for the EU), that is 169 sub-regions, and the temporal resolution (\textit{i.e.,} time-step) is 10 years.
Consistently with 1.5$^{\circ}$C scenarios, we assume that the worldwide economy transitions towards net-zero, particularly the electricity and the transport/fuel sectors, as presented previously \cite{Fajardy2018, Fajardy2020b}. This is illustrated in Figure \ref{fig: ESD}. 

The MONET framework, developed initially for BECCS, has been presented previously \cite{Fajardy2017,Fajardy2018,Fajardy2020b}. 
Appendix A briefly describes the key characteristics of the BECCS archetype, as it has been already implemented in the MONET framework, and further details the key characteristics of AR and DACCS archetypes.
Appendix B presents the mathematical formulation of the optimisation model, adapted from previous publications to include AR and DACCS archetypes, and Appendix C describes the recently added (or updated) datasets (\textit{i.e.,} land and geological CO$_2$ storage availabilities) used to constraint the optimisation model.

\subsection{Key optimisation constraints}

\subsubsection{Long-term CDR targets.}
\label{sec: CDR Targets}

Cumulative CDR targets consistent with the IPCC scenarios limiting global warming to 1.5$^{\circ}$C are selected in this study as follows \cite{IPCC2018,Huppmann2018}. 
The P3 scenario --- a middle-of-the-road scenario, in which societal and technological development follows historical trends --- is used in our reference scenarios.
The P4 scenario --- a fossil-fueled development scenario, in which economic growth and globalisation lead to the widespread adoption of greenhouse-gas-intensive lifestyle --- is used in sensitivity analysis scenarios, in which higher CDR targets are imposed (See Appendix E).
\newline
\indent However, because of the complexity and sophistication of IAMs, the spatial resolution of climate mitigation scenarios is necessarily limited, \textit{i.e.} the world is usually represented with a limited number of regions. 
Particularly, these regions don't exactly coincide with the ones considered in this study. 
Moreover, the different levels of CDR deployed in IAMs are the result of global and cost-optimal climate mitigation pathways, and therefore, they don't reflect on the responsibility for climate change of each nation,  nor on its capability to address it.
\newline
\indent For these reasons, we apply here a responsibility-based burden-sharing principle to allocate regional CDR targets, \textit{i.e.} to each region considered in the MONET framework \cite{Raupach2014}. 
In both IPCC scenarios (P3 or P4), global CDR targets are distributed in proportion to each region's cumulative historic GHG emissions \cite{Gutschow2016,Gutschow2021}.
This is presented in detail in Table \ref{tab: Cumu CDR targets}. 
Note that we don't intend to be prescriptive in our selection of the burden-sharing principle, but rather provide a proxy for a socio-economically fair regional distribution of 1.5$^{\circ}$C-consistent CDR targets.
Recognising that the distribution of the global CDR burden will likely be decided upon \textit{via} international negotiations rather than \textit{via} deterministic analytical approaches, we direct interested readers to Pozo \textit{et al.} \cite{Pozo2020} and references therein for a broader discussion of burden-sharing principles in the context of CDR.

\begin{table}[h]
\small
  \caption{Implications of the responsibility-based burden-sharing principle --- based on cumulative historic GHG emissions --- on the regional allocation of the IPCC P3 and P4 CDR targets in this study. The USA and the EU are the two largest GHG emitters here, on a cumulative-basis. They are allocated 21.3\% and 19.9\% of the IPCC CDR targets, respectively. Conversely, Brazil's historical GHG emissions are very low, and is allocated only 1.8\% of the IPCC CDR targets.}
  \label{tab: Cumu CDR targets}
  \begin{tabular*}{0.48\textwidth}{@{\extracolsep{\fill}}lllll}
    \hline
    Nations & \makecell[l]{Cumulative \\ GHG emissions \\ 1850-2019 \\(GtCO$_2$)$^a$} & \makecell[l]{Proportion \\ of CDR\\  targets\\ (\%)} & \makecell[l]{Cumulative \\P3 target\\ 2100\\ (GtCO$_2$)} & \makecell[l]{ Cumulative \\P4 target \\ 2100 \\(GtCO$_2$)} \\
    \hline
    Brazil & 47 & 1.8 & 7 & 21 \\
    China & 357 & 13.7 & 56 & 161 \\
    EU--28 & 521 & 19.9 & 81 & 235 \\
    India & 128 & 4.1 & 20 & 58 \\
    USA & 557 & 21.3 & 87 & 252 \\
    \hline
    \makecell[l]{Total \\MONET \\nations} & 1,610 & 61.6 & 251 & 727 \\
    \hline
    World & 2,612 & 100 & 408 & 1,179 \\
    \hline
    \multicolumn{5}{c}{\makecell[l]{$^a$ Cumulative historic GHG emissions excluding Land Use, Land Use\\ Change, and Forestry (LULUCF) between 1850--2019\cite{Gutschow2021}, as categorised \\ by the IPCC 2006 Guidelines for National Greenhouse Gas Inventories\cite{IPCC2006a}}}\\
  \end{tabular*}
\end{table}

Particularly, the 5 regions considered in this study are responsible for 62.4\% of the cumulative historic GHG emissions \cite{Gutschow2019}. Together, they also accounted for 50\% of the global population and 68\% of the global GDP in 2018\cite{WorldBank2020e}. Therefore, the case-study presented here can be reasonably considered representative of the international landscape, as well as the insights obtained here can be found valuable for policymakers in climate change mitigation. 

\subsubsection{CDR deployment rates.}
\label{sec: CDR Expansion}

The deployment of CDR options is limited here by lifetime-operating conditions and deployment rates.
We assume that BECCS and DACCS plants have a lifetime of 30 years. 
Following previous work, they are operating base-load \cite{Daggash2019,daggash2019higher,daggash2019implications,daggash2019structural}.
Conversely, AR has a "perpetual" lifetime, \textit{i.e.} once established, forests need to be maintained in perpetuity in order to avoid any reversal of CO$_2$ emissions back to the atmosphere.
\newline
\indent We assume a maximum build rate for BECCS plants of 2 GW/yr at the sub-region scale, based on the literature surveyed on energy system and climate mitigation strategy modelling \cite{Heaton2014}.
Note that if BECCS plants were maximally-deployed (\textit{i.e.,} as much is built as allowed by the build rate constraints), given an average BECCS CO$_2$ capture capacity of 4.2 MtCO$_2$/yr/plant\footnote{The CO$_2$ capture capacity of a BECCS plant is calculated here for a 500 MW dedicated biomass power plant, with a capture rate of 90\%, as presented previously \cite{Fajardy2020b}.}, this would be equivalent to 16.8 MtCO$_2$/yr at the sub-regional scale, and 2.8 GtCO$_2$/yr at the MONET scale.
\newline
\indent Because of the relative immaturity of the DAC technology, little build rate estimates can be found for DACCS in the literature.
To ensure fair comparison across CDR technologies, a maximum build rate for DAC plants of 16.8 MtCO$_2$/yr at the sub-regional scale is also used, \textit{i.e.} the same rate as BECCS. 
If both BECCS and DACCS were maximally-deployed, the maximum CO$_2$ capture capacity would thus be equivalent to 5.7 GtCO$_2$/yr at the MONET scale. 
\newline
\indent Based on a maximum worldwide deployment rate of 47 Mha/yr for AR reported in the IPCC SR15\footnote{After comparing all scenarios of the IPCC SR15 (P1, P2, P3 and P4), we found that the maximum deployment rate for AR was 47 Mha/yr, between 2020--2030, in the IPCC P2 scenario --- a sustainable development scenario\cite{Huppmann2018}.}, we downscaled this number to 8.5 Mha/yr at the MONET scale, then 50 kha/yr at sub-regional scale (equal sub-regional rates), using forest areas at both the global and MONET scales \cite{IPCC2018,MODIS_LC}.
For context, note that historical rates between 1990--2020 reported by the FAO are usually much lower, with afforestation rates of 2,095 kha/yr in China, 470 kha/yr in the EU, 274 kha/yr in India, and 245 kha/yr in the USA, and with a deforestation rate of 3,076 kha/yr in Brazil \cite{FAO2020}. 
\newline
\indent Recognising that the assumptions made here on maximum deployment rates are relatively optimistic in comparison to historical afforestation/deforestation rates, as well as owing to the highly speculative and non-commercial status of CDR options, we also run a sensitivity analysis on higher deployment rates (See Appendix E).

\subsubsection{Land \& Biomass availabilities.}

Sustainability criteria are also considered here, particularly for the deployment of land-based CDR solutions, such as AR and BECCS.
AR is limited by the availability of ecologically appealing areas with a potential for reforestation \cite{Griscom2017} (RP) (See Appendix C.2 for a detail overview of the dataset used here).
Biomass for BECCS is restricted to dedicated-energy crops (DEC) cultivated on marginal agricultural lands \cite{Cai2011} (MAL), and agricultural residues , particularly wheat straw, collected from harvested wheat areas \cite{Yu2020}. 
Finally, to avoid exacerbating potential water stress and creating or intensifying water scarcities, the cultivation of biomass for BECCS is further limited to areas with low water stress, \textit{i.e.} areas wherein the overall water risk is less than or equal to 3 on a 5-point scale \cite{GASSERT2015} as described previously \cite{Fajardy2018, Fajardy2020a, Fajardy2020b}. 
Therefore, the production of biomass for BECCS in our study has no negative impacts on the agricultural sector and its associated food supply.

\subsubsection{Geological CO$_2$ storage availability.}

Regional geological CO$_2$ storage availability and capacity are used here to constrain the deployment of geological CDR options, such as BECCS and DACCS.
Quantitative assessments of varying levels of detail were available for the USA \cite{USDOE2015} and China \cite{Li2009,Dahowski2009} at the sub-regional scale, and for the EU \cite{Vangkilde-Pedersen2009a,Vangkilde-Pedersen2009b,Poulsen2014,Gammer} at the national scale. 
However, with the exception of one quantitative study on the Campos Basin oil fields in Brazil \cite{Rockett2013}, only qualitative national assessments were identified for Brazil \cite{Ketzer2014} and India \cite{Holloway2009}.
Therefore, the reference scenarios presented in this study are based exclusively on quantitative data on geological CO$_2$ storage capacity (See Appendix C.1 for a detail overview of the CO$_2$ storage capacity datasets used here). 
\newline
\indent Recognising the current uncertainty surrounding CO$_2$ storage capacity and availability, especially the strong probability for CO$_2$ storage sites to exist both in Brazil and India, in spite of not being identified yet, we also run a sensitivity analysis on higher CO$_2$ storage availability and capacity, based on both quantitative and qualitative data (See Appendix E).

\subsection{Key metrics}
We used different metrics in this study to describe the cost-efficiency of the CDR pathways deployed in the different policy scenarios.

\subsubsection{Cumulative total net cost.}

The cumulative total net cost --- CTNC --- quantifies the total net investment --- for the deployment of any CDR pathway, as shown in Eq. \ref{eq: CTNC}. For the BECCS archetype, the CNTC is equal to BECCS total cost minus the revenues from electricity generation, For AR and DACCS archetypes, the CNTCs are equal to their total costs only.

\begin{flalign} 
\label{eq: CTNC}
\! CTNC(t) = CTC^{AR}(t) + CTNC^{BECCS}(t) + CTC^{DACCS}(t) & \qquad \forall \; t
\end{flalign}

\noindent where: $CTNC(t)$ is the cumulative total net cost of the CDR pathway until the year $t$ (\$); $CTC^{AR}(t)$ is the cumulative total cost of AR until the year $t$ (\$); $CTNC^{BECCS}(t)$ is the cumulative total net cost of BECCS -- total cost minus revenues from electricity generation -- until the year $t$ (\$); and $CTC^{DACCS}(t)$ is the cumulative total cost of DACCS until the year $t$ (\$). Note that $t \in \{2020, 2030, ..., 2100\}$, and by default, $CTNC(2020) = 0$.

\subsubsection{Cumulative net cost of CDR.}

The cumulative net cost of CDR -- CNC -- quantifies the averaged cost of deploying CDR, as shown in Eq \ref{eq: CNC}:

\begin{flalign} 
\label{eq: CNC}
CNC(t) = \frac{CTNC(t)}{CRCO2(t)} & \qquad \forall \; t
\end{flalign}

\noindent where: $CNC(t)$ is the cumulative net cost of CDR until the year $t$ (\$/tCO$_2$); $CTNC(t)$ is the cumulative total net cost of the CDR pathway deployed until the year $t$ (\$); and $CRCO2(t)$ is the cumulative total CDR until the year $t$ (tCO$_2$). Note that $t \in \{2020, 2030, ..., 2100\}$, and by default, $CTNC(2020) = 0$ and $CRCO2(2020) = 0$, therefore $CNC(2020) = 0$. 

\subsubsection{Marginal net cost of CDR.}

The marginal net cost of CDR -- MNC -- quantifies the actual/real cost of deploying CDR, as shown in Eq \ref{eq: MNC}:

\begin{flalign} 
\label{eq: MNC}
MNC(t) = \begin{dcases}
 \frac{CTNC(t)}{CRCO2(t)} & \forall \; t=2020 \\
\frac{CTNC(t) - CTNC(t-1)}{CRCO2(t) - CRCO2(t-1)} & \forall \; t>2020
\end{dcases}
\end{flalign}


\noindent where: $MNC(t)$ is the marginal net cost of CDR until the year $t$ (\$/tCO$_2$); $CTNC(t)$ is the cumulative total net cost of the CDR pathway deployed until the year $t$ (\$); and $CRCO2(t)$ is the cumulative total CDR until the year $t$ (tCO$_2$). Note that $t \in \{2020, 2030, ..., 2100\}$, and by default, $CTNC(2020) = 0$ and $CRCO2(2020) = 0$, therefore $MNC(2020) = 0$. 

\subsection{Alternative policy scenarios}
\label{subsec: Opt Scenarios}

The MONET framework is used here to determine the cost-optimal co-deployment of AR, BECCS and DACCS to deliver the Paris Agreement's 1.5$^{\circ}$C-consistent CDR objectives --- here, the IPCC P3 CDR targets \cite{IPCC2018, Huppmann2018} --- subject to the following alternative policy options:

\begin{list}{\labelitemi}{\leftmargin=1em}

    \item \textbf{International cooperation policy scenario:} 
    In this scenario (referred as COOPERATION scenario), CDR targets are pursued in an international cooperation paradigm.
    We assume that an international policy instrument (such as one of the carbon market approaches defined in Articles 6.2 and 6.4 of the Paris Agreement \cite{UNFCCC2015}) has been developed, allowing regions to share the effort to meet 1.5$^{\circ}$C-consistent CDR targets.
    Therefore, the regions considered in this study can meet the cumulative CDR targets together, based on their shared (as opposed to individual) responsibility for climate change (See Section \ref{sec: CDR Targets}).
    They can also trade bio-geophysical resources, particularly biomass, and therefore deploying inter-regional biomass supply chains for BECCS.
    
    \item \textbf{"Current policy" scenario:} 
    In this scenario (referred as CURRENT POLICY scenario), CDR targets are pursued in a climate policy paradigm envisaged by the current policy landscape.
    For context, domestic emissions trading systems (ETS), such as the EU ETS, the UK ETS, or the California (USA) ETS, are currently creating incentives to reduce CO$_2$ emissions \textit{via} a "cap-and-trade" principle. 
    However, these ETS are not linked, \textit{i.e.} there are no bilateral or multilateral transfers between them, and negative emissions are not yet integrated within them.
    In light of this, we assume that the regions considered in this study must meet individual cumulative CDR targets, based on their respective responsibilities for climate change (See Section \ref{sec: CDR Targets}). 
    Bio-geophysical resources, particularly biomass, can still be traded inter-regionally (\textit{i.e.,} local or imported biomass for BECCS).
    This scenario is equivalent to an international climate policy landscape in which CDR has been incorporated into domestic ETS, but cannot be transferred from one to another.
    
    \item \textbf{"National isolation policy" scenario:} 
    In this scenario (referred as ISOLATION scenario), the 1.5$^{\circ}$C-consistent CDR targets are pursued in a national isolation paradigm.
    We assume that no international policy instrument framework has been developed to distribute the effort to meet the 1.5$^{\circ}$C-consistent CDR targets (\textit{i.e.,} individual CDR targets), and that there is no  inter-regional trading of bio-geophysical resources (\textit{i.e.,} only local biomass supply chains for BECCS).
\end{list}

\section{Optimal co-deployment of CDR options}
\label{sec: Opt CDR}

Here, we identify the deployment of cost-optimal CDR pathways under the 3 policy scenarios described in Section \ref{subsec: Opt Scenarios} in order to deliver CDR targets that are consistent with the Paris Agreement's 1.5$^{\circ}$C objective. 
As discussed, we use the IPCC P3 CDR targets \cite{IPCC2018,Huppmann2018} for these reference scenarios. 
The composition of these CDR pathways --- AR, BECCS and/or DACCS --- and their spatio-temporal evolution --- between 2020--2100 and across Brazil, China, the EU, India and the USA --- are discussed in this section.

\begin{figure*}[t!]
\centering
  \includegraphics[width=17.1cm]{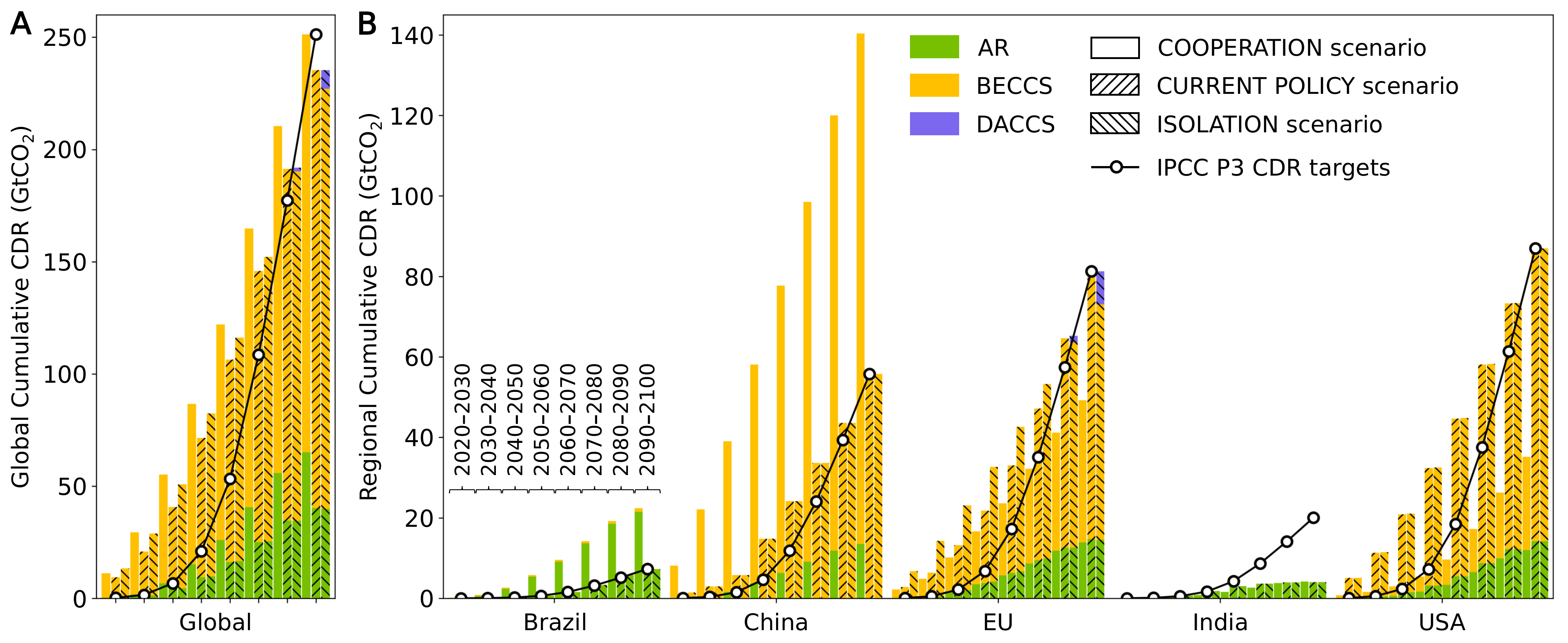}
  \includegraphics[width=17.1cm]{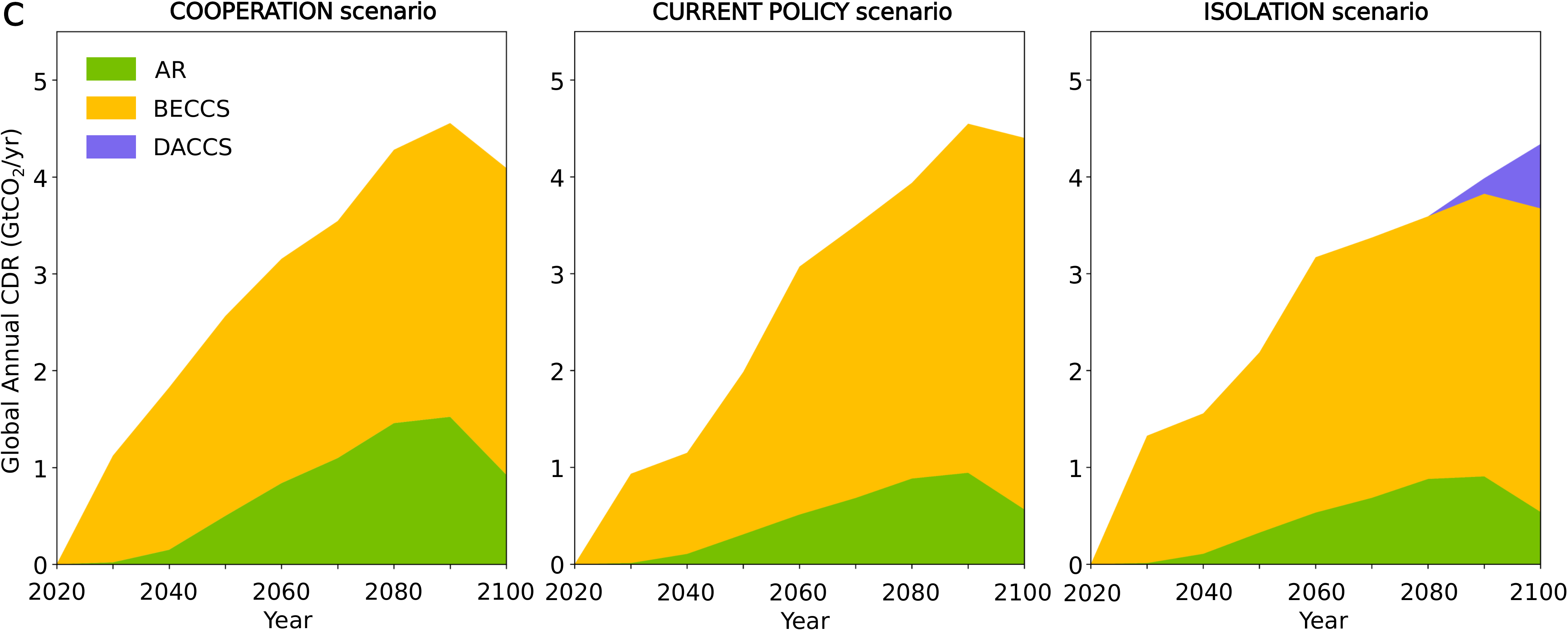}
  \caption{Cost-optimal CO$_2$ removal from 2020 to 2100, for each CDR option and for each region, under alternative P3-consistent policy scenarios: A) globally and cumulatively, B) regionally and cumulatively, and C) globally and annually. With international cooperation in climate mitigation policy, CDR options (\textit{i.e.,} BECCS and AR) are aggressively deployed in most cost-efficient regions --- in this study, in China. As a result, the deployment of CDR differs significantly from the regional distribution of the IPCC P3 CDR targets (consistent with the Paris Agreement's 1.5$^{\circ}$C objective) based on the responsibility-based burden-sharing principle, used here as a proxy for fair allocation. With little or no cooperation, CDR options are deployed in regions where individual targets are the greatest --- in the USA, the EU and China --- but the global 2100 CDR target is missed by 14 GtCO$_2$ in India, owing to the lack of available and appropriate bio-geophysical resources (\textit{i.e.,} geological and biogenic CO2 sinks). With no cooperation at all, DACCS is deployed in the EU and the USA to overcome the exhausted local biomass supply for BECCS.}
  \label{fig: Opt P3 CDR cumu}
\end{figure*}

\subsection{The international cooperation policy paradigm} 

In the COOPERATION scenario, CDR is successfully delivered at the Paris Agreement’s 1.5$^{\circ}$C scale by 2100. This is achieved \textit{via} BECCS mainly, with 186 GtCO$_2$ (74\%), and AR, with 65 GtCO$_2$ (26\%) (Fig \ref{fig: Opt P3 CDR cumu}-A). 
\newline
\indent Given the anticipated large scale of CO$_2$ removal over the century (\textit{i.e.,} high CDR targets increasing over time), constrained here by maximum deployment rates of the different CDR options, we find that the prompt deployment of the CDR pathway, starting in the 2020s, is required to deliver the Paris Agreement's CDR objectives by 2100. Particularly, the amount of CDR achieved is systematically greater than pre-2100 CDR targets.
As illustrated in Fig. \ref{fig: Opt P3 CDR cumu}-C, BECCS starts delivering CO$_2$ removal straightaway, and increasingly up to 3.2 GtCO$_2$/yr in 2100. 
This is equivalent to 154 GW of BECCS capacity.
For context, this is 4\% of the current electricity capacity of China, the EU and the USA, all together (2,200 GW in China in 2020, 1,117 GW in the USA in 2020 and 946 GW in the EU-28 in 2019) \cite{CEC2022,EIA2022,EUComission2022}. 
\linebreak
\indent Importantly, AR is also deployed in the early 2020s, but its CO$_2$ removal is delayed owing to the period of time required for trees to grow (See Appendix A for a description of the AR model used here). 
Then, because of a combination of CO$_2$ sinks saturation, \textit{i.e.} trees reach maturity and hit their maximum CDR potential, and optimisation edge effect, \textit{i.e.} trees planted after 2070 would only play an important role in the 22$^{\mathrm{nd}}$ century but not before, AR’s CO$_2$ removal peaks at a rate of approximately 1.5 GtCO$_2$/yr in 2090 and falls substantially thereafter. 
\linebreak
\indent Note that in the COOPERATION scenario (as well as in any other reference scenario), whilst BECCS plants are rarely maximally-deployed, \textit{i.e.} as much is built as allowed by the build rate constraints, AR deployment is constrained by its maximum deployment rate, assumed here to be 50 kha/yr/sub-region. 
However, the sensitivity analysis carried out in Appendix E showed that higher AR deployment rates would only increase its CO$_2$ removal moderately, due to the exhaustion of land availability.
 \linebreak
\indent Overall, forward planning and strategic deployment of the different CDR options is thus key to deliver the Paris Agreement's 1.5$^{\circ}$C ambition. 
Whilst all CDR options have specific techno-economic and sustainability characteristics, which influence the rate and scale at which they can be deployed, they can also be distinguished by when they start to capture and remove CO$_2$, and how long they remove and store CO$_2$.
CO$_2$ removal efficiency, timing and permanence will certainly have to be carefully and clearly accounted for, when deploying the different CDR options.
\newline
\newline
\indent We also find that the spatial deployment of the CDR pathway differs from a CDR option to another.
As shown in Fig. \ref{fig: Opt P3 CDR cumu}-B, there is no silver-bullet to meet the Paris Agreement's CDR objectives, as the optimal portfolio of CDR options within a given region, or even sub-region, varies around the world (See Appendix D.1 for a detailed overview of the CDR pathway at the sub-regional scale). 
\linebreak
\indent AR can be deployed in most parts of the world, given the combination of available land and appropriate climate --- a balance between warm temperature and humidity. 
Specifically, between 12--21.5 GtCO$_2$ by 2100 are removed \textit{via} AR in most regions (Brazil, China, the EU and the USA) whereas only 4 GtCO$_2$ by 2100 are removed in India, where, in spite of the availability of land and good climates, they usually do not coincide.
\linebreak
\indent Conversely, because of its complex value chain, we observe that BECCS deployment is more localized than AR deployment, owing to the combination of several bio-geophysical and economic factors: well-characterised CO$_2$ storage capacity; cost-effective biomass supply (\textit{i.e.,} high MAL availability and DEC yields, high agricultural residues availability, and low-cost biomass production); and affordable CO$_2$ transport \& storage (T\&S) infrastructures.
Moreover, regions or sub-regions where electricity prices are anticipated to be high (as assumed here) also prove to be more advantageous for BECCS deployment, as they benefit from higher revenues from electricity generation. 
Specifically, by 2100, 68\% of CDR achieved \textit{via} BECCS is deployed in China, with a further 19\% and 12.5\% deployed in the EU and USA, respectively, and 0.5\% in Brazil.
\linebreak
\indent As illustrated in Fig. \ref{fig: Opt P3 Cumu pellet}, where possible, local biomass supply chains are prioritised --- 100 Gt$_{\mathrm{DM}}$ of pellets (85\%) by 2100, of which 63 Gt$_{\mathrm{DM}}$ are in China alone.
Remaining imported biomass supply chains mostly originate in regions with limited or no CO$_2$ storage capacity (as identified here), such as Brazil and India. 
These regions contribute thus indirectly to the delivering of the Paris Agreement's CDR objectives through the supply of biomass to other regions, similarly to the current and incumbent global biomass trade.
\linebreak
\indent Note that DACCS is not deployed in this scenario, owing to its significant higher cost, compared to AR and BECCS. For DACCS to become cost-competitive, current average costs should decrease to below \$100/tCO$_2$, which is equivalent to a cost reduction of around 60--70\% for liquid solvent DAC technologies\cite{Keith2018}, and 90\% for solid sorbent technologies\cite{Climeworks2018} (See Appendix D.2).
\linebreak
\indent Overall, adapting the spatial deployment of the CDR pathway to each CDR option is thus also key to deliver the Paris Agreement's 1.5$^{\circ}$C ambition.
\newline
\newline
\indent Finally, we find that CO$_2$ removal is achieved at lowest cost in the COOPERATION scenario, with a CNC of \$57/tCO$_2$ by 2100.
This is because AR and BECCS can be deployed most cost-efficiently, \textit{i.e.} without being restricted regionally by individual CDR targets, such as in Brazil and China in the COOPERATION scenario, or conversely without being "over-deployed", \textit{i.e.} imposed by individual CDR targets and therefore less cost-efficiently, such as in the EU and the USA in the CURRENT POLICY and ISOLATION scenarios.

\subsection{The current policy paradigm}

\begin{figure*}[t]
\centering
  \includegraphics[width=17.1cm]{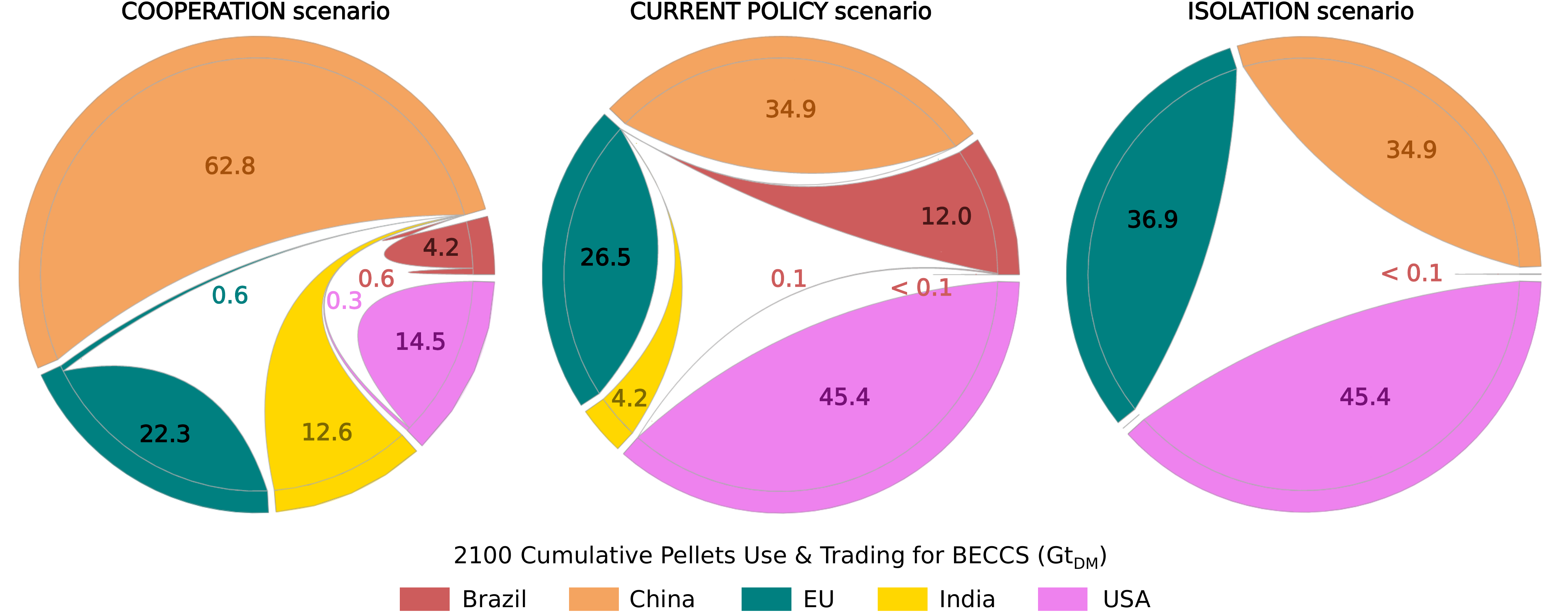}
  \caption{Cost-optimal cumulative pellet trading for BECCS by 2100, in all P3-consistent policy scenarios. Local biomass supply chains are prioritised, even when biomass trading is allowed (in the COOPERATION and CURRENT POLICY scenarios).  When international biomass is allowed, imported biomass supply chains mostly originate in Brazil and India -- where little or no CO$_2$ storage sites have been identified (and considered here).  With international cooperation in climate mitigation policy (in the COOPERATION scenario), imported biomass is shipped to China, where most of the CO$_2$ is removed from the atmosphere by 2100.  With limited international cooperation (in the CURRENT POLICY scenario), imported biomass is shipped to the EU , or used locally in China, where individual CDR targets must be met domestically by 2100. Without cooperation (in the ISOLATION scenario), biomass is used locally in the USA, the EU and China, where individual CDR targets must be met domestically by 2100. Where there is limited or no CO$_2$ storage, such as in Brazil and India, limited or no biomass can be used for BECCS.}
  \label{fig: Opt P3 Cumu pellet}
\end{figure*}

In the CURRENT POLICY scenario, the likelihood of delivering the Paris Agreement is reduced because the extent to which each region can remove domestically CO$_2$ from the atmosphere is limited by its own bio-geophysical sinks for CO$_2$, \textit{i.e.} geological sites for BECCS and DACCS and land for AR. 
As in the COOPERATION scenario, the CDR pathway deployed by 2100 is still composed of BECCS mainly, whose contribution increased to 195 GtCO$_2$ (83\%), and of AR, whose contribution decreased to 40 GtCO$_2$ (17\%), but the 2100 CDR target is missed by 16 GtCO$_2$.
\linebreak
\indent As shown in Fig \ref{fig: Opt P3 CDR cumu}-B, most regions successfully reach their individual CDR targets, \textit{via} exclusively AR in Brazil or BECCS in China, or \textit{via} both in the EU and the USA for instance.
Note that DACCS is also not deployed in this scenario. For it to become cost-competitive with AR and BECCS, current average costs should decrease to below \$170/tCO$_2$ (See Appendix D.2).
The EU, and in a much lesser extend the USA as well, rely partly on biomass supply originating in Brazil and India in order to achieve CDR \textit{via} BECCS (Fig. \ref{fig: Opt P3 Cumu pellet}).
Specifically, 38\% and 0.2\% the pellet use cumulatively by 2100 in the EU and the USA, respectively, are imported from Brazil mostly, and India.
This is because domestic biomass supply becomes critical and almost exhausted, and imported biomass proves then less costly.
\linebreak
\indent However, a few regions, such as India, are much less well endowed with domestic CO$_2$ sinks, yet required to meet individually CDR targets, and therefore do not meet their individual CDR targets by the end of the century.
By lack of international cooperation or other alternative CDR solutions, individual CDR targets are missed in India by 16 GtCO$_2$ by 2100.
Note that, given the size of India, the existence of CO$_2$ storage sites within its borders is very likely. 
Such CO$_2$ storage sites would first need to be identified and assessed before deploying geological CDR options, such as BECCS or DACCS, but then, they could increase India's likelihood to reach its individual CDR targets (See Appendix E for a sensitivity analysis on higher CO$_2$ storage availability).
\linebreak
\indent Finally, because of the mismatch between a socio-economically fair regional distribution of the Paris Agreement’s CDR objectives, and the regional availability of bio-geophysical sinks for CO$_2$, we find that the CNC of CO$_2$ removal by 2100 increases by 45\% relatively to the COOPERATION scenario.
\linebreak
\indent Thus, this scenario highlights that delivering the Paris Agreement's 1.5$^{\circ}$C ambition in a climate mitigation policy paradigm similar to what is envisioned in the current policy landscape is less likely than in an international cooperation policy paradigm, and will certainly increase the Paris Agreement's financial burden.

\subsection{The national isolation policy paradigm}

In the ISOLATION scenario, the likelihood of the Paris Agreement's CDR objectives is reduced even further than in the CURRENT POLICY scenario.
Not only is the 2100 CDR target still missed by 16 GtCO$_2$, but the CNC of CO$_2$ removal by 2100 increases by 69\% relatively to the COOPERATION scenario.
As shown in Fig. \ref{fig: Opt P3 CDR cumu}-A, 80\% of CDR is removed from the atmosphere \textit{via} BECCS, 17\% \textit{via} AR, and also 4\% \textit{via} DACCS.
\linebreak
\indent Because BECCS is deployed exclusively using local biomass, domestic biomass supply becomes rapidly exhausted in the EU and the USA. 
Specifically, 83\% and 100\% of the available land in the USA and the EU (MAL and harvested wheat areas), respectively, are already allocated to biomass production for BECCS by 2030. 
This leads therefore to the deployment of DACCS in the EU from 2080 and up to 0.7 GtCO$_2$/yr in 2100.
\linebreak
\indent Besides increasing the Paris Agreement’s financial burden, the energy consumed annually to deploy DACCS by 2100 is approximately 265 TWh in the ISOLATION scenario.
This is equivalent to 8\% of the EU current electricity production (3,275 GWh in the EU-28 in 2018) \cite{IEA2018}.
\linebreak
\indent Thus, this scenario shows that delivering the Paris Agreement's 1.5$^{\circ}$C ambition in an isolation policy paradigm is highly unlikely. It will not only be more expensive than in an international cooperation policy paradigm, but also certainly more energy-intensive, and could thus compromise the sustainability of the CDR pathway deployed.

\begin{figure*}[t]
\centering
  \includegraphics[width=17.1cm]{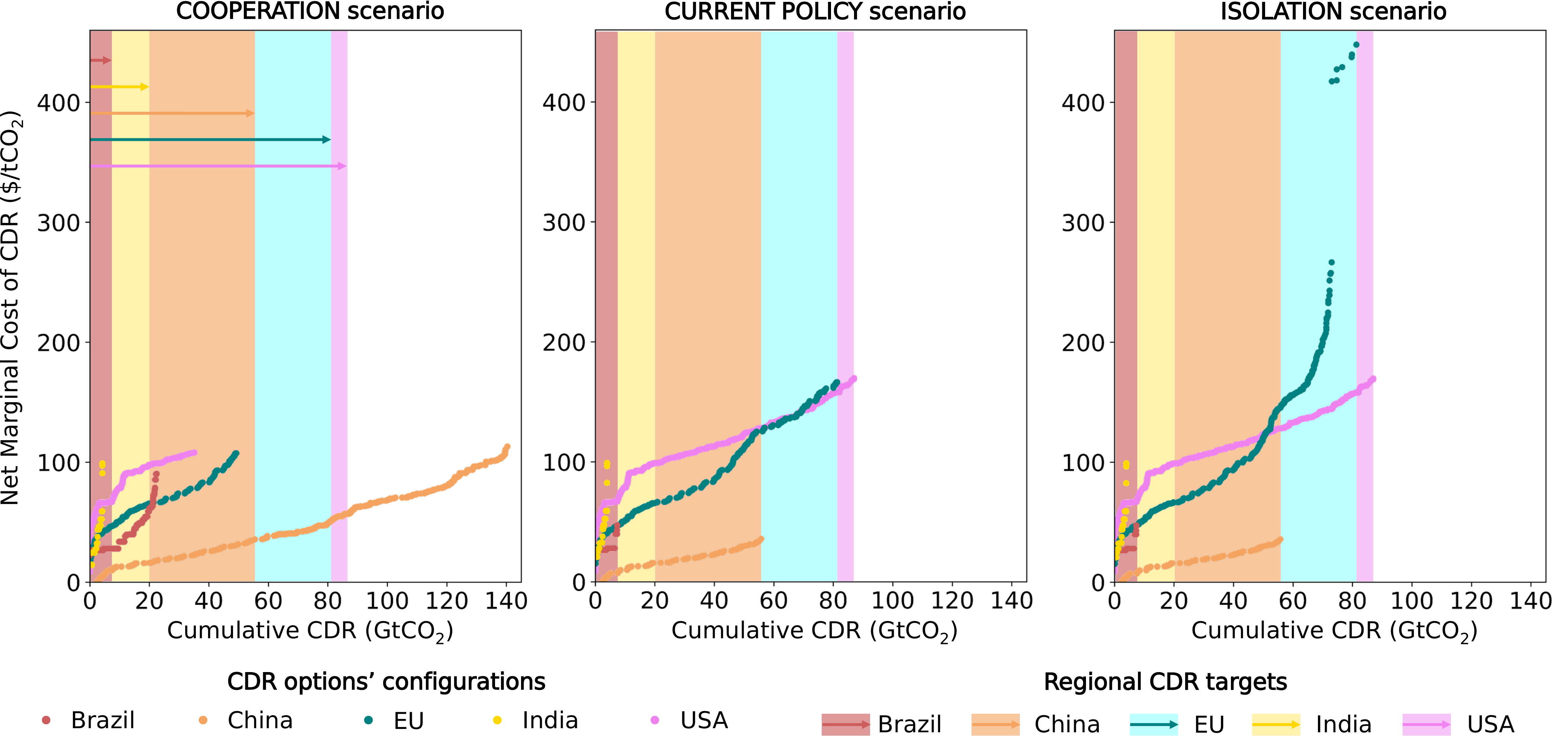}
  \caption{Regional cost supply curves and 2100 CDR targets under alternative P3-consistent policy scenarios. The marginal net cost of CDR appears on the vertical axis, while the CO$_2$ removal achieved and the 2100 CDR targets appear on the horizontal axis. 2100 CDR targets consistent with the IPCC P3 scenario are ordered from the smallest --- 7 GtCO$_2$ in Brazil --- to the greatest --- 87 GtCO$_2$ in the USA. China delivers the most cost-efficient CDR -- China's cost supply curve is below all the other regions' cost supply curves --- and therefore the most CDR surplus in the COOPERATION scenario. Brazil also provides CDR surplus, whereas the EU, the USA ad India benefit from CDR surplus in this scenario. In the EU and the USA's cases, this results from a national cost-efficiency strategy -- the EU and the USA can successfully meet their national CDR targets in the CURRENT POLICY and ISOLATION scenarios, yet at higher (marginal net) cost --- whereas in the case of India, this is because of a lack of national bio-geophysical resources --- India cannot meet its national CDR targets in the CURRENT POLICY and ISOLATION scenarios.}
  \label{fig: Opt P3 Cost Supply}
\end{figure*}

\subsection{The benefits of international cooperation policy}
Fig. \ref{fig: Opt P3 CDR cumu}-B showed that the socio-economically fair distribution of the IPCC P3 CDR targets at the regional scale, based here on the responsibility for climate change, differs greatly from the deployment of the most cost-optimal CDR pathway achieved with international cooperation.
\linebreak
\indent In the COOPERATION scenario, more than half of the CO$_2$ removal by 2100 is delivered in China, in this study the most cost-effective region to achieve CDR. 
Specifically, 140 GtCO$_2$ (56\% of the global CDR achieved by 2100) are removed in China by 2100 in this scenario, \textit{via} BECCS and AR, whereas China's individual CDR target is only 56 GtCO$_2$ (22\% of the global 2100 CDR target).
Brazil also provides an additional 15 GtCO$_2$ over and above its individual CDR target by 2100, almost exclusively \textit{via} AR. 
\linebreak
\indent Establishing of a policy instrument that enables international cooperation while fulfilling the Paris Agreement's CDR objectives, \textit{via} the international trade of negative emissions for instance, is therefore found to substantially reduces the financial burden of the Paris Agreement by 51--69\%.  
Whilst AR's CNC of CDR remains approximately constant in all scenarios (See Appendix D.2 for a detailed overview of the MNC of CDR achieved \textit{via} all CDR options, in all scenarios) --- this CDR solution is used to its full extent in all scenarios ---, BECCS becomes significantly more costly in the CURRENT POLICY and ISOLATION scenarios.
This is because the most cost-efficient BECCS value chains become more difficult to deploy, owing to the requirement to meet individual CDR targets domestically.
In the ISOLATION scenario, the complete lack of international cooperation forces some regions to resort to more costly CDR options such as DACCS, and thus drives the CNC of the CDR pathway deployed the most up.
\linebreak
\indent Thus, international cooperation policy is key in delivering the Paris Agreement's 1.5$^{\circ}$C ambition in the most cost-efficient manner as it allows for the deployment of a CDR pathway in least-cost and most "CDR-efficient" regions while still delivering individual CDR targets.

\section{Assessing the value of a negative emissions trading system}
\label{sec: Neg ETS}

Section \ref{sec: Opt CDR} highlighted that, despite the feasibility of the Paris Agreement's 1.5$^{\circ}$C ambition, CDR pathways reflecting either current policy or national isolation policy paradigms may not only be more challenging and expensive, but also less likely to meet the Paris Agreement's CDR objectives than CDR pathways deployed in an international cooperation policy paradigm.
\linebreak
\indent Because CO$_2$ sinks are unevenly distributed across the world, the most cost-optimal CDR pathway, here in the COOPERATION scenario, differs greatly from individual CDR targets. 
It can therefore be considered unfair, particularly towards regions achieving CDR over and above their individual CDR targets.
Conversely, fair CDR pathways aligned with individual CDR targets, here in the CURRENT POLICY and ISOLATION scenarios, must generate negative emissions domestically in each region, regardless of the cost or the availability of CO$_2$ sinks within the region's borders. 
\linebreak
\indent As a result, they are less cost-efficient, and therefore more costly.
Integrating CDR options within an international negative emissions trading system would address the trade-off between cost-efficiency and equity by allowing regions that are well endowed with bio-geophysical CO$_2$ sinks, and therefore CDR potentials, to trade CDR surplus (\textit{i.e.,} the additional CO$_2$ removal achieved over and above any given CDR target) with other regions for which delivering CDR is more difficult, or more costly. 

\subsection{Introducing the concept of a negative emissions trading system}

Conceptually, a negative emissions trading system would work in a reverse way to any current existing ETS.
An inter-regional or international negative emissions target would be set to be met, and increased over a given time-period, following the IPCC P3 cumulative CDR targets for instance.
Regions would then be allocated a share of this global negative emissions target, based on responsibility for climate change for instance. 
In such trading system, "verified" \footnote{The monitoring, reporting, and verification challenges implicit in delivering this verification step are non-trivial, and will vary for each CDR option. For simplicity, we do not address this point further in this study, leaving it for future work.} 
CO$_2$ removal would generate negative emissions credits (NECs) that could be traded between regions as required to meet individual negative emissions targets.
"NECs provider" regions could therefore provide CDR surplus and sell NECs to "NECs beneficiary" regions that, themselves, could benefit from CDR surplus --- either because they could not meet their individual CDR targets domestically, or because they would find it less expensive --- and buy NECs from "NECs provider" regions.
\linebreak
\indent The negative emissions price (NEP) --- the price at which NECs are traded between regions --- could, in theory, be as low as the marginal net cost of generating CDR surplus in "NECs provider" regions. 
Admittedly optimistic, this approach is used here to demonstrate the potential role and value of a NE trading system. 
Investigating further cost-sharing approaches in the context of CDR, while fulfilling the Paris Agreement’s 1.5$^{\circ}$C ambition, is left for future work.
\linebreak
\indent In this study, we assume the existence of such above-mentioned negative emissions trading system, within which all CDR surplus can be traded at a unique NEP, calculated as the averaged marginal net cost of CDR surplus. 
This is shown in Eq. \ref{eq: NEP}:

\begin{equation} 
\label{eq: NEP}
NEP = \frac{\sum_{i, k} MNC(i,k) \times CDRSurplus(i,k)}{\sum_{i, k} CDRSurplus(i,k)}
\end{equation}

\noindent where $CDRSurplus(i,k)$ is the CDR surplus (tCO$_2$) achieved by the k$^{th}$ configuration of a given CDR option deployed in a region i, and MNC$_{i,k}$ is the marginal net cost (\$/tCO$_2$) at which this configuration is deployed.

\subsection{The value of a negative emissions trading system}

Fig. \ref{fig: Opt P3 Cost Supply} shows how the marginal net cost of CDR in all regions increases as more CDR is achieved over the century, for all scenarios.
China and Brazil provide CDR surplus on behalf of the EU, the USA, and India --- 85 and 15 GtCO$_2$, respectively --- in the COOPERATION scenario. 
Particularly, in the case of China, this is twice as much as China's individual CDR target by 2100.
Assuming that the cheapest CDR options are first deployed to meet individual CDR targets and then the more expensive ones are developed to provide CDR surplus, the MNCs of CDR surplus range between \$36--113/tCO$_2$ in China and \$28--90/tCO$_2$ in Brazil.
Using Eq. \ref{eq: NEP}, this results in a NEP of \$64/tCO$_2$, paid by the USA, the EU and India to Brazil and China in order to benefit from CDR surplus --- 52, 32 and 16 GtCO$_2$, respectively --- in the COOPERATION scenario. 
\newline
\begin{figure}[t]
\centering
 \includegraphics[width=8.3cm]{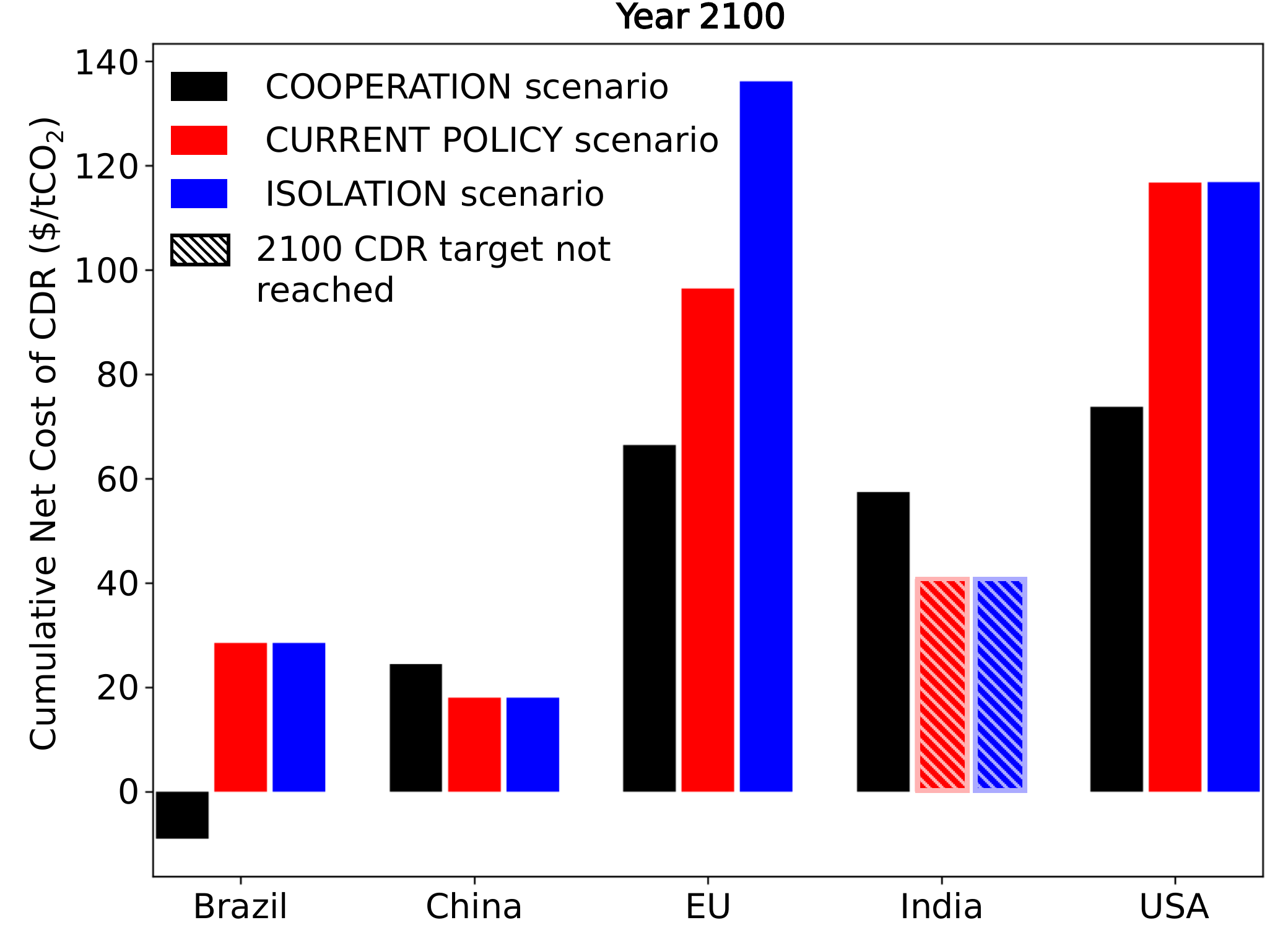}
  \caption{Regional cumulative total net costs (CNCs) by 2100, including NECs trading, for all P3-consistent policy scenarios. NECs trading allows for the most cost-efficient distribution of the global CDR targets by 2100, in all regions, with the exception of India (in the COOPERATION scenario). This is because India can't reach its individual CDR targets in the CURRENT POLICY and ISOLATION scenarios.}
  \label{fig: Opt P3 Cost Value}
\end{figure}
\indent Fig. \ref{fig: Opt P3 Cost Value} compares the regional CNCs of CO$_2$ removal by 2100 under the different policy options --- in the COOPERATION scenario, including NECs trading, and in the CURRENT POLICY and ISOLATION scenarios.
We find that regions benefiting from CDR surplus with international cooperation, yet capable of meeting domestically their individual CDR targets, deliver the Paris Agreements' CDR objectives at least cost in the COOPERATION scenario, owing to NECs trading between regions.
Specifically, the CNCs of CDR by 2100 are \$66/tCO$_2$ and \$74/tCO$_2$ in the EU and the USA, respectively, in the COOPERATION scenario (including NECs trading). 
In the CURRENT POLICY and ISOLATION scenarios, these CNCs increase by 45--105\% in the EU and 58\% in the USA, respectively.
The EU and the USA are thus, here, archetypal "independent NECs beneficiary" regions, as trading NECs result from a cost-optimal national strategy.
\linebreak
\indent However, we observe that regions relying on CDR surplus to meet individual CDR targets deliver the Paris Agreement's CDR objectives at highest cost.
This is because CDR objectives are missed with less or no international cooperation, and require therefore additional costs from the purchase of NECs to be delivered with international cooperation.
Specifically, the CNC of CDR by 2100 is \$57/tCO$_2$ in India in the COOPERATION scenario (including NECs trading). 
India is thus an archetypal "dependent NECs beneficiary" region as it relies essentially, here, on NECs trading with "NECs provider" regions, such as Brazil or China, to meet its individual CDR targets.
\linebreak
\indent Nonetheless, we recognise that regions such as India contribute to the delivering of the Paris Agreement in the COOPERATION scenario, particularly by producing and exporting biomass for BECCS's deployment in China (See Fig. \ref{fig: Opt P3 Cumu pellet}).
This contribution is neglected economically here, but would be expected to be remunerated appropriately in reality, and therefore constitutes an incentive to participate to such international negative emissions trading system.
\newline
\indent Finally, we observe that regions providing CDR surplus can deliver the Paris Agreement's CDR objectives at similar cost or benefit from a great cost reduction.
For instance, because China contributes mostly to the global CDR surplus provided and based on the assumption made here on a unique NEP, the NEP calculated here is very close to the averaged MNC of all CDR surplus delivered in China (67\$/tCO$_2$ on average). Therefore, China amortizes its additional costs for providing CDR surplus, and its CNC of CDR by 2100 of \$24.5/tCO$_2$ in the COOPERATION scenario (including NECs trading) is similar to the CNCs in the other scenarios.
Brazil also provide CDR surplus, but in a lesser extent and less cost-efficiently. 
Therefore, it benefits from a NEP that is 27\% lower than the averaged MNC of all CDR surplus delivered in Brazil (\$46/tCO$_2$ on average), leading therefore to negative net costs, \textit{i.e.}  net revenues, by 2100. 
The CNC of CDR by 2100 in Brazil of -\$9/tCO$_2$ is 132\% lower than in the CURRENT POLICY and ISOLATION scenarios, respectively.
\linebreak
\indent Importantly, acknowledging that the current policy landscape is characterised by international competition, higher NEPs could reduce even more the cost of CDR for "NECs provider" regions, leading to higher negative net revenues, and therefore making the delivering of CDR surplus an economically viable climate mitigation action.
\newline
\indent Thus, these results emphasise the value of international cooperation \textit{via} the deployment of an international policy instrument, such as a negative emissions trading system, and the need for robust institutions to enable monitoring, verification and accreditation of NECs.

\section{Does time matter?}
\label{sec: Switch Policy}

\begin{figure}[t]
\centering
 \includegraphics[width=8.3cm]{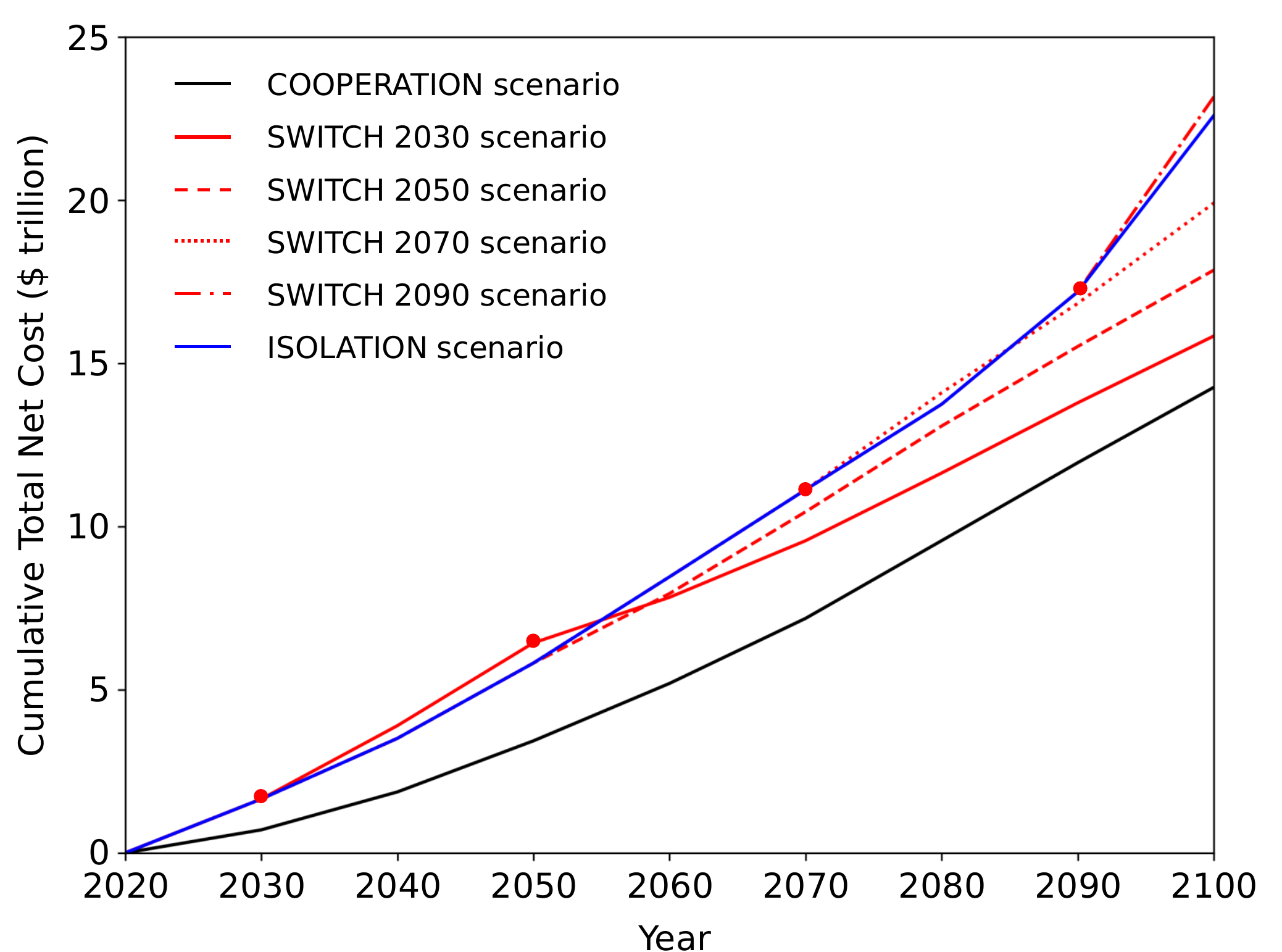}
  \caption{Cumulative total net costs (CTNCs) between 2030 and 2100 for alternative P3-consistent policy scenarios. SWITCH 20XX scenarios involve the adoption of an international cooperation policy in 20XX, \textit{i.e.} a “switch” from isolation policy to cooperation policy in 20XX, with 20XX being 2030, 2050, 2070 and 2090. Delaying international cooperation policy always results in the deployment of more costly CDR pathways.}
  \label{fig: P3 CNCs Switch}
\end{figure}

The previous sections established that international cooperation is key in delivering the Paris Agreement's 1.5$^{\circ}$C ambition most cost-efficiently.
Because the current international policy landscape on climate change mitigation is closer to an "isolation" policy paradigm than a "cooperation" policy paradigm, we investigate here the implications of developing such international cooperative approach, and we discuss the impacts of delaying its development.
\newline
\indent Figure \ref{fig: P3 CNCs Switch} shows how the total cost of CDR deployment at the Paris Agreement's 1.5$^{\circ}$C scale evolves, as the adoption of an international policy is further delayed. 
The COOPERATION scenario and the ISOLATION scenario, described previously, represent respectively the "earliest" and "latest" scenarios.
We find that delaying the adoption of an international cooperation policy always results in a more expensive deployment of CDR options --- between 11\% and 62\% more than the CTNC by 2100 in the COOPERATION scenario, as the delay increases.
\linebreak
\indent Whilst we acknowledge that it is already too late to deploy CDR most cost-efficiently, as suggested in the "earliest" scenario (the COOPERATION scenario) here --- it implies the possibility to trade internationally negative emissions as early as in the 2020s ---, there is an imperative to develop the necessary geopolitical and economic instruments as soon as possible. 
\newline
\indent Aspiring to synchronise the establishment and development of such instruments around the world is highly ambitious, but starting from existing ETS at the inter-regional, national, or even regional scales, such as the EU ETS, the UK ETS or the California (USA) ETS, could show the way and set the basics for a future inter-regional, and possibly international instrument for negative emissions trading.
\linebreak
\indent Importantly, the Article 6 of the Paris Agreement established a framework for market-based approaches \cite{UNFCCC2015} (through Article 6.2 and Article 6.4), within which direct references to negative emissions (\textit{i.e.,} “emission removals”) are made. 
Negative emissions could thus, for instance, be included directly within domestic ETS and then transferred between domestics ETS, as suggested by Article 6.2, or directly included and traded within an international market, as suggested by Article 6.4. Note that these examples are not intended to be prescriptive, nor exhaustive.
\newline
\indent These results highlight thus that, the longer we wait to establish and develop geopolitical and economic instruments for fostering international cooperation in climate change mitigation, particularly in deploying of CDR, the greater the costs will be. 
Given the decreased likelihood of meeting the Paris Agreement’s CDR objectives without international cooperation, as set out here, the greater the associated adaptation costs will be, and thus the overall challenges to be addressed in the near-future.

\section{Conclusions}
\label{sec: Conclusions}

In line with the Paris Agreement, limiting global warming to 1.5$^{\circ}$C will require CDR to be deployed at large-scale between now and the end of the century, as indicated in IAMs scenarios.
Beside NDCs, as most UNFCCC Parties have pledged to reach net-zero by mid-century, it is important to understand what CDR pathways might look like, as well as when and where they might be sustainably and affordably deployed. 
Moreover, as might be anticipated, the climate mitigation policy landscape will significantly impact the spatio-temporal composition and evolution of the CDR pathways deployed.
As the Article 6 of the Paris Agreement established a framework for carbon market-based approaches within the context of climate change mitigation, it is all the more crucial to obtain insights on how international policy would impact such deployment.
\newline 
\indent 
This study has shown that international cooperation in climate change policy is key for deploying the most cost-optimal CDR pathway to deliver the Paris Agreement 1.5$^{\circ}$C ambition.
With international cooperation, this CDR pathway is preferably composed of BECCS, mainly (78\%) and AR (22\%). 
Given the large scale of CO$_2$ removal over the century, we found that CDR options must be promptly deployed --- particularly in the case of AR, owing to the time required for trees to grow.
Thus, for sustainability and feasibility safeguards, forward planning and strategic deployment of the different CDR options is crucial to deliver the Paris Agreement's 1.5$^{\circ}$C ambition.
\linebreak
\indent More broadly, the issue of time emerges as a contingent to the discussion on CDR options, such as biochar or EW. 
Whilst in the case of biochar, the carbon content slowly decays with time --- issue of permanence --- as a function of a soil conditions and biochar characteristics, there will typically be a delay, in the case of EW, between the time that the minerals are exposed to the atmosphere and when the carbonation reaction is completed --- issue of timing.
In this case, this delay is a function of soil conditions, such as temperature or pH, or minerals supply chain, such as particle size. 
These cases are intended to illustrate the importance of prompt action to support the commercial deployment of CDR options, particularly those that enable to prompt removal of CO$_2$ from the atmosphere, such as BECCS and DACCS.
\newline
\indent With international cooperation, we found that the spatio-temporal evolution of the CDR pathway, most cost-efficiently deployed, differs greatly from the regional allocation of the Paris Agreement's CDR objectives --- based on responsibility for climate change, here used as a proxy for their socio-economically fair distribution.
Although this is not the case with less or no international cooperation in climate mitigation policy --- the amount of CO$_2$ removal deployed regionally, within each region's borders, must meet their share of the Paris Agreement's CDR objectives ---, we found that the resulting CDR pathways are likely to become more challenging and more costly to deploy.
This is due to a more difficult access to cost-efficient bio-geophysical CO$_2$ sinks, and therefore the use of more costly alternatives, such as DACCS.
Importantly, not only does the likelihood to meet the Paris Agreement's CDR objectives thus decreases, but associated costs also increase by 51--69\% relatively to the CDR pathway deployed with international cooperation.
\newline 
\indent To overcome this challenge --- the trade-off between cost-efficiency and equity ---, this study has shown that international cooperation could be implemented \textit{via} an international market for negative emissions trading, in which regions most well-endowed with CDR potentials could generate CDR surplus, \textit{i.e.} additional CO$_2$ removal, over and above their individual CDR targets, and provide it as a remunerative service to other less well-endowed regions.
In such market, CDR surplus would generate negative emissions credits (NECs), that could be traded between regions, and thus enabling the delivery of the Paris Agreement most cost-efficiently and equitably.
\linebreak
\indent Particularly, we found that such market would allow "dependant NECs beneficiary" regions --- regions that could not meet their national CDR targets domestically --- to successfully deliver their share of the Paris Agreement's CDR objectives.
It would also decrease the financial burden of the Paris Agreement for "independent NECs beneficiary" regions --- regions that could meet their national CDR targets domestically --- by up to 51--90\%.
The design of such market --- trading mechanisms such as the allocation of NECs between participating regions, or the determination of the price of NECs --- is out of the scope of this study, but remains to be investigated further.
\newline 
\indent Finally, recognising that current policies on climate change mitigation are far from an "international cooperation" paradigm, this study has also argued that the later such market for negative emissions trading would be implemented, the more expensive delivering the Paris Agreement's CDR objectives would be, and subsequently imposing more of the associated financial effort on to future generations.
Imminent action towards the establishment and deployment of multi-regional, or possibly international, geopolitical and economic instruments for negative emissions trading, and robust institutions to enable their monitoring, verification and accreditation, will therefore be key in delivering the Paris Agreement's CDR objectives consistent with a global warming of 1.5$^{\circ}$C.

\section*{Conflicts of interest}
There are no conflicts to declare.

\section*{Appendix A ~~ CDR options in MONET}

The MONET framework used in this study has been extended to include AR and DACCS. 
Their models are described here.

\subsection*{A.1 ~~ Afforestation/Reforestation (AR) model}

\subsubsection*{A.1.1 ~~ Overview.}

Afforestation/reforestation (AR) refers to the process of planting or facilitating the natural regeneration of trees.
Although afforestation and reforestation can be distinguished by the period of time during which the land has not been forested --- commonly for a period of at least 50 years ---, or by the climato-ecological suitability of the land --- forest, shrubland versus grassland, savannah --- , AR is often jointly categorized in the context of land use change and associated (biogenic) CO$_2$ emissions/sequestration accounting \cite{IPCC2006a}.
This is also the case in this study.

We have developed an explicit spatio-temporal model of AR's whole-system, in which 5 sub-models are integrated: 1) a forest growth model, 2) a forest management cycle model, 3) a biogenic carbon (C) (and CO$_2$) sequestration model and 4) its associated "fire-penalty" model, and 5) a forestry operations model. 
Specifically, energy, CO$_2$ (and N$_2$O) and cost balances are carried out for each step of the forestry operations model.
Spatial resolution of AR's whole-system model is at the climato-ecological level --- ecological zones\footnote{An ecological zone is defined as “a zone or area with broad yet relatively homogeneous natural vegetation formations, similar (not necessarily identical) in physiognomy” \cite{FAO2001}.} --- within each State/Province in Brazil, China, India and the USA and within each country in the EU (EU-27 \& UK).
Temporal resolution of AR's whole-system model is double: 1) 10 years (decadal), ranging between 2020--2100, and 2) 1 year (annual), over a default 100 years time-period. 
The decadal timescale is used for determining the establishment of newly afforested stands, then the annual timescale is used for computing and evaluating AR's whole-system model.

\begin{figure*}[t]
\centering
  \includegraphics[width=17.1cm]{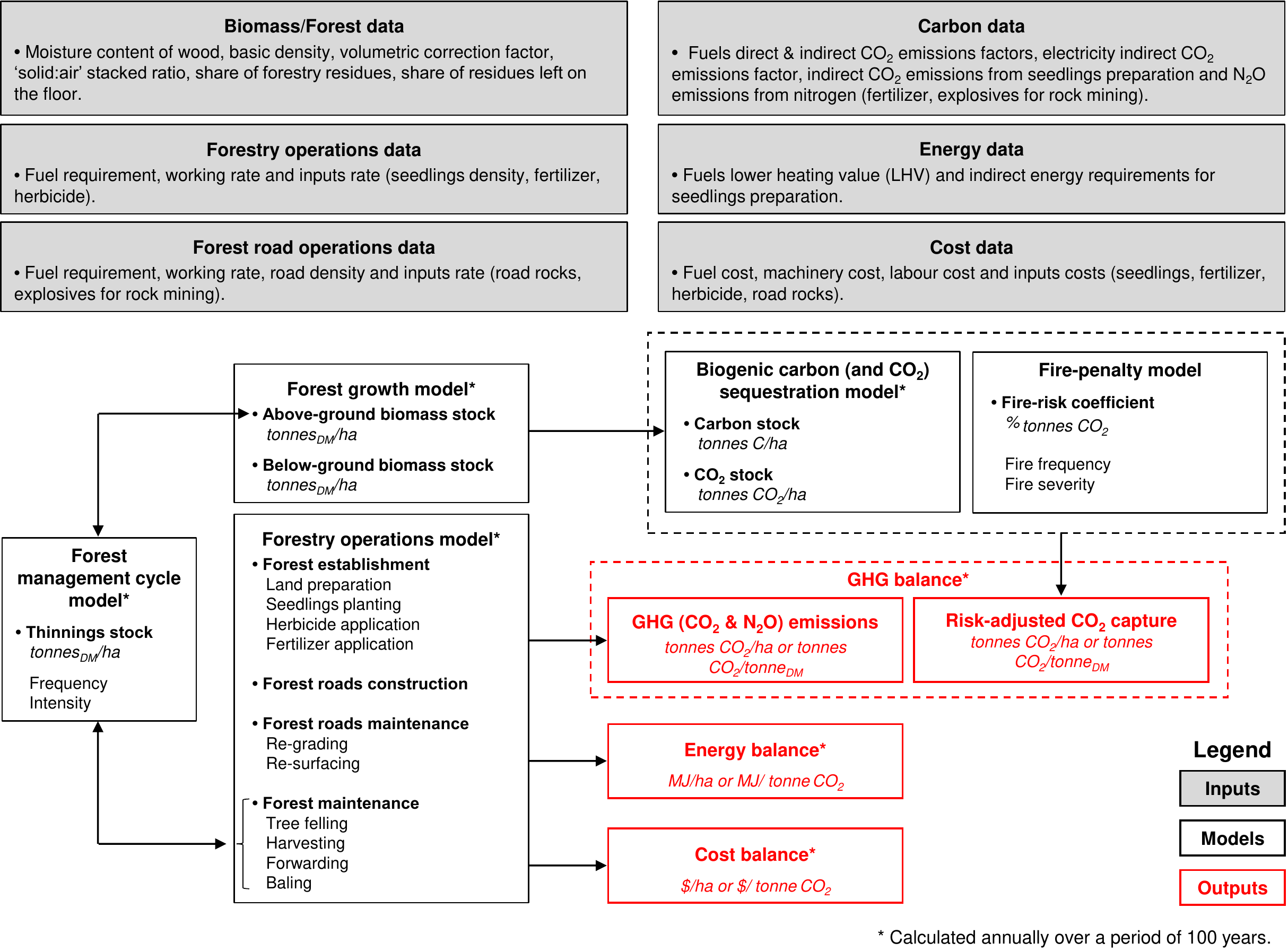}
  \caption{Schematic of the AR's whole-system model, outlining the interactions between 5 integrated sub-models: 1) a forest growth model, 2) a forest management cycle model, 3) a biogenic carbon (C) (and CO$_2$) sequestration model, and 4) its associated 'fire-penalty' model, and 5) a forestry operations model.}
  \label{fig: AR_Diagram}
\end{figure*}

The different interactions between the AR sub-models are outlined in Fig. \ref{fig: AR_Diagram}.

\subsubsection*{A.1.2 ~~ Forest growth \& forest management cycle models.}

Here, we describe in details the forest growth model and its associated forest management cycle model.
Within the forest growth model, forest growth curves are characterised by ecological zone and forest type (broadleaves/conifers), to account for geographic, climatic and ecological variations \cite{FAO2001,Resources2012}.
Both the above-ground biomass --- the vegetation above the soil, such as stems, branches, foliage or bark ---, without (of reference) and with forest management (managed), and the below-ground biomass --- the roots --- are included in the forest growth model.

Within the forest management cycle model, we assume that forest stands are subject to a non-intensive forest management --- with reduced or minimum human intervention. 
The purpose of this forest management is to maximise and maintain the C (and CO$_2$) sequestration potential of the forest by clearing the forest of old and/or sick trees in order to let younger trees grow, more vigorously and with more space.
Although the forest management cycle model consists in determining directly the proportion of above-ground biomass, only, that needs to be thinned for the aforementioned reasons, it also impacts indirectly the proportion of below-ground biomass that remains after.

\paragraph*{A.1.2.1 ~ Above-ground biomass (of reference).}
We define the above-ground biomass stock of reference $B^{AG}_{Ref}$ as a sigmoid curve, which is typical in even-aged stands in the absence of forest management (without human intervention), as shown in Eqs. \ref{eq: AGBio Ref 1}--\ref{eq: AGBio Ref 4}.
Eqs. \ref{eq: AGBio Ref 1}--\ref{eq: AGBio Ref 4} are written as follows:

\begin{equation}
    \label{eq: AGBio Ref 1}
    B^{AG}_{Ref}(yr) = \frac{B_{Ref}}{1+exp(-k_{Ref}(yr-x_{0,Ref})} \qquad \forall \; t
\end{equation}

\begin{equation}
    T_{Ref} = \frac{B_{Ref}}{G_{Ref}}
\end{equation}

\begin{equation}
    x_{0,Ref}= \frac{T_{Ref}}{2}
\end{equation}

\begin{equation}
    \label{eq: AGBio Ref 4}
    k_{Ref} = \frac{ln(99)}{x_{0,Ref}}
\end{equation}

\noindent where: 
\begin{list}{\labelitemi}{\leftmargin=1em}
\item $T_{Ref}$ is the growing period of reference (yrs),
\item $B_{Ref}$ is the maximum biomass stock of reference (t$_{DM}$/ha),
\item $G_{Ref}$ is the average annual biomass growth of reference (t$_{DM}$/ha/yr),
\item $x_{0,Ref}$ is the mid-point of the reference sigmoid curve (yrs),
\item and $k_{Ref}$ is the slope coefficient of the reference sigmoid curve (t$_{DM}$/ha/yr).
\end{list}

$B^{AG}_{Ref}$ is calculated using the IPCC default values for biomass stock $B_{Ref}$ and biomass growth $G_{Ref}$ of natural forests \cite{IPCC2006c}, as provided/given in Table \ref{tbl IPCC above-ground biomass}).

\begin{table}[h]
\small
  \caption{IPCC default biomass stock, biomass growth and growing period of natural forests, characterised by ecological zone \cite{IPCC2006c}. 
}
  \label{tbl IPCC above-ground biomass}
  \begin{tabular*}{0.48\textwidth}{@{\extracolsep{\fill}}llll}
    \hline
    \makecell[l]{Ecological zone} & \makecell[l]{Biomass\\ stock\\ (t$_{DM}$/ha)} & \makecell[l]{Biomass\\ growth\\ (t$_{DM}$/ha/yr)} & \makecell[l]{Growing\\ period$^a$\\ (yrs)} \\
    \hline
    Tropical rainforest & 300 & 7.0 & 43\\
    Tropical moist deciduous forest & 180 & 5.0 & 36\\
    Tropical dry forest & 130 & 2.4 & 54\\
    Tropical shrubland & 70 & 1.0 & 70\\
    Tropical mountain systems & 140 & 1.0 & 140\\
       \hline
     Subtropical humid forest & 220 & 5.0 & 44\\
     Subtropical dry forest & 130 & 2.4 & 54\\
     Subtropical steppe & 70 & 1.0 & 70\\
     Subtropical mountain systems & 140 & 1.0 & 140\\
         \hline
     Temperate oceanic forest & 180 & 4.4 & 41\\
     Temperate continental forest & 120 & 4.0 & 30\\
     Temperate mountain systems & 100 & 3.0 & 33\\
         \hline
     Boreal coniferous forest & 50 & 1.0 & 50\\
     Boreal tundra woodland & 15 & 0.4 & 38\\
     Boreal mountain systems & 30 & 1.0 & 30\\
    \hline
    \multicolumn{4}{l}{\makecell[l]{$^a$ Calculated with the IPCC default values for biomass stock and\\ biomass growth \cite{IPCC2006c}.}}\\
  \end{tabular*}
\end{table}

\paragraph*{A.1.2.2 ~ Managed above-ground biomass.}

The managed above-ground biomass stock derives from the above-ground biomass stock of reference, subject to a forest management cycle.
Here, we introduce the concept of forest management cycle (FMC) and its associated phases, as developed within the forest management cycle model, and describes how it impacts the forest growth model.

\subparagraph*{A.1.2.2.1 ~ Forest growth phases.}

In the context of timber production, a forest growth is usually broken down into 5 phases --- the establishment, initial, full-vigour, mature, and old-growth phases.
These 5 phases constitute the FMC, in which harvesting (and thinning) operations' characteristics --- the frequency and the intensity --- are set, and specific to each phase.

Because timber production is evaluated in terms of merchantable biomass volume, forest growth phases are usually determined based on the Mean Annual Increment (MAI) --- the average rate of merchantable volume of biomass growth --- and its Maximum Mean Annual Increment (MMAI).
In the context of (biogenic) CO$_2$ removal, however, both merchantable and non-merchantable biomass stocks are considered, including both above-ground biomass and below-ground biomass stocks, and evaluated in terms of total biomass dry-mass.

The FMC modelled here is developed accordingly to the aforementioned adaptations. 
We introduce and define the Mean Annual Growth (MAG) --- the average rate of dry-mass of above-ground biomass growth --- and the Maximum Mean Annual Growth (MMAG), which replace, respectively, the MAI and the MMAI in the determination of the forest growth phases. 
These forest growth phases are described in Table \ref{tbl FMC phases}.

\begin{table}[h]
\small
  \caption{Description of the forest growth phases.}
  \label{tbl FMC phases}
  \begin{tabularx}{0.48\textwidth}{@{\extracolsep{\fill}}lL}
    \hline
    Phase & Description \\
    \hline
    Establishment & Seedlings are planted to create a new stand of trees. In this study, this phase is defined as lasting for the first five years after planting.\\
    \hline
    Initial & Once established, young trees grow from seedlings and their AG increases. This phase is defined here as lasting from age 5 up until (and including) the age of the first thinning, as specified here at 10 years after planting (age 5 + 10 = 15 years). This phase has zero thinning.\\
    \hline
    Full-vigour & During this period, trees grow with the highest rate of AG. This phase is defined as lasting from the age of the first thinning (excluding) until (and including) the time at which the MMAG occurs. Thinning operations occur every 5 years, with an intensity of 10\% of the above-ground biomass stock.\\
    \hline
    Mature & During this phase, the rate of AG declines progressively from its maximum value. The phase is defined as lasting from the time at which the MMAG is reached (excluding) up until (and including) the time at which the MAG has dropped to 50\% of its maximum value. Thinning operations occur every 10 years with an intensity of 10\% of the above-ground biomass stock.\\
    \hline
    Old-growth & During this last phase, the biomass accumulation rate reaches its peak and stabilises --- the biomass stock saturates ---, and the AG slowly levels off to zero. This phase is defined as lasting indefinitely from the time at which the MAG declines from half of the maximum (excluding). Stands in this phase also shift from an even-aged composition to a diverse structure of ages and sizes. Thinning operations occur every 15 years with an intensity of 10\% of the above-ground biomass stock.\\
    \hline
    \multicolumn{2}{l}{\makecell[l]{AG: annual growth; MMAG: maximum mean annual growth; and\\ MAG: mean annual growth.}}
  \end{tabularx}
\end{table}

Here, the annual (above-ground) growth (AG) $AG^{AG}_{Ref}$, the MAG $MAG^{AG}_{Ref}$ and the MMAG $MMAG^{AG}_{Ref}$ are derived from the above-ground biomass stock of reference $B^{AG}_{Ref}$, as shown in Eqs. \ref{eq: AG Ref}--\ref{eq: MMAG Ref} below:

\begin{equation}
\label{eq: AG Ref}
  AG^{AG}_{Ref}(yr) = \begin{dcases} B^{AG}_{Ref}(yr) & \quad \forall \; yr, \; yr = 1 \\
B^{AG}_{Ref}(yr) - B^{AG}_{Ref}(yr-1) & \quad \forall \; yr, \; yr > 1 \end{dcases} \\   
\end{equation}

\begin{equation}
\label{eq: MAG Ref}
 MAG^{AG}_{Ref}(yr) = \frac{AG^{AG}_{Ref}(yr)}{yr} \qquad \forall \; yr
 \end{equation}

\begin{equation}
\label{eq: MMAG Ref}
  MMAG^{AG}_{Ref} = \max_{yr} MAG^{AG}_{Ref}(yr)
\end{equation}

\begin{figure}[t]
\centering
  \includegraphics[width=8.3cm]{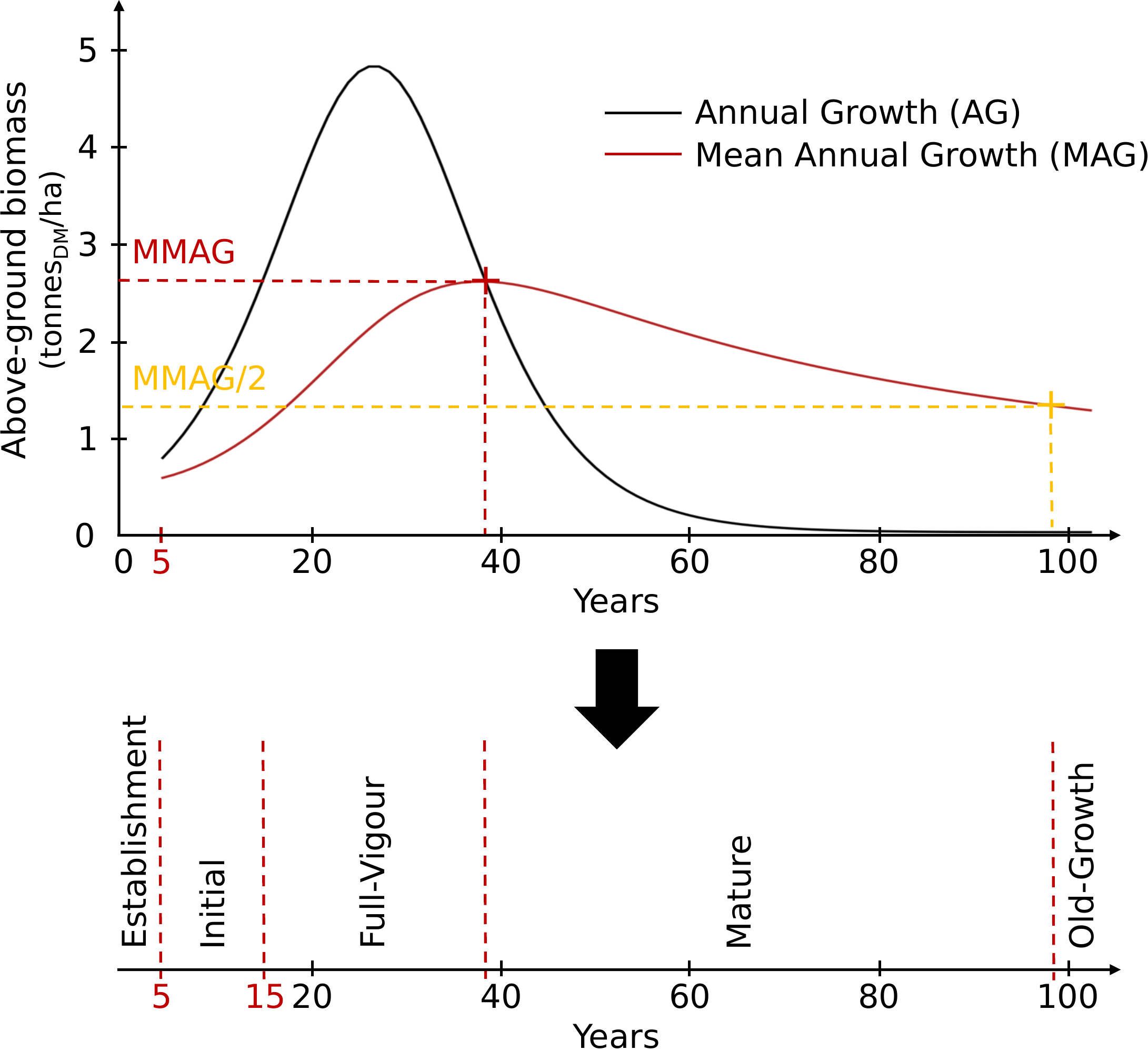} 
  \caption{Illustration of a forest annual growth curve (AG), its associated mean annual growth curve (MAG) and maximum mean annual growth point (MMAG), and its resulting forest growth phases.}
  \label{fig: FMC Phases}
\end{figure}

Fig. \ref{fig: FMC Phases} illustrates a forest AG curve, its associated MAG curve and MMAG point, and its resulting forest growth phases.
 
\subparagraph*{A.1.2.2.2 ~ Forest management cycle (FMC).}

In the context of climate mitigation, AR is deployed while prioritising C (and CO$_2$) sequestration potential over timber production.
Therefore, the FMC model introduced here is only comprised of thinning operations (no harvesting operations), in order to maximise and maintain the forest C (and CO$_2$) stock. 
We assume that: 1) the frequency of thinning operations decreases with time --- from every 5 years during the full-vigour phase to every 15 years during the old-growth phase ---, and 2) the intensity is set as 10\% of the above-ground biomass stock at any time.

A schematic of the FMC workflow FMC workflow is described in Fig.\ref{fig: FMC Algo}, where: 

\begin{figure*}
\centering
    \includegraphics[width=17.1cm]{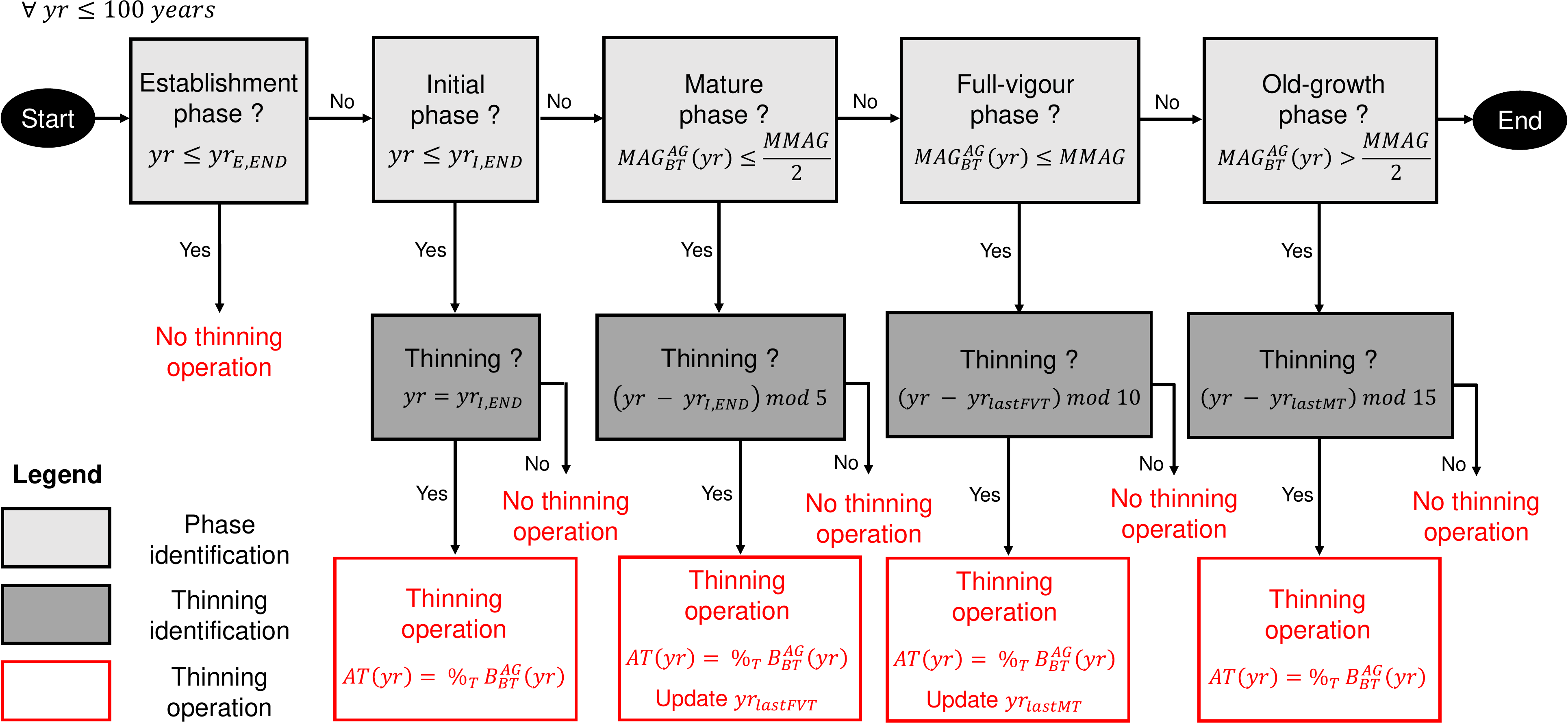}
  \caption{Schematic of the forest management cycle (FMC) workflow used for determining thinning operations' frequency and intensity.}
  \label{fig: FMC Algo}
\end{figure*}

\begin{list}{\labelitemi}{\leftmargin=1em}
\item $yr_{E,END}$ is the last year of the establishment phase (yrs);
\item $yr_{I,END}$ is the last year of the initial phase (yrs);
\item $MAG^{BT}_{ha}(yr)$ is the MAG (before thinning) (t$_{DM}$/ha/yr),
\item $MMAG$ is the MMAG (t$_{DM}$/ha/yr);
\item $AT(yr)$ is the annual thinning stock (t$_{DM}$/ha/yr);
\item $\%_T$ is the thinning share of the above-ground biomass stock (\%), $\%_T$ default value is 10\%;
\item $B^{AG}_{BT}(yr)$ is the above-ground biomass stock, before thinning (t$_{DM}$/ha);
\item $yr_{lastFVT}$ is the time at which the last thinning during the full-vigour phase occurs (yrs);
\item and $yr_{lastMT}$ is the time at which the last thinning during the mature phase occurs (yrs), 
$y_{E,END}$ and $yr_{I,END}$ default values are respectively 5 and 15 yrs.
\end{list}

\begin{figure}[t]
\centering
\includegraphics[width=8.3cm]{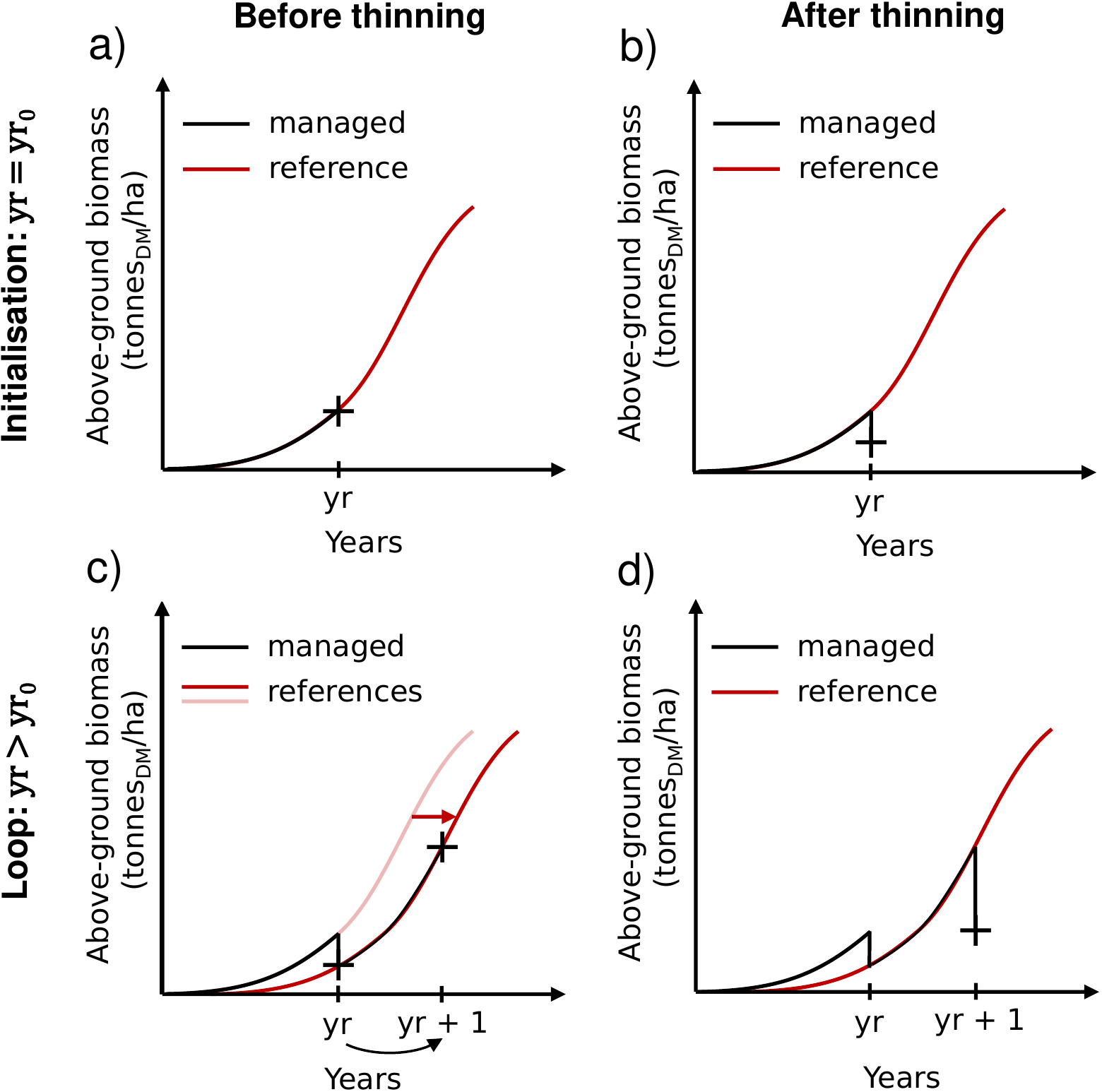}
 \caption{Illustration of the two-step workflow used for evaluating the managed above-ground biomass stock.}
\label{fig: growth 2 steps}
\end{figure}

\subparagraph*{A.1.2.2.3 ~ Managed above-ground biomass.}
The managed above-ground biomass stock derives from the above-ground biomass stock of reference $B^{AG}_{Ref}$ and the annual thinning stock $AT$.
Here, we evaluate the managed above-ground biomass stock in two steps --- before thinning $B^{AG}_{BT}$ and after thinning $B^{AG}_{AT}$, as illustrated in Fig.\ref{fig: growth 2 steps}.

\noindent \textbf{Initialisation ($yr=yr_0$):} 
The year of initialisation $yr_0$  is defined as the year of the first thinning.
In step a), $B^{AG}_{BT}$ is equal to $B^{AG}_{Ref}$, as shown in Eq. \ref{eq: growth 2 steps a} (Fig.\ref{fig: growth 2 steps}-a):

\begin{equation}
\label{eq: growth 2 steps a}
   B^{AG}_{BT}(yr_0) = B^{AG}_{Ref}(yr_0)
\end{equation}

Then, in step b), $B^{AG}_{AT}$ is obtained by subtracting $AT$ to $B^{AG}_{BT}$, as shown below in Eqs. \ref{eq: growth 2 steps b1}--\ref{eq: growth 2 steps b2} (Fig.\ref{fig: growth 2 steps}-b):

\begin{equation}
\label{eq: growth 2 steps b1}
   B^{AG}_{AT}(yr_0) = B^{AG}_{BT}(yr_0) - AT(yr_0)
\end{equation}

\begin{equation}
\label{eq: growth 2 steps b2}
   AT(yr_0) = \%_T \times B^{AG}_{BT}(yr_0))
\end{equation}

\noindent \textbf{Loop ($ \forall yr > yr_0$):}
In step c), the following year, $B^{AG}_{BT}(yr_0+1)$ has increased by $\Delta B^{AG}(yr_0+1)$, following $B^{AG}_{Ref}$, as shown in Eq. \ref{eq: growth 2 steps c1}. 
$\Delta B^{AG}(yr_0+1)$ is obtained as shown in Eqs. \ref{eq: growth 2 steps c2}--\ref{eq: growth 2 steps c4} (Fig.\ref{fig: growth 2 steps}-c). 
Eqs. \ref{eq: growth 2 steps c1}--\ref{eq: growth 2 steps c4} are written as follows: 

\begin{equation}
    \label{eq: growth 2 steps c1}
   B^{AG}_{BT}(yr_0+1) = B^{AG}_{AT}(yr_0) + \Delta B^{AG}(yr_0+1)
\end{equation}

\begin{equation}
     \label{eq: growth 2 steps c2}
     yr_{Ref,1} = B^{AG-1}_{Ref}(B^{AG}_{BT}(yr_0))
\end{equation}

\begin{equation}
    \label{eq: growth 2 steps c3}
    yr_{Ref,2} = yr_{Ref,1} + 1
\end{equation}

\begin{equation}
    \label{eq: growth 2 steps c4}
    \Delta B^{AG}(yr_0+1) = B^{AG}_{Ref}(yr_{Ref,2}) - B^{AG}_{Ref}(yr_{Ref,1})
\end{equation}

Lastly, in step d) the above-ground biomass stock, after thinning, $B^{AG}_{AT}(yr_0+1)$ is obtained as previously, in step b) (Fig.\ref{fig: growth 2 steps}-d).
Steps c) and d) are repeated for each $yr > yr_0$.
During the years of thinning operations, $\%_T=10\%$, $\%_T=0\%$ otherwise.

\paragraph*{A.1.2.3 ~ Below-ground biomass.}

The below-ground biomass stock can be estimated from the above-ground biomass stock with the use of a "root-to-shoot" ratio. 
A "root-to-shoot" ratio usually depends on climate, tree species, soil type and declines with stand age and/or productivity. 
Extreme range values of 0.09–1.16 t root$_{DM}$/t shoot$_{DM}$ have been reported in the literature, although average range values of 0.20–0.56 t root$_{DM}$/t shoot$_{DM}$ might be more likely \cite{IPCC2006c}.

Here, the "root-to-shoot" ratio $R_{RS}$ evolves with the amount of above-ground biomass stock (before thinning) $B^{AG}_{BT}$.
Specifically, $R_{RS}$ is interpolated from the IPCC default values \cite{IPCC2006c} provided in Table \ref{tbl IPCC root-to-shoot ratio}, as shown below in Eq. \ref{eq: R root to shoot}.

\begin{equation}
    \label{eq: R root to shoot}
    R_{RS}(yr) = \begin{dcases} R_1 \times ln(R_2 \times B^{AG}_{BT}(yr)+1) & \quad \forall \; yr, \; R_2 \neq 0 \\
R_1 & \quad \forall \; yr, \; R_2 = 0 \end{dcases}
\end{equation}

\noindent where:
\begin{list}{\labelitemi}{\leftmargin=1em}
\item $R_{RS}(yr)$ is the "root-to-shoot" ratio (t root$_{DM}$/t shoot$_{DM}$),
\item and $R_1$ (-) and $R_2$ (-) are the coefficients interpolated from the IPCC
\footnote{$R_1$ and $R_2$ were obtained by solving a non-linear curve-fitting (data-fitting) problem in least-squares sense in Python 3.7 (function scipy.optimize.leastsq).} (See Table \ref{tbl Interpolated root-to-shoot ratios}).
\end{list}

The below-ground biomass stock $B^{BG}$ derives from the managed above-ground biomass stock (after thinning) $B^{AG}_{AT}$ and the "root-to-shoot" ratio $R_{RS}$, as shown below in Eq \ref{eq: BGBio}:

\begin{equation}
\label{eq: BGBio}
    B^{BG} = B^{AG}_{AT}(yr) \times R_{RS}(yr) \qquad \forall \; yr
\end{equation}

\begin{table}[h]
\small
  \caption{IPCC default "root-to-shoot" ratio, characterised by ecological zone, forest type and above-ground biomass stock \cite{IPCC2006c}.}
  \label{tbl IPCC root-to-shoot ratio}
  \begin{tabular*}{0.48\textwidth}{@{\extracolsep{\fill}}llll}
    \hline
    \makecell[l]{Ecological zone} & \multicolumn{2}{l}{\makecell[l]{Above-ground\\ biomass\\ (t$_{DM}$/ha)}} & \makecell[l]{"Root-to-\\shoot"\\ ratio (-)} \\
    \hline
    Tropical rainforest & - & & 0.37\\
    Tropical moist deciduous forest & \multicolumn{2}{l}{< 125} & 0.20\\
     & \multicolumn{2}{l}{> 125} & 0.24\\
    Tropical dry forest & \multicolumn{2}{l}{< 20} & 0.56\\
     & \multicolumn{2}{l}{> 20} & 0.28\\
    Tropical shrubland & - & & 0.40\\
    Tropical mountain systems & - & & 0.27\\
    \hline
    Subtropical humid forest & \multicolumn{2}{l}{< 125} & 0.20\\
     & \multicolumn{2}{l}{> 125} & 0.24\\
     Subtropical dry forest & \multicolumn{2}{l}{< 20} & 0.56\\
     & \multicolumn{2}{l}{> 20} & 0.28\\
    Subtropical steppe & -& & 0.32\\
    Subtropical mountain systems$^a$ & - & & 0.27\\
     \hline
    \multirow{6}{*}{\makecell[l]{Temperate oceanic forest,\\ Temperate continental forest,\\
     Temperate mountain systems }} & \multirow{3}{*}{Conifers} & < 50 & 0.4\\
     & & 50--150 & 0.29\\
     & & > 150 & 0.20\\
     & \multirow{3}{*}{Broadleaves} & < 75 & 0.46\\
     & & 50--150 & 0.23\\
     & & > 150 & 0.24\\
         \hline
    \multirow{3}{*}{\makecell[l]{Boreal coniferous forest,\\ Boreal tundra woodland,\\
     Boreal mountain systems}} & \multirow{3}{*}{\makecell[l]{ < 75\\ > 75}} & & \multirow{3}{*}{\makecell[l]{ 0.39\\ 0.24}}\\
     & & &\\
     & & &\\
         \hline
    \multicolumn{4}{l}{\makecell[l]{$^a$ Used IPCC tropical mountain systems values\cite{IPCC2006c}.}}\\
  \end{tabular*}
\end{table}

\begin{table}[h]
\small
  \caption{Interpolated "root-to-shoot" ratios $R_1$ and $R_2$ values.}
  \label{tbl Interpolated root-to-shoot ratios}
  \begin{tabular*}{0.48\textwidth}{@{\extracolsep{\fill}}llll}
    \hline
    Ecological zone & Forest type & $R_1$ & $R_2$ \\
    \hline
    Tropical rainforest$^a$ & - & 0.37 & - \\
    Tropical moist deciduous forest & - & 801 & 0.00029 \\
    Tropical dry forest & - & 33 & 0.01113 \\
    Tropical shrubland$^a$ & - & 0.40 & - \\
    Tropical mountain systems$^a$ & - & 0.27 & - \\
    \hline
    Subtropical humid forest & - & 848 & 0.00028 \\
    Subtropical dry forest & - & 474 & 0.00126 \\
    Subtropical steppe & - & 0.32& - \\
    Subtropical mountain systems & - & 0.27 & - \\
    \hline
    \multirow{3}{*}{\makecell[l]{Temperate oceanic forest,\\ Temperate continental forest,\\
     Temperate mountain systems }} & \multirow{3}{*}{\makecell[l]{Conifers\\ Broadleaves}} & \multirow{3}{*}{\makecell[l]{25\\ 24}} & \multirow{3}{*}{\makecell[l]{0.02159\\ 0.02275}}\\
     & & \\
     & & \\
    \hline
    \multirow{3}{*}{\makecell[l]{Boreal coniferous forest,\\ Boreal tundra woodland,\\
     Boreal mountain systems}} & \multirow{3}{*}{\makecell[l]{-}} & \multirow{3}{*}{\makecell[l]{25}} & \multirow{3}{*}{\makecell[l]{ 0.04466}}\\
     & & \\
     & & \\
         \hline
    \multicolumn{4}{l}{\makecell[l]{$^a$ Used IPCC "root-to-shoot" value \cite{IPCC2006c}.}}\\
  \end{tabular*}
\end{table}

\paragraph*{A.1.2.4 Total biomass.}

The total biomass stock $B^{Total}$ is evaluated as shown below in Eq. \ref{eq: TotalBio}:

\begin{equation}
\label{eq: TotalBio}
    B^{Total}(yr) = B^{AG}_{AT}(yr) + B^{BG}(yr) \qquad \forall \; yr
\end{equation}

\subsubsection*{A.1.2 Biogenic C (and CO$_2$) sequestration model.}

Here, we describe in details the biogenic C (and CO$_2$) sequestration model and its associated C pools.

\paragraph*{A.1.2.1 Biogenic C pools.}

Growing forests capture CO$_2$ from the atmosphere via photosynthesis. 
The sequestrated C --- the CO$_2$ is sequestrated in the form of C --- contained in the above-ground biomass is then partially transferred to the below-ground biomass, dead organic matter and soil.
During harvesting or thinning operations, timber and forest residues are extracted from the forest stands, and are considered as "harvested wood products". 
All together constitute biogenic C pools \cite{IPCC2006b,IPCC2006c}.
Fig. illustrates the C flow into and out of the AR's whole system, as modelled here, as well as between the 5 aforementioned C pools. 

\begin{figure}[h]
\centering
\includegraphics[width=8.3cm]{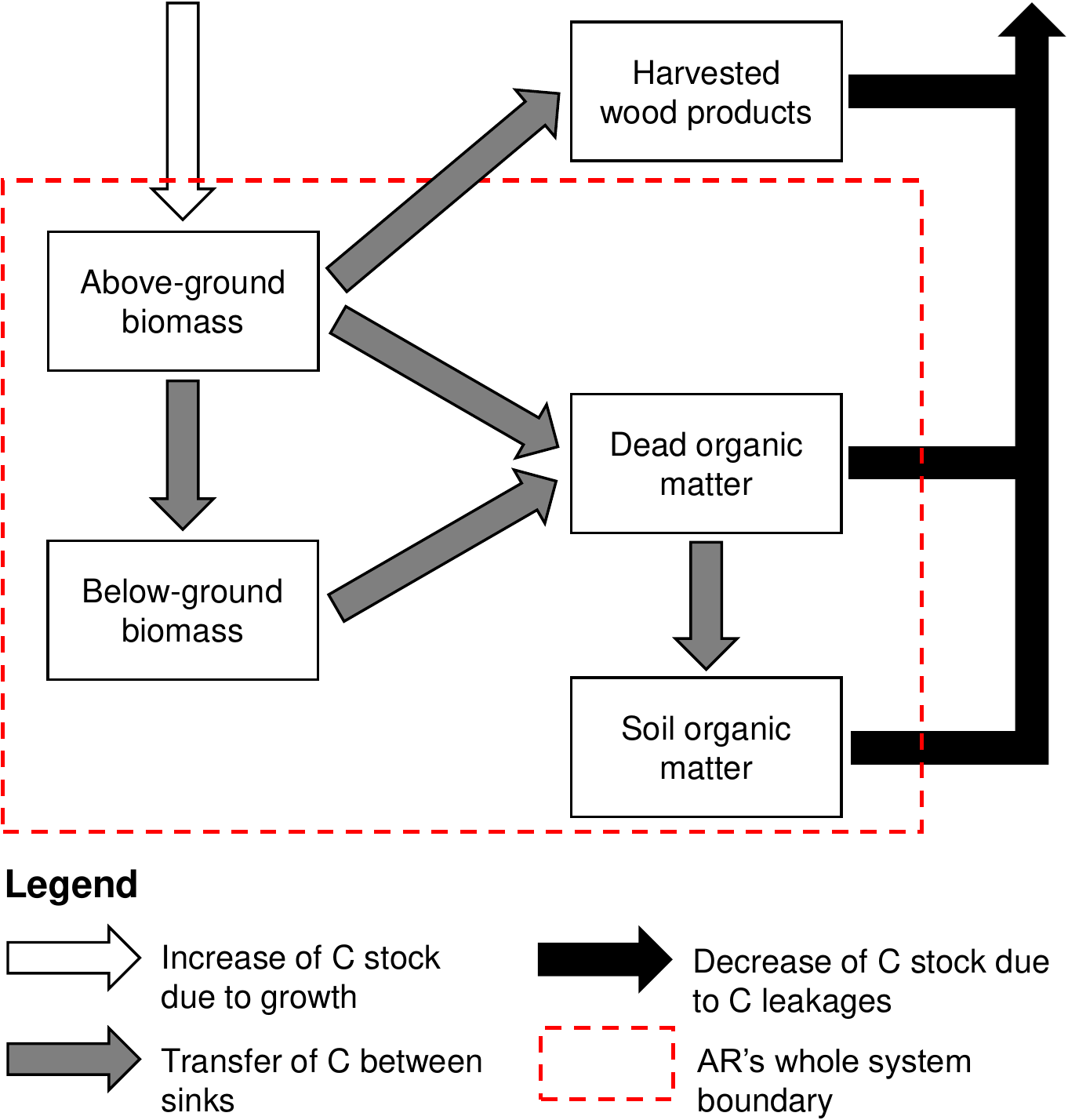}
 \caption{Illustration of the C flow into and out of the AR's whole system, as modelled here, as well as between the 5 main carbon pools: 1) the above-ground biomass, 2) the below-ground biomass, 3) the dead organic matter, 4) the soil organic matter and 5) the harvested wood products pools. Adapted from the 2006 IPCC Guidelines for National Greenhouse Gas Inventories \cite{IPCC2006b}.
}
\label{fig: C flow}
\end{figure}

\paragraph*{A.1.2.2 Dead organic matter C pool.}

Dead organic matter is comprised of litter and dead wood. 
The dead organic matter C pool is highly variable, site-specific and time-evolving, depending specifically on forest characteristics, forest management and disturbance history, such as fire, hurricanes or pest. 
For instance, dead organic matter decay rates range from high in warm and moist climates to low in cold and dry climates \cite{IPCC2006b}.
Litter decay has also been observed to be much faster in deciduous (broadleaves) forests than in evergreen (conifers) forests, owing to humus formation, higher acidity of coniferous litter, and different canopy density resulting in different amount of light and water reaching the forest floor \cite{Schulp2008}.

A range value of 15–-150 tCO$_2$/ha, with an average value of 66 tCO$_2$/ha can be found in the literature \cite{Berg2007}. 
Although the C dynamics of the dead organic matter pool are qualitatively well understood, it is currently difficult to obtain complete set of data at the national or regional scales.
Therefore, we conservatively simplify the dead organic matter C pool by assuming that all C transferring from biomass to dead organic matter is directly emitted to the atmosphere.

\paragraph*{A.1.2.3 Soil C pool.}

Both organic and inorganic forms of C are found in soil, but the soil organic matter C pool is more largely affected by land-use and management activities, and therefore mostly investigated in the literature.
As the dead organic matter C pool, the soil organic matter C pool is highly variable and site-specific.
Depending on forest characteristics, management regime and disturbance history, the soil organic matter C pool is also time-evolving, due to differences between C inputs and losses over time.

The organic C content of mineral forest soils has been reported between 20--300 tC/ha in the literature \cite{Jobbagy2000, IPCC2006c}, but current available data remain largely site- and study-specific, and are therefore still incomplete and highly uncertain at national or regional scales.
Although the conversion of non-forested lands to forested lands would be expected to increase the organic C content of newly afforested soils during the first decades, we assume conservatively that the soil organic matter C pool remains constant with AR.

\paragraph*{A.1.2.4 Biomass C pool.}

The above-ground biomass C pool is comprised of all C that is contained in the vegetation above the soil, such as stems, branches, foliage or bark, and the below-ground  biomass C pool is comprised of the C contained in the roots.
Together, they constitute the biomass C pool, of which the C stock can be estimated from the biomass stock with the use of a carbon content factor $C_f$.
$C_f$ depends on climate, tree specie, such as conifers or broadleaves, and tree characteristics, such as age, size or tree parts. 
Average values have been reported within the range of 0.43–0.55 tC/t$_{DM}$\cite{IPCC2006c}.

Here, the (total) biomass C stock $C^{Total}$ derives from the total biomass stock $B^{Total}$.
We use the IPCC default values for carbon content factor $C_f$ \cite{IPCC2006c} given in Table \ref{tbl IPCC carbon content}, as shown below in Eq. \ref{eq: C Total}.

\begin{equation}
\label{eq: C Total}
    C^{Total}(yr) = B^{Total}(yr) \times C_f \qquad \forall \; yr
\end{equation}

\begin{table}[h]
\small
  \caption{IPCC default carbon content factor, characterised by climate and forest type \cite{IPCC2006c}.}
  \label{tbl IPCC carbon content}
  \begin{tabular*}{0.48\textwidth}{@{\extracolsep{\fill}}lll}
    \hline
    Climate &  \multicolumn{2}{c}{Carbon content (tC/t$_{DM}$)}\\
    & Broadleaves & Conifers \\
    \hline
    Tropical & 0.47 & 0.47 \\
    Subtropical & 0.47 & 0.47\\
    Temperate & 0.48 & 0.51\\
    Boreal & 0.48 & 0.51\\
    \hline
  \end{tabular*}
\end{table}

Then, the total biomass CO$_2$ stock $CO_2^{Total}$ derives from the conversion of C to CO$_2$, based on the ratio of molecular weights (44/12), as shown below in Eq. \ref{eq: CO2 Total}:

\begin{equation}
\label{eq: CO2 Total}
CO_2^{Total}(yr) = C^{Total}(yr) \times \frac{44}{12} \qquad \forall \; yr
\end{equation}

\begin{figure*}[h]
\centering
\includegraphics[width=18.5cm]{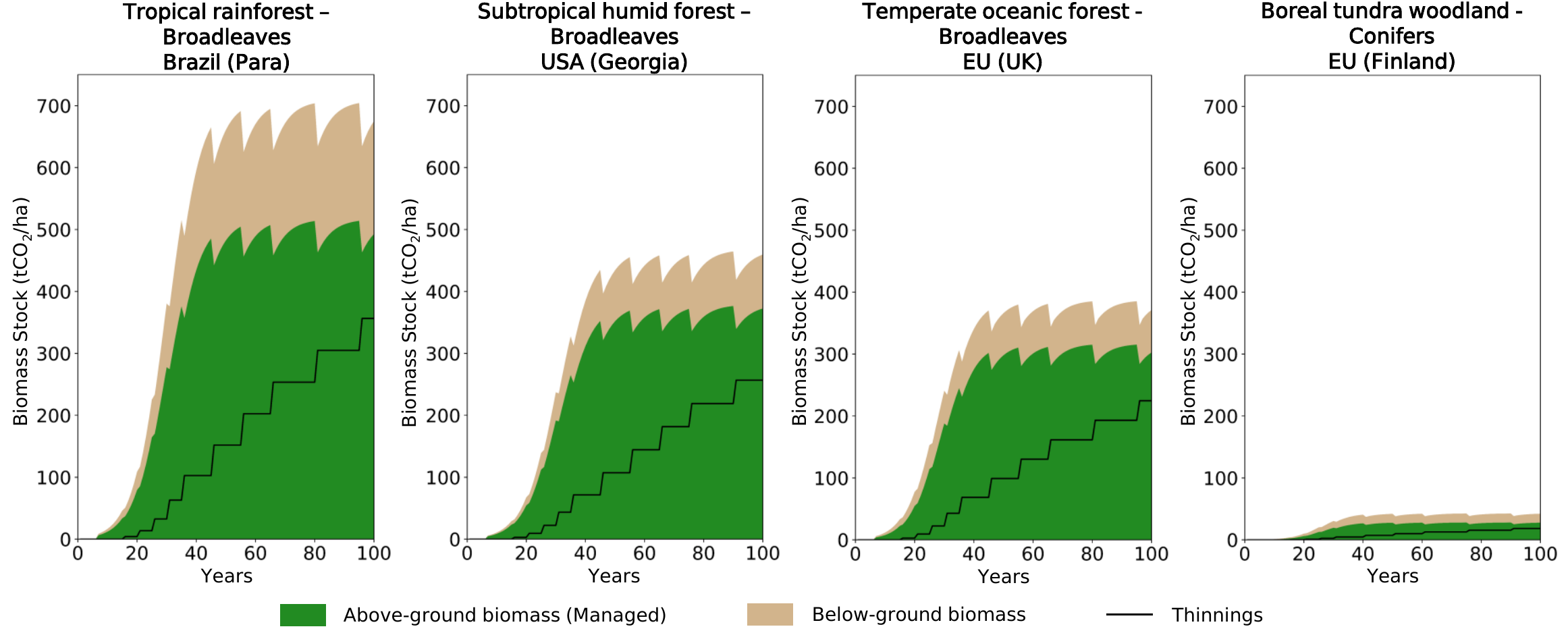}
\caption{Examples of total (above-ground and below-ground) biomass CO$_2$ stocks and thinnings CO$_2$ stocks for different ecological zones and forest types over a default period of 100 years.}
\label{fig: CO2 Growth}
\end{figure*}

Examples of biomass CO$_2$ stocks show the typical sigmoid pattern of growth in even-aged forest stands and the impact of thinning in Fig. \ref{fig: CO2 Growth}.
We observe that: 1) not only there is a delay of approximately 10--20 years between the establishment of new forest stands (in the establishment phase) and their effective CO$_2$ sequestration potential (in the full-vigour phase), 2) but these new forest stands also saturates after approximately 40--80 years, which implies that the annual rate of CO$_2$ sequestration is reduced to approximately zero.

Moreover, because of the higher biomass stock (on a per hectare basis) in warm climates --- tropical climates --- compared to cold climates --- boreal climates --- (300 t$_{DM}$/ha in tropical rainforests compared to 15 t$_{DM}$/ha in boreal tundra woodlands, as indicated in Table \ref{tbl IPCC above-ground biomass}), the CO$_2$ sequestration potential of AR is greatest in Brazilian States, such as Para than in Northern EU countries, such as Sweden.
By the time forest stands reach maturity (in the old growth-phase), we find that the maximum CO$_2$ sequestration potential of AR ranges between 40–-709 t CO$_2$/ha.

\paragraph*{A.1.2.5 "Harvested wood products" C pool.}

The duration of the C contained in the "harvested wood products" pool varies with the product itself and its uses \cite{Pingoud2006}.
As a results, there are currently several different approaches to account for C (and CO$_2$) sequestration in wood products.

Here, all C contained in extracted above-ground biomass (resulting from thinning operations) --- both thinnings and forest residues --- is included into the "harvested wood products" C pool.
Because thinning operations are carried out in the context of climate mitigation rather than timber production, the "harvested wood products" C pool is excluded from the AR's whole-system boundaries. 
All C contained in this pool can therefore be considered as a "C leakage".
This is equivalent to assume conservatively that all C contained in the "harvested wood products" pool is directly emitted to the atmosphere, as it is the case in the dead organic matter or soil organic matter pools.

Moreover, the below-ground biomass stock, proportional to the above-ground biomass stock, is also affected by thinning operations.
Because all C contained in the affected below-ground biomass pool should be transferred to the dead organic matter pool, it is therefore simplified accordingly, and assumed to be directly emitted to the atmosphere.

The thinning C stock $C^{Thinning}$ and CO$_2$ stock $CO_2^{Thinning}$ are estimated with the use of the carbon content factor $C_f$ and the ratio of molecular weights between C and CO$_2$.
This is shown in Eqs. \ref{eq: C Thinning} and \ref{eq: CO2 Thinning} below:

\begin{equation}
\label{eq: C Thinning}
C^{Thinning}(yr) = \sum_1^{yr} AT(yr) \times C_f \qquad \forall \; yr
\end{equation}

\begin{equation}
\label{eq: CO2 Thinning}
CO_2^{Thinning}(yr) = C^{Thinning}(yr) \times \frac{44}{12} \qquad \forall \; yr
\end{equation}

From Fig. \ref{fig: CO2 Growth}, we find that CO$_2$ leakages, \textit{via} the extraction of thinnings and forest residues, are estimated to be 43--70\% of the total CO$_2$ sequestrated in forest stands (above-ground and below-ground biomass CO$_2$) over a 100 years time-period.
Although such time-period is greater than the averaged human life span, managing intelligently such forest residues supply chain could increase the overall CO$_2$ sequestration potential of AR. 
Examples include the use of forest residues as a feedstock for BECCS.

\subsubsection*{A.1.3 ~~ Fire-penalty model.}

Here, we discuss the permanence of CO$_2$ sequestration \textit{via} AR, subject specifically to the risk of wildfires, and describe in details the resulting fire-penalty model.

\paragraph*{A.1.3.1 ~ Permanence of biogenic CO$_2$ sequestration.}

Forests are vulnerable to natural disturbances, such as drought, hurricanes, forest fires and pests, or to human-induced reversals, such as active deforestation.
Consequently, the permanence of biogenic CO$_2$ sequestration is less reliable than the one of geological CO$_2$ sequestration, such as in the cases of BECCS or DACCS.

Because the impact of natural disturbances on forest stands can be catastrophic, both in terms of biodiversity or financial losses --- specifically in the context of timber production ---, how the risk of natural disturbances should be anticipated and integrated in forest management has been increasingly investigated.
However, the focus has been predominantly set on maximising timber productivity and economic value, with or without carbon sequestration benefits, such as carbon price or carbon tax, and scarcely on minimising biomass and resulting CO$_2$ sequestration losses \cite{Stainback2004,Hu2016,Couture2011}. 

A few risk-accounting methods have been introduced, specifically for hurricanes or wildfires \cite{McNulty2002,Hurteau2009,Chuvieco2014,Szpakowski2019}, although most of the literature focuses on wildfires.
In spite of the increasing widespread use of remote-sensing --- the use of satellites to search for and collect geo-spatial data ---, these risk-accounting methods remains site-, region-, or at most, country-specific.
For instance, the Landscape Fire and Resource Management Planning Tools (LandFire) Program in the USA has been providing national geo-spatial datasets (partially or completely based on remote-sensing) on vegetation distribution, fire regime and other fuel characteristics \cite{Landfire2014}. 
The NASA Land Use and Land Cover Program has also been releasing the MODIS Active Fire Products and Burned Area Products, ones of the most complete datasets at the global scale, but insufficient for the evaluation of wildfires' risk \cite{ Giglio2016}.

Here, we model the risk of wildfires in the form of a penalty coefficient in order to evaluate the impact of such wildfires on AR's CO$_2$ sequestration potential.

\begin{figure}[t]
\centering
\includegraphics[width=8.5cm]{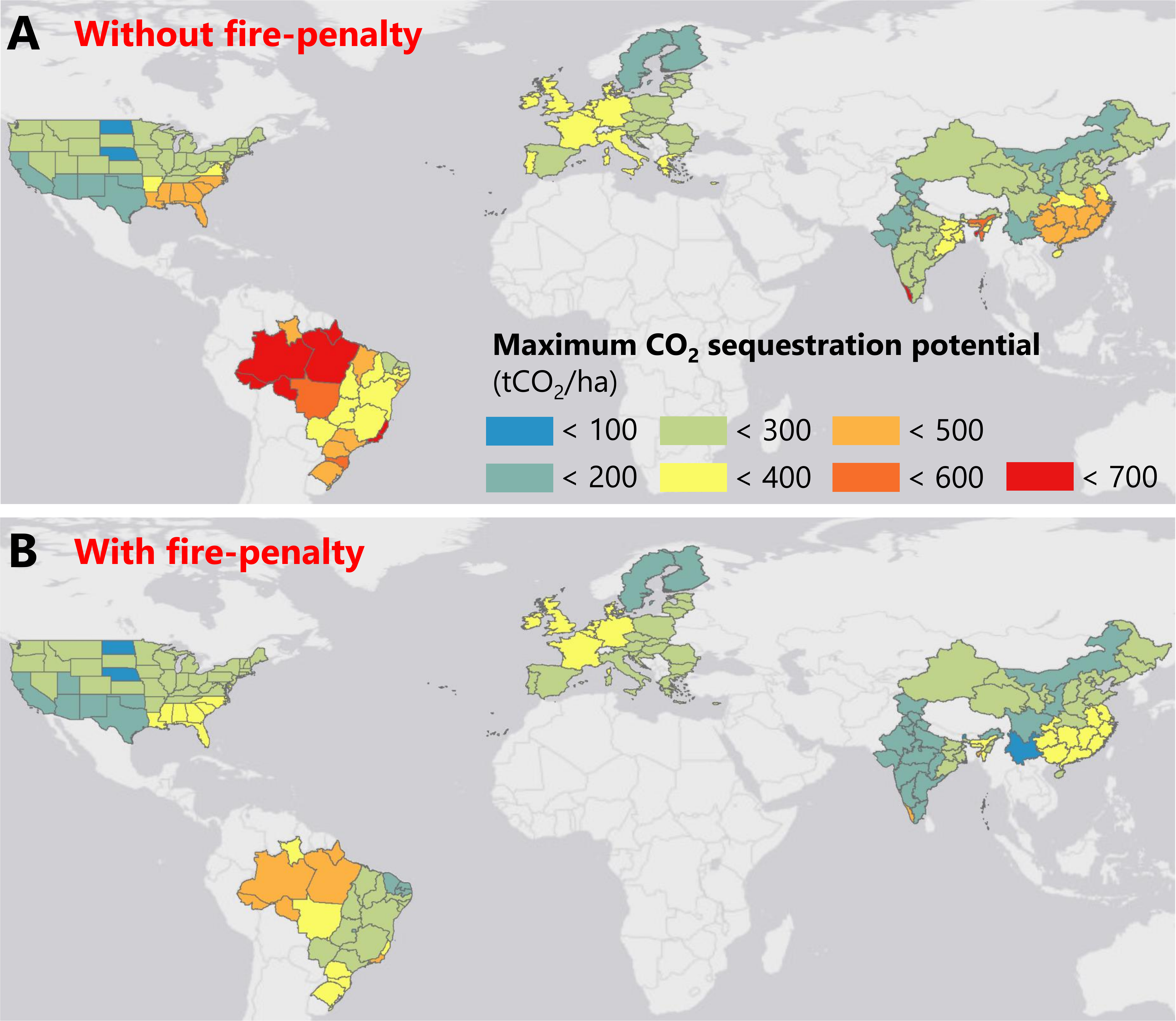}
    \caption{AR's maximum CO$_2$ sequestration potential: A) without considering the risk of wildfire, and B) with considering the risk of wildfire.}
    \label{fig: CO2 seq fire}
\end{figure}

\paragraph*{A.1.3.1 ~ Wildfire-penalty coefficient.}

We adapt the risk-accounting methodology developed in Hurteau \textit{et al.} \cite{Hurteau2009} to define a wildfire-penalty coefficient $R_{fire}$, characterised by ecological zones, and therefore applicable at the global scale.
$R_{fire}$ is built upon wildfires' severity --- the potential biomass loss given a fire occurrence --- and periodicity --- the probability of a fire event occurring during a specified time period, as shown in Eq. \ref{eq: fire penalty}. 
As in Hurteau \textit{et al.} \cite{Hurteau2009}, wildfires' severity and periodicity are respectively quantified by the mean fire return interval (mFRI) $mFRI$ --- ranging from 0 years (very frequent) to 1,000 years (very rare) --- and the vegetation departure index (VDep) $VDep$ --- ranging from 0\% (zero potential biomass loss) to 100\% (complete potential biomass loss).
Eq. \ref{eq: fire penalty} is written as follows:

\begin{equation}
\label{eq: fire penalty}
    R_{fire} = \begin{dcases} VDep \times (1 - \frac{mFRI}{P}) & \mbox{if} \; mFRI \leq P \\
0 &\mbox{if} \; mFRI > P \end{dcases}
\end{equation}

\noindent with: 
\begin{list}{\labelitemi}{\leftmargin=1em}
\item $R_{fire}$ is the fire-penalty (-), 
\item $mFRI$ is the mean fire return interval (yrs),
\item $VDep$ is the vegetation departure index (-),
\item and $P$ is the permanence (yrs), which represents the length of time that biogenic carbon must be sequestered to be accounted as permanent here. $P$ default value is 100 years.
\end{list}

mFRI and VDep are 2 geo-spatial datasets initially developed by the LandFire Program, and cover the US territory with a 30m geographic resolution \cite{Landfire2014}. 
Here, the mFRI and VDep datasets were aggregated at the ecological scale\footnote{$mFRI$ and $VDep$ are processed in ArcGIS 10.6 (ESRI).}
First, the existing vegetation type (EVT) dataset, also developed by the LANDFIRE project \cite{Landfire2014}, was used to restrict the mFRI and the VDep datasets to forest vegetation types.
Then, the global ecological zone (GEZ) dataset, developed by the FAO \cite{FAO2001}, was used to aggregate the restricted mFRI and VDep datasets with ecological zones and finally extrapolate the aggregated and restricted mFRI and VDep datasets at the global scale.
Because all ecological zones are not present on the US territory, values from “Tropical moist deciduous forest” were used for the “Tropical rainforest” ecological zone (closest ecological zone), and default values of 1000 years and 0 were assigned to $mFRI$ and $VDep$, respectively, in boreal ecological zones (extremely low risk of wildfire in such cold and humid climate). 
Resulting values for $mFRI$, $VDep$ and $R_{fire}$ are provided in Table \ref{tbl mFRI VDep}.

\begin{table}
\small
\caption{Mean fire return interval mFRI and vegetation departure index VDep values.}
\begin{tabular*}{0.48\textwidth}{@{\extracolsep{\fill}}llll}
    \hline
    Ecological zone & \makecell[l]{mFRI \\ (yrs)} & \makecell[l]{VDep \\ (\%)} & \makecell[l]{Wildfire-penalty \\ coefficient (\%)} \\
    \hline
    \multirow{5}{*}{\makecell[l]{Tropical rainforest$^a$,\\
    Tropical moist deciduous forest,\\
    Tropical dry forest$^a$,\\
    Tropical shrubland$^a$,\\
    Tropical mountain systems$^a$} }
    & \multirow{5}{*}{46} & \multirow{5}{*}{65} & \multirow{5}{*}{65} \\
    & & &\\
    & & &\\
    & & &\\
    & & &\\
    \hline
    Subtropical humid forest & 50 & 57 & 72 \\
    Subtropical dry forest & 24 & 30 & 77 \\
    Subtropical steppe & 52 & 47 & 77 \\
    Subtropical mountain systems & 40 & 43 & 74\\
    \hline
    Temperate oceanic forest & 424 & 61 & 100 \\ 
    Temperate continental forest & 165 & 61 & 100 \\
   Temperate mountain systems & 158 & 51 & 100\\ 
    \hline
    \multirow{3}{*}{\makecell[l]{Boreal coniferous forest$^b$,\\ Boreal tundra woodland$^b$,\\
     Boreal mountain systems$^b$}} 
     & \multirow{3}{*}{\makecell[l]{1,000}} & \multirow{3}{*}{\makecell[l]{0}} &
     \multirow{3}{*}{\makecell[l]{0}}\\
     & & & \\
     & & & \\
         \hline
    \multicolumn{4}{l}{\makecell[l]{$^a$ Used tropical moist deciduous values for mFRI and VDep.}}\\ 
    \multicolumn{4}{l}{\makecell[l]{$^b$ Used default values of 1,000 and 0 for mFRI and VDep, respectively.}}\\
    \end{tabular*}
    \label{tbl mFRI VDep}
\end{table}

\paragraph*{A.1.3.2 ~ CO$_2$ sequestration potential.}

AR's CO$_2$ sequestration potential $CO_2^{Seq}$ is evaluated as shown below in Eq. \ref{eq: CO2 seq fire}:

\begin{equation}
\label{eq: CO2 seq fire}
CO_2^{Seq}(yr) = CO_2^{Total}(yr) \times (1- R_{fire}) \qquad \forall \; yr
\end{equation}

The maximal CO$_2$ sequestration potential of AR in Brazil, China, the EU, India and the USA, at the regional scale is illustrated in Fig. \ref{fig: CO2 seq fire}.
We find that AR's CO$_2$ sequestration potential is not affected by wildfires in boreal and temperate climates, whereas it decreases by 23--29\% in subtropical climates, and up to 35\% in tropical climates.

\subsubsection*{A.1.4. ~ Forestry operations model.}

Here, we describes in details the forestry operations model.
Specifically, AR requires the establishment and the on-going maintenance of the forest to maximise and maintain CDR. 
These include site establishment, forest roads construction and annual maintenance, and annual forestry (thinning) operations \cite{Whittaker2011,Morison2012,Roder2015}. 

\paragraph*{A.1.4.1 ~ Site establishment.}

The forest is established by land preparation and planting of new seedlings. 
For land preparation, mounding is carried out by an excavator \cite{Whittaker2010,Whittaker2011, Morison2012}, and herbicide and fertiliser are applied using a tractor \cite{Whittaker2011, Morison2012}. 
Tree seedlings are prepared in nurseries \cite{Aldentun2002}, then planted by hand \cite{Timmermann2014,Whittaker2011,Morison2012}.

\paragraph*{A.1.4.2 ~ Forest road construction \& maintenance.}

The access and the maintenance of forests requires forest roads, that are classified according to the frequency of their usage. 
Specifically, type A roads are the principal routes, used very frequently and type B road are secondary routes, only used during specific activities, such as thinning operations.
Here, we assume that forest roads construction (for types A and B roads) involves spreading blasted rock on top of the soil, and then covering with a layer of finer, crushed aggregate \cite{Whittaker2010,Whittaker2011, Morison2012}.
Forest roads maintenance depends on the road type.
Type A roads maintenance involves the re-grading of the road every year, or the re-surfacing --- the re-application of the top layer of aggregate --- before thinning operations, whereas type B roads maintenance only involves the re-grading and rolling of the remaining aggregate layer before thinning operations \cite{Whittaker2010,Whittaker2011, Morison2012}.
Mining and crushing of road rock and aggregate are also included \cite{Whittaker2010,Whittaker2011}, as contributing to forest road construction \& maintenance's indirect GHG emissions.

\paragraph*{A.1.4.3 ~ Annual forestry (thinning) operations.}

As part of the FMC, a selection of trees is thinned using a cut-to-length logging system. 
This involves the felling and the extraction of trees from the forest site using a combination of harvesters and forwarders \cite{Whittaker2011, Morison2012,Berg2003,Kenney2014}. 
Here, we assume that this selection of thinned forest biomass is composed of 80\% thinnings (whole tree thinnings and roundwood) and 20\% forest residues, such as branches, foliage or bark \cite{Roder2015}.
Early whole tree thinnings involves tree felling by harvesters, followed by whole tree removal from the site to the roadside by forwarders. 
Harvesting roundwood requires the use of harvesters that cut and top the trees, leaving branches and other forest residues on the forest floor. 
The roundwood is then transported to the roadside by forwarders. 
Lastly, 35\% of the residues are left in the forest for ecological reasons \cite{Roder2015} --- forest residues are left on the forest floor to maintain the nutrient and soil carbon balance ---, and the remainder is collected and extracted by forwarders that compress the forest residues into bundles.
All thinned and extracted forest biomass are then stored at the roadside to allow for natural drying from 50\% to 30\% moisture content \cite{Whittaker2011,Roder2015}.

Dry matter losses are also included at every step of the forestry operations --- tree felling, harvesting, forwarding and storage --- resulting in a total loss percentage of 11.6\%. 
This value is consistent with the literature \cite{Roder2015, IPCC2006c}. 
As a reference, the IPCC defaults values for harvest loss are 10\% for broadleaves and 8\% for conifers \cite{IPCC2006c}.

\subsubsection*{A.1.5 Energy balance.}

AR's total energy requirements include:

\begin{list}{\labelitemi}{\leftmargin=1em}
\item direct energy (energy density) of the combustion of fuels \cite{BEIS2020} or the consumption of electricity;
\item indirect energy (embodied energy) due to the production of these fuels \cite{Edward2014} and the generation of electricity \cite{Eurelectric2003};
\item and indirect energy (embodied energy) due to the production of materials, such as agrochemicals \cite{Camargo2013,Lewandowski1995,Bullard2001,Elsayed2003}, seedlings \cite{Camargo2013,Lewandowski1995,Elsayed2003,Smeets2009} or road rock and aggregate \cite{Whittaker2010,Whittaker2011,BEIS2020}. 
\end{list}

\subsubsection*{A.1.6 CO$_2$ (and N$_2$O) balance model.}

AR's total CO$_2$ (and N$_2$O) emissions $CO_2^{Emissions}$ include:

\begin{list}{\labelitemi}{\leftmargin=1em}
\item direct CO$_2$ emissions from the combustion of fuels \cite{BEIS2020}, 
\item indirect CO$_2$ emissions due to the production of these fuels \cite{BEIS2020} and the generation of electricity \cite{BEIS2020,IEA2018};
\item indirect CO$_2$ emissions due to the production of materials, such as agrochemicals \cite{Camargo2013,Lewandowski1995,Bullard2001,Elsayed2003}, seedlings \cite{Camargo2013,Lewandowski1995,Elsayed2003} or road rock and aggregate \cite{Whittaker2010,Whittaker2011,BEIS2020}. 
\item direct N$_2$O emissions \cite{IPCC2006d,Parajuli2014,Murphy2013,Mekonnen2011} arising from the application of nitrogen-based fertiliser during the forest establishment and from the use of ammonium nitrate-based explosive for road rocks extraction;
\item and direct and indirect CO$_2$ emissions arising from land-use change (LUC).
\end{list}

For safeguards of sustainability and biodiversity, we assume that AR's deployment is restricted to lands with reforestation potential (LRP), as identified in Griscom \textit{et al.} \cite{Griscom2017} (See Appendix C).
Although the climato-ecological characteristics of LRP are well detailed, there are little information on the current use of LRP.
Specifically, because LRP are characterised by low tree cover (and therefore low biomass density) and current croplands are excluded from LRP, we assume that: 1) the quality of LRP is similar to the one of marginal agricultural lands (MAL), and therefore the direct LUC is equal to 25 kgCO$_2$/ha; and 2) no human activities are displaced by AR's deployment on LRP, and therefore the indirect LUC is null.
The methodology used for the accounting of N$_2$O emissions and LUC (direct \& indirect) emissions has been presented previously \cite{Fajardy2017,Fajardy2020b} and will not be repeated here.

Overall, AR's CO$_2$ balance --- AR's CDR potential --- $CO_2^{Removal}$ is the difference between AR's CO$_2$ sequestration potential (discounted by a "fire-penalty" coefficient) $CO_2^{Seq}$ and AR's CO$_2$ emissions $CO_2^{Emissions}$, as shown in Eq. \ref{eq: CO2 Removed}:

\begin{equation}
\label{eq: CO2 Removed}
CO_2^{Removal}(yr) = CO_2^{Seq}(yr) - CO_2^{Emissions}(yr) \qquad \forall \; yr
\end{equation}

\begin{figure}[t]
\centering
\includegraphics[width=8.5cm]{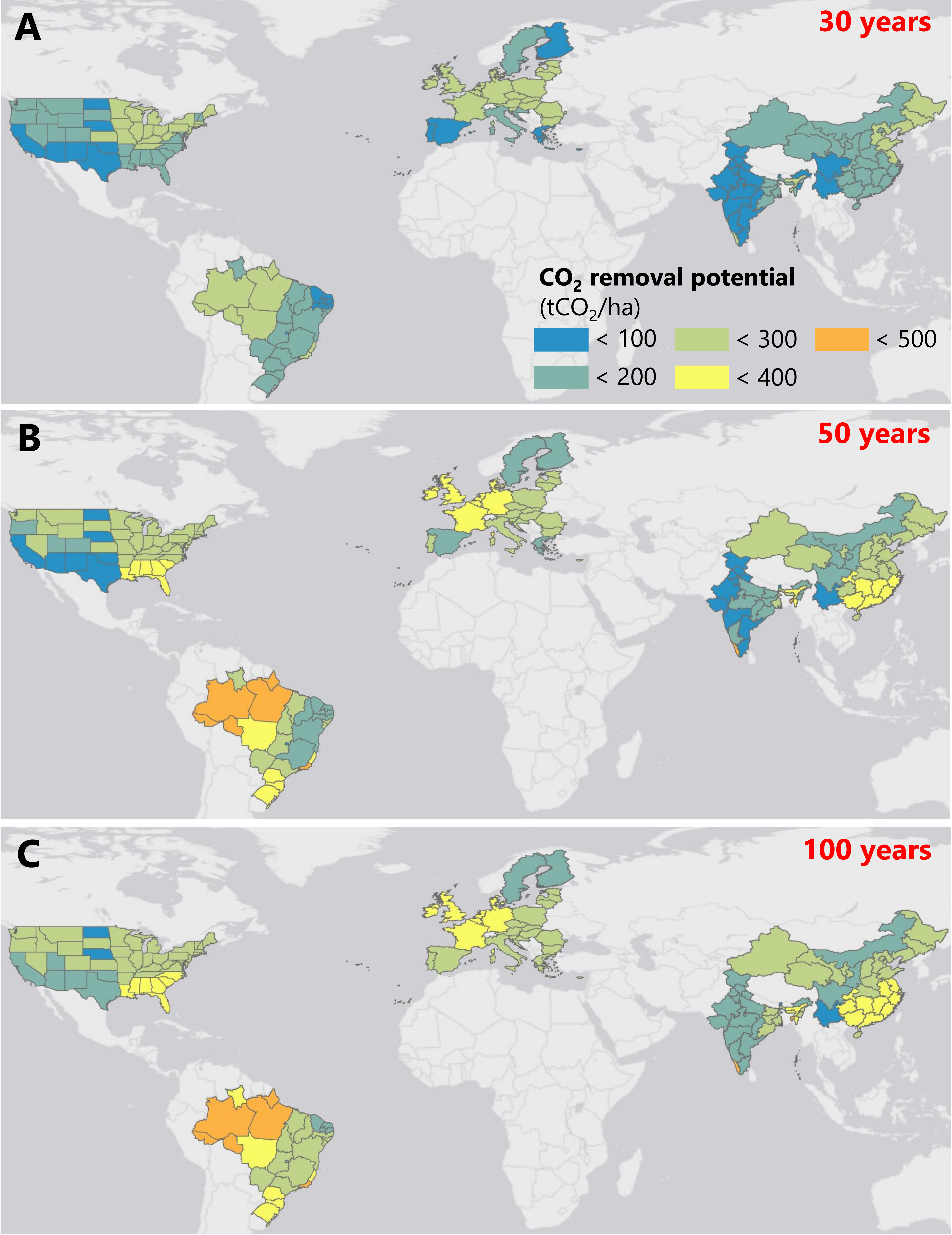}
    \caption{AR's CO$_2$ removal potential: A) over 30 years, B) over 50 years, and C) over 100 years.}
    \label{fig: CDR 30 50 100 yrs}
\end{figure}

Fig. \ref{fig: CDR 30 50 100 yrs} illustrates AR's CDR potential over time-periods of 30, 50 and 100 years in Brazil, China, the EU, India and the USA.
AR's CDR potential slowly increases during the first 30 years, reaching up between 200--250 GtCO$_2$ in Central EU, North-East Brazil, North-West USA and North-West China.
After only 50 years, AR is deployed up to its maximum CDR potential in most regions of the world, reaching up to 400--430 GtCO$_2$ in North-East Brazil or South-West China.
After a 100 years time-period, AR's maximum CDR potential is reached everywhere across the world, greatest in Brazil owing to warm and humid climates, and lowest in Northern EU countries owing to boreal climates, or in India owing to warm, yet dry climates.
Moreover, we find that AR's total CO$_2$ emissions are negligible compared to AR's CDR sequestration potential over a 100 years time-period, ranging between 1.2--4.4 tCO$_2$/ha.
Overall, although AR: 1) needs at least 10--20 years before effectively removing CO$_2$ from the atmosphere, and 2) saturates after 40--80 years, AR is found to be significantly efficient at removing CO$_2$ from the atmosphere, as much as 88.8--95.8\% over 30 years, and 97.5--99.6\% over 100 years.

\subsubsection*{A.1.7 Cost balance model.}

AR's total costs include:

\begin{list}{\labelitemi}{\leftmargin=1em}
\item the cost of energy, such as fuels \cite{WorldBank2020d} and electricity \cite{MME2018,CEIC2020,Eurostat2020a,GovIndia2014,EIA2020a}; 
\item the cost of machinery, such as trucks or excavator for land preparation, harvester and forwarder for harvesting operations, or other machinery for road construction and maintenance \cite{Brinker2002,Johansson2006};
\item the cost of labour, such as ground worker, forest worker, road operator, etc \cite{BLS2019}; 
\item and the cost of feedstocks and materials, such as agrochemicals \cite{Duffy2008,deWit2010}, seedlings \cite{AlbaTree2020} or road rocks \cite{HomeGuide2020}.
\item and the cost of land.
\end{list}

In this study, costs are expressed in 2018 US \$, but disaggregated at the national level.
When available in another currency or another year, costs of energy, machinery and feedstocks are converted with the use of exchange rate and inflation factors respectively \cite{WorldBank2020a,WorldBank2020b}.
When only available for one country (often the USA), costs of energy, machinery and feedstocks are converted with the use of purchasing power parity (PPP) \cite{WorldBank2020c}, and costs of labour are converted proportionally to national hourly compensation costs from the Conference Board, as described previously \cite{Fajardy2020a,Fajardy2020b}. 

Because AR's deployment is restricted to LRP, of which the climato-ecological characteristics and the current usage can be assimilated to the ones of MAL (low quality of land, unused for crop production), we assume, similarly to MAL, that the cost of LRP is null.

AR's initial investment, due to the establishment of the forest and the construction of forest roads, is levelised with the use of a capital recovery factor (CRF), itself calculated with an interest rate of 8\% and a financial lifetime of 30 years. 

\begin{figure}[h]
\centering
\includegraphics[width=8.5cm]{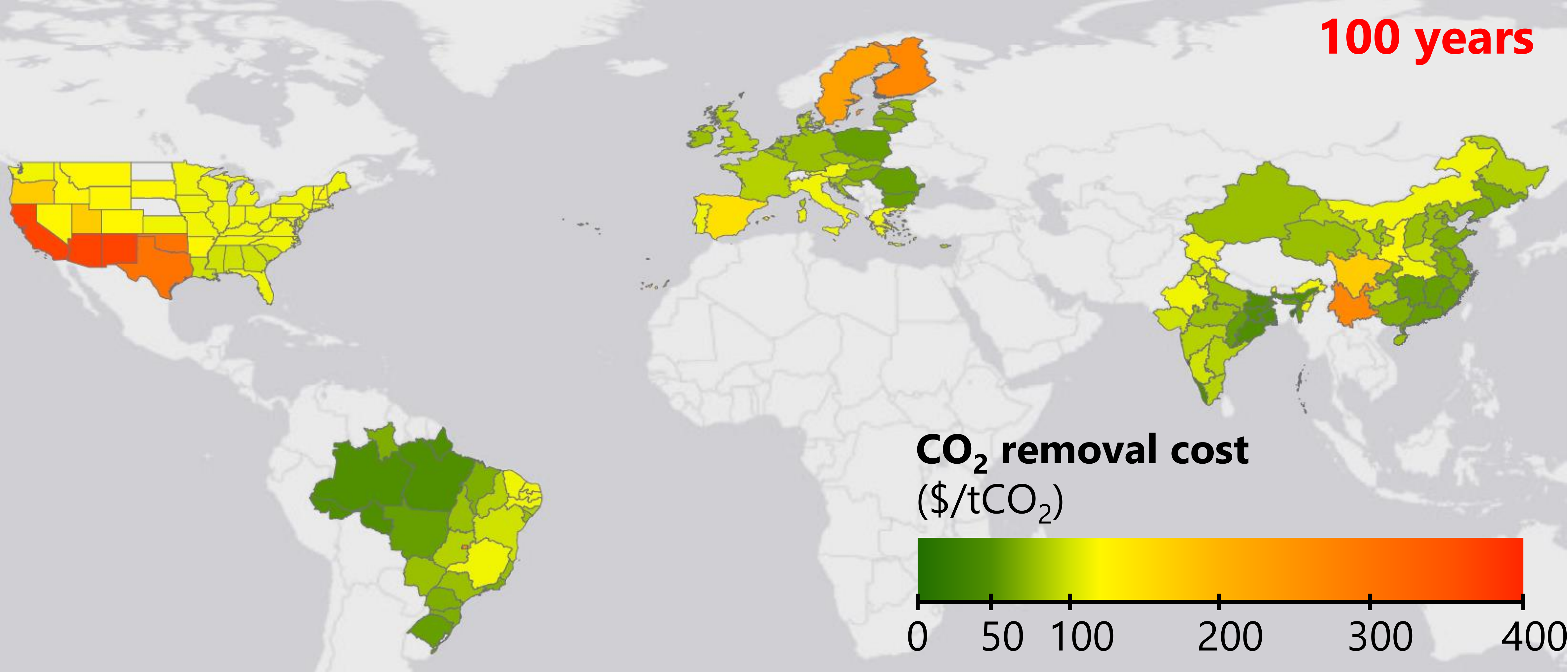}
    \caption{AR's CO$_2$ removal cost over a 100 years time-period. Because of different economies across the world and different AR's CDR potentials within these economies, AR's CO$_2$ removal cost varies at the global and national scale. }
    \label{fig: Cost 100 yrs}
\end{figure}

Fig. \ref{fig: Cost 100 yrs} illustrates AR's CO$_2$ removal costs over a 100 years time-period in Brazil, China, the EU, India and the USA.
Because of different economies across the world and different AR's CDR potentials within these economies, AR's CO$_2$ removal cost varies at the global --- between the 5 regions considered in MONET --- and national scale --- among each of the 5 regions considered in MONET.
Overall, CO$_2$ removal \textit{via} AR is the most expensive in the USA, ranging between \$77-395/tCO$_2$ whereas it is the cheapest in India, ranging between \$22--139/tCO$_2$, and in Brazil, ranging between \$44--242/tCO$_2$.
Finally, AR's CO$_2$ removal costs between \$52--247/tCO$_2$ in the EU, and between \$59--265/tCO$_2$ in China.

\subsection*{A.2 ~~ Bioenergy and Carbon Capture and Storage (BECCS) model}

Combining bioenergy production with CCS can provide net negative emissions as the CO$_2$ captured by photosynthesizing biomass growth is geologically sequestered rather than released to the atmosphere during its conversion into energy.
BECCS can involve various biomass feedstocks, such as trees, crops, algae, organic wastes, and biomass conversion pathways, such as power, liquid fuel, hydrogen. 
Usually, BECCS simply refers to: 1) the integration of growing trees and/or crops that extract CO$_2$ from the atmosphere, 2) their combustion in power plants to produce electricity, and the application of CCS \textit{via} CO$_2$ capture at the power plant and 3) the transport and injection of CO$_2$ into geological formations.
This study focuses on biomass --- woody and herbaceous crops, agricultural residues and forest residues --- combustion for power.

BECCS's CDR potential has been shown to highly depend on biomass supply chain management and LUC \cite{Fajardy2017}, and to not not always ultimately result into negative emissions.
Biomass supply chain --- LUC, biomass cultivation, processing and transport to the power plant --- requires energy and emits GHG, and must therefore be considered in the evaluation of BECCS's CO$_2$ balance, in order to ensure BECCS's carbon negativity.
The power plant technology assumed in this study is a 500 MW dedicated pulverised biomass thermal power plant, combined with post-combustion amine-based carbon capture with a CCS efficiency of 90\%.
The CO$_2$ capture capacity of the BECCS plant is approximately 4.2 MtCO$_2$/yr \cite{Fajardy2017}. 
The BECCS plant is assumed to be in the vicinity of geological formations, with a distance of 100 km for CO$_2$ transport.
Finally, the production of low-carbon electricity by BECCS is assumed to generate revenues that can, in some configurations, be greater than BECCS total cost.
These revenues from electricity sales are received at the region-specific wholesale price of electricity.
BECCS's whole-system model has been developed and presented in previous work \cite{Fajardy2017, Fajardy2018, Fajardy2020a, Fajardy2020b}, and will not be repeated here.

\subsection*{A.3 ~~ Direct Air CO$_2$ Capture and Storage (DACCS) model}

With DACCS, CO$_2$ is directly captured from the atmosphere using a range of sorbents or solvents, and then transported and injected into geological formations.
Because of the low atmospheric concentration of CO$_2$ (approximately 412 ppm \cite{GML2020}), DACCS is very energy intensive --- DACCS's total energy requirements, including the separation of CO$_2$ from the air and the compression of CO$_2$, have been reported 4 times greater than conventional CCS's total energy requirements \cite{Herzog2021}.
Because of this, and the low maturity of the technology, DACCS's cost is still uncertain and expensive, ranging between \$30-1,000/tCO$_2$ \cite{Sanz-Perez2016,Fuss2018}.

Two archetypal DAC plants are currently being developed at the demonstration-scale. 
The first one, developed by Carbon Engineering Ltd., captures CO$_2$ directly from the air with a potassium hydroxide (KOH) sorbent in the air contactor and stores it as a carbonate (K$_2$CO$_3$) \cite{Keith2018, Socolow2011}.
The sorbent is then regenerated by reacting K$_2$CO$_3$ with a calcium hydroxide (Ca(OH)$_2$) in the pellet reactor. 
Ca(OH)$_2$ is obtained in the slacker by hydrating calcium oxide (CaO), which itself is produced by calcining calcium carbonate (CaCO$_3$) in a kiln (also called calciner), at 900$^{\circ}$C. 
The high-temperature heat required by the regeneration process is currently supplied by the combustion of natural gas, of which the CO$_2$ emissions are also captured in a CO$_2$ absorber.
The second one, developed by Climeworks, captures CO$_2$ with an amine-functionalised sorbent on a filter \cite{Wurzbacher2017,Lackner2009}. 
Once the filter is saturated, it is heated to 100$^{\circ}$C, and the CO$_2$ is released and collected. 
The low-temperature heat of the regeneration process can be provided by the electricity grid.
For both DAC plant archetypes, fans, liquid pumping and CO$_2$ compression also requires power from the electricity grid.
Both archetypes --- Carbon Engineering (DACCS-CE) and Climeworks (DACCS-CW) --- have been implemented in this study.

Similarly to BECCS plants, we assume that DAC plants are built and operating in the vicinity of geological formations that are suitable for CO$_2$ storage, at a distance of 100 km.

In this study, we have developed an explicit spatio-temporal model of DACCS whole-system model, integrating a DAC plant and CO$_2$ transport and storage (T\&S).
For each step of the model, energy, CO$_2$ and cost balances are carried out.
Spatial resolution of DACCS's whole-system model is at the State/Province in Brazil, China, India and the USA and at the country in the EU (EU-27 \& UK). 
Temporal resolution is 10 years (decadal), ranging between 2020--2100.

\paragraph*{A.3.1 ~ Energy balance model.}

DACCS's total energy requirements include:

\begin{list}{\labelitemi}{\leftmargin=1em}
\item direct energy requirements (energy density) from the combustion of natural gas in the case of Carbon Engineering archetype or from the consumption of electricity in both archetypes' cases \cite{Keith2018,Climeworks2018};
\item and indirect energy requirements (embodied energy) from the production of natural gas \cite{Edward2014} in the case of DACCS-CE archetype, the production of electricity \cite{Eurelectric2003}, and the T\&S of CO$_2$ \cite{McCollum2006}.
\end{list}

\paragraph*{A.3.2 ~ CO$_2$ balance model.}

DACCS's CDR potential is equal to the difference between DACCS's CO$_2$ captured at the DAC plant -- both from the atmosphere and from the combustion of natural gas in the case of DACCS-CE archetype -- and DACCS's total CO$_2$ emissions.
DACCS's total CO$_2$ emissions include:

\begin{list}{\labelitemi}{\leftmargin=1em}
\item direct CO$_2$ emissions from the combustion of natural gas \cite{BEIS2020} in the case of DACCS-CE archetype,
\item and indirect CO$_2$ emissions from the production of natural gas \cite{Balcombe2017,Schuller2017} in the case of DACCS-CE archetype, the production of electricity \cite{BEIS2020,IEA2018}, and the T\&S of CO$_2$ \cite{Smith2013}.
\end{list}

\begin{figure*}[t]
\centering
\includegraphics[width=17.1cm]{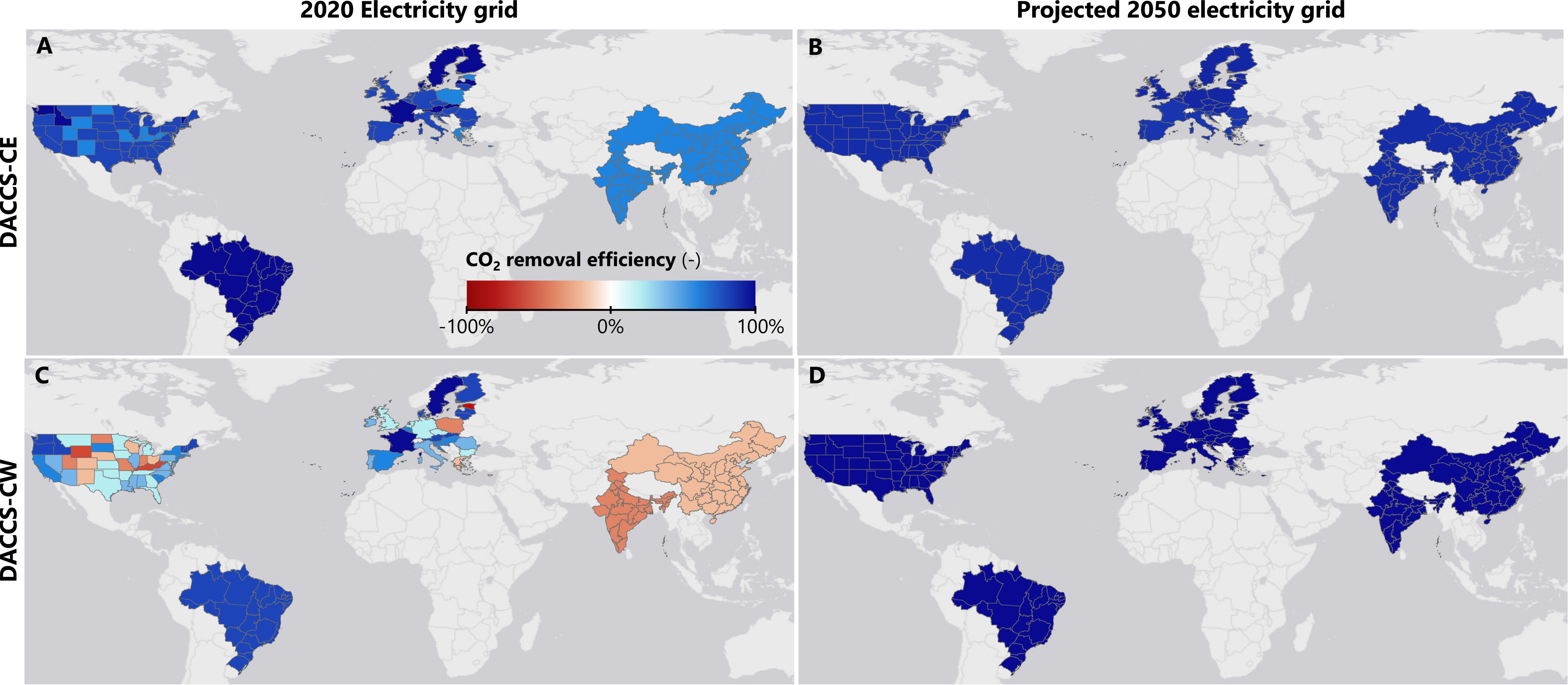}
    \caption{DACCS's CO$_2$ removal efficiency for Carbon Engineering and Climeworks archetypes, in 2020 and after 2050.}
    \label{fig: DACCS CDR eff}
\end{figure*}

Fig. \ref{fig: DACCS CDR eff} shows the impact of the energy sector (specifically the power sector) decarbonisation on the DACCS's CDR efficiency for both DACCS archetypes (DACCS-CE and DACCS-CW). 
Considering the current --- 2020 --- carbon intensity of national (and sub-national) electricity grids, DACCS-CE archetype's efficiency ranges between 44--88\%, whereas the one of DACCS-CW archetype ranges between -79--92\%.
Specifically, DACCS-CE archetype not only fails at removing CO$_2$ from the atmosphere in China, in  India, and in a few EU countries and US States, but it even emits more CO$_2$ into the atmosphere.
However, as the power sector follows a complete decarbonisation by 2050, both archetypes's CDR efficiencies increase, up to 86--88\% in DACCS-CE's case and 94\% in DACCS-CW's case.
The carbon intensity of DACCS's energy plays therefore a significant and determinant role in DACCS's CDR efficiency.
Crucially, the decarbonisation of the power sector appears like a necessary prerequisite for the deployment of DACCS.

\paragraph*{A.3.3 ~ Cost balance model.}

DACCS's total cost includes:

\begin{list}{\labelitemi}{\leftmargin=1em}
\item the cost of energy such as electricity \cite{MME2018,CEIC2020,Eurostat2020a,GovIndia2014,EIA2020a} and natural gas \cite{MME2018,Hu2015,Eurostat2020b,GovIndia2014,EIA2020b};
\item the CAPEX and OPEX (including labour, operating and maintenance costs) of the DAC plant \cite{Keith2018,Climeworks2018},
\item the cost of CO$_2$ T\&S \cite{Johnson2014}.
\end{list}

For DACCS-CE archetype, DACCS's CO$_2$ capture cost has been broken down into energy (natural gas and electricity), CAPEX and OPEX in Keith \textit{et al.} \cite{Keith2018}.
Specifically, after excluding the energy cost, the levelised CAPEX and the OPEX account respectively for 69\% and 31\%, in all configurations\footnote{In Keith et al. \cite{Keith2018}, two DAC plant configurations are investigated. In the first, natural gas is used both for heat and power, whereas in the second, natural gas is replaced by electricity for power. In this study, we only consider the second DAC plant configuration.}
In this study, we conservatively assume that DACCS's CO$_2$ capture cost, including the energy cost, is twice more than suggested around \$235/tCO$_2$, while maintaining the ratio of CAPEX/OPEX as it is.
Assuming an interest rate of 8\% and a financial lifetime of 30 years, this results into a (non-levelised) CAPEX of \$1,600/tCO$_2$ captured, and an OPEX of \$65/tCO$_2$/yr captured.
For DACCS-CW archetype, DACCS's CO$_2$ capture cost has been reported around \$600/tCO$_2$ \cite{CarbonBrief2017}.
Conversely to DACCS-CE archetype, DACCS-CW archetype is a modular process, for which the DAC plant size is smaller (lower CAPEX), and is operated in a two time-steps, which requires more maintenance (higher OPEX).
Therefore, we assume that, after excluding the energy (electricity only) cost\footnote{Because Climeworks's first pilot plant was operated in Switzerland, we used the electricity price of Switzerland.}, the (levelised) CAPEX and the OPEX account both for 50\% of the remaining cost of CO$_2$ capture.
With an interest rate of 8\% and a financial lifetime of 30 years, this results into a (non-levelised) CAPEX of \$1,815/tCO$_2$ captured, and an OPEX of \$160/tCO$_2$/yr captured.
Finally, the methodology for evaluating the cost of CO$_2$ T\&S has already been presented previous and will not be repeated here.

\section*{Appendix B ~~ Equations of the MONET optimisation model}

The MONET framework used in this study has been extended to include AR and DACCS. 
The initial mathematical formulation of the MONET optimisation model, described previously \cite{Fajardy2020a,Fajardy2020b}, has been adapted accordingly. 
This is described here.

\subsection*{B.1 ~~ Overview}

The MONET framework is a linear optimisation problem (LP), that determines the optimal co-deployment of CDR pathways --- AR, BECCS and DACCS --- in line with the Paris Agreement's objectives of 1.5$^{\circ}$C, subject to the constraints below:

\begin{list}{\labelitemi}{\leftmargin=1em}
    \setlength\itemsep{-0.5em}
    \item \textbf{CDR targets constraints:}
    Cumulative CDR targets are specified throughout the whole planning horizon, according to the Paris Agreement's 1.5$^{\circ}$C objectives. This is discussed in Section \ref{sec: CDR Targets}.
    
    \item \textbf{Expansion of CDR solutions constraints:} 
    Operation conditions, specific to each CDR pathway, and including lifetime, maximum built rates for BECCS and DACCS and maximum expansion rate for AR, are specified.
   
    \item \textbf{Sustainability \& Land constraints:} 
    Region-specific bio-geophysical limits --- land and biomass supply availabilities, water stress --- are also specified.

    \item \textbf{Geological CO$_2$ storage constraints:} BECCS and DACCS plants are deployed in the vicinity of geological CO$_2$ storage --- a distance of 100 km --- with sufficient CO$_2$ storage capacity for the lifetime of the technology.
    CO$_2$ is assumed to be safely and permanently stored.
    Maximum CO$_2$ storage capacity for BECCS and DACCS are specified, owing to region-specific limits.
   
\end{list}

In this study, the MONET framework minimises the total net cost of the co-deployment of CDR pathways --- total capital costs (CAPEX) and operating costs (OPEX) minus total revenues ---.
It assumes perfect foresight over a 2020--2100 planning horizon, with a 10 years time-step.

\subsection*{B.2 ~~ Main equations}

Here, we describe in details the main equations of the MONET optimisation model, including AR, BECCS and DACCS.

\subsubsection*{B.2.1 ~~ Objective function.}

The objective function is the total cumulative net cost of CDR $tcb$ (\$).
Eq. \ref{eq: obj} denotes the objective function, which is minimised in the model formulation.

\begin{align}
\label{eq: obj}
& tcb = \sum\limits_{sr', ez, t='2100', a} cCost^{AF}(sr', ez, t, td) \times lu^{AF}(sr', ez, t, a) \nonumber \\
    &  + \sum\limits_{sr,sr', b, l, t} aCostB^{BECCS}(sr, sr', b, l, t)
      \times arco2^{BECCS}(sr, sr', b, l, t) \times 10 \nonumber \\
    &  + \sum\limits_{tk, sr', t} aCost^{DACCS}(tk, sr', t)  
      \times arco2^{DACCS}(tk, sr', t) \times 10
\end{align}

\noindent where:
\begin{list}{\labelitemi}{\leftmargin=1em}
\setlength\itemsep{-0.5em}
\item parameter $cCost^{AF}(sr', ez, t, a)$ is the cumulative cost per hectare of AR in sub-region \textit{sr'} and ecological zone \textit{ez}, in year \textit{t}, and deployed from year \textit{a} (\$/ha);

\item parameter $aCostB^{BECCS}(sr, sr', b, l, t)$ is the annual cost balance per tonne of CO$_2$ removed of BECCS with biomass \textit{b}, cultivated on land type \textit{l} in sub-region \textit{sr}, and transported to sub-region \textit{sr'} in year \textit{t} (\$/tCO$_2$ removed);
\item parameter $aCost^{DACCS}(tk, sr', t)$ is the annual cost per tonne of CO$_2$ removed of DACCS with technology \textit{tk}, in sub-region \textit{sr'} in year \textit{t} (\$/tCO$_2$ removed);

\item decision variable $lu^{AF}(sr', ez, t, a)$ is the land use of AR in sub-region \textit{sr'} and ecological zone \textit{ez}, in year \textit{t}, and deployed from year \textit{a} (ha); 

\item decision variable $arco2^{BECCS}(sr, sr', b, l, t)$ is the annual CO$_2$ removed \textit{via} BECCS with biomass \textit{b}, cultivated on land type \textit{l} in sub-region \textit{sr}, and transported to sub-region \textit{sr'} in year \textit{t} (tCO$_2$ removed/yr); 

\item and decision variable $arco2^{DACCS}(tk, sr', t)$ is the annual CO$_2$ removed \textit{via} DACCS with technology \textit{tk}, in sub-region \textit{sr'} in year \textit{t} (tCO$_2$ removed/yr).
\end{list}

\subsubsection*{B.2.2 ~~ Cumulative CDR targets constraint.}

Eqs. \ref{eq: cumu global CDR targets 1}--\ref{eq: cumu national CDR targets 2} introduce the cumulative CDR target constraints on the deployment of a portfolio of CDR pathways. 
In the COOPERATION scenario, the total cumulative CO$_2$ removed \textit{via} AR, BECCS and DACCS must be greater than or equal to the global CDR targets $gCDRT(t)$ in year $t$ (tCO$_2$ removed).
This is written below:
\\

\noindent $\forall \; t \leq \mbox{2100},$
\begin{align}
\label{eq: cumu global CDR targets 1}
& \sum\limits_{sr', ez, a} cRCO2^{AF}(sr', ez, t, a) \times lu^{AF}(sr', ez, t, a) \nonumber \\
&  + \sum\limits_{sr, sr', b, l, t} arco2^{BECCS}(sr, sr', b, l, t) \times 10 \nonumber \\
&  + \sum\limits_{sr',tk} arco2^{DACCS}(sr',tk,t) \times 10 \geq gCDRT(t)
\end{align}

\noindent $\forall \; t = \mbox{2100},$
\begin{align}
\label{eq: cumu global CDR targets 2}
& \sum\limits_{sr',ez,a} cRCO2^{AF}(sr',ez,t,a) lu^{AF}(sr',ez,t,a) \nonumber \\
&  + \sum\limits_{sr,sr',b,l} arco2^{BECCS}(sr, sr', b, l, t) \times 10 \nonumber\\
&  + \sum\limits_{sr',tk} arco2^{DACCS}(sr',tk,t) \times 10 = gCDRT(t)
\end{align}

\noindent where: parameter $cRCO2^{AF}(sr', ez, t, a)$ describes the cumulative CO$_2$ removed \textit{via} AR, in sub-region \textit{sr'} and ecological zone \textit{ez}, in year \textit{t}, and deployed from year \textit{a} (tCO2 removed/ha).

In the CURRENT POLICY and ISOLATION scenarios, national cumulative CO$_{2}$ removed \textit{via} AR, BECCS and DACCS must be greater than or equal to the national CDR targets $CDRT(c, t)$ in region \textit{c} and at time \textit{t} (tCO$_2$ removed).
This is written below:

\noindent $\forall \; c, \; t \geq \mbox{2100} ,$
\begin{align}
\label{eq: cumu national CDR targets 1}
  &  \sum\limits_{sr' \in c, ez, a} cRCO2^{AF}(sr',ez, t, a) \times lu^{AF}(sr', ez, t, a) \nonumber \\
   &  + \sum\limits_{sr, sr' \in c, b, l, t} arco2^{BECCS}(sr, sr', b, l, t) \times 10 \nonumber \\
   &  + \sum\limits_{tk, sr' \in c, t} arco2^{DACCS}(tk, sr', t) \times 10 = CDRT(c, t)
\end{align}

\noindent $\forall \; c, \; t = \mbox{2100},$
\begin{align}
\label{eq: cumu national CDR targets 2}
 &  \sum\limits_{sr' \in c, ez, a} cRCO2^{AF}(sr',ez, t, a) \times lu^{AF}(sr', ez, t, a) \nonumber \\
   & + \sum\limits_{sr, sr' \in c, b, l, t} arco2^{BECCS}(sr, sr', b, l, t) \times 10 \nonumber \\
   & + \sum\limits_{tk, sr' \in c, t} arco2^{DACCS}(tk, sr', t) \times 10 = CDRT(c, t)
\end{align}

In the scenarios discussed in Section \ref{sec: Switch Policy}, in which CDR targets "switch" from national to global from 2030, 2050, 2070 and 2090, Eqs. \ref{eq: cumu national CDR targets 1}--\ref{eq: cumu national CDR targets 2} are used before the "switch", and Eqs. \ref{eq: cumu global CDR targets 1}--\ref{eq: cumu global CDR targets 2} are used after.

\subsubsection*{B.2.3 ~~ CO$_2$ storage constraint.}

Eq. \ref{eq: CO2 storage} ensures that the cumulative CO$_2$ removed \textit{via} BECCS and DACCS never  exceeds  the  maximum  theoretical capacity  of  the  sites  in  which  it  is  stored:
\\

\noindent $\forall \; sr', \; t,$
\begin{align}
\label{eq: CO2 storage}
& \sum\limits_{sr, b, l} aSCO2^{BECCS}(sr, sr', b, l, t) \times arco2^{BECCS}(sr, sr', b, l, t) \times 10 \nonumber \\
& + \sum\limits_{tk} aSCO2^{DACCS}(tk, sr',t) \times arco2^{DACCS}(tk, sr',t) \times 10 \nonumber \\
& \leq CO2Smax(sr')
\end{align}

\noindent where:
\begin{list}{\labelitemi}{\leftmargin=1em}
\setlength\itemsep{-0.5em}
\item parameter $CO2Smax(sr')$ is the theoretical CO$_2$ storage capacity of geological formations in sub-region $sr'$;

\item parameter $aSCO2^{BECCS}(sr, sr', b, l, t)$ is the annual CO2 stored per tonne of CO$_2$ removed \textit{via} BECCS with biomass \textit{b}, cultivated on land type \textit{l} in sub-region \textit{sr}, and transported to sub-region \textit{sr'} in year \textit{t} (tCO$_2$ stored/tCO$_2$ removed/yr);

\item and parameter $aSCO2^{DACCS}(tk, sr', t)$ is the annual CO2 stored per tonne of CO$_2$ removed \textit{via} DACCS with technology \textit{tk}, in sub-region\textit{sr} in year \textit{t} (tCO$_2$ stored/tCO$_2$ removed/yr).
\end{list}

\subsubsection*{B.2.4 ~~ Sustainability/Land availability constraint.}

AR and BECCS deployments are constrained by maximum land availabilities, specific to the type of biomass grown --- forests in the case of AR, dedicated-energy crops (DEC) or agricultural residues in the case of BECCS.
Specifically, AR deployment is bounded by the availability of land with reforestation potential (LRP) as introduced in Griscom \textit{et al.} \cite{Griscom2017} (Eq. \ref{eq: LRP land}) whereas BECCS deployment is bounded by the availability of marginal agricultural land (MAL) \cite{Cai2011} for dedicated-energy crops (DEC) (Eq. \ref{eq: MAL land}), and by the availability of land with harvested wheat (LHW) \cite{Yu2020} for wheat agricultural residues (Eq. \ref{eq: WH land}).
Because LRPs and MALs are not incompatible, they can sometimes overlap, and Eq. \ref{eq: LRP MAL land} ensures that there is no double counting.
Moreover, as a safeguard against water stress and scarcity, LRPs and MALs subject to high or extremely high overall water risk (OWR) \cite{GASSERT2015} --- the OWR is greater than or equal to 3 out of a 5-scale ---  are excluded. 
This is written below:
\\

\noindent $\forall \; sr', \; ez, \; t,$
\begin{align}
\label{eq: LRP land}
\!\!\!\!\sum\limits_a lu^{AF}(sr', ez, l, a) \leq LRP^{owr3}(sr', ez) + LRP\&MAL^{owr3}(sr', ez)
\end{align}

\noindent $\forall \; sr, \; t, \; b \neq \mbox{wheat} \; or \; \mbox{forest residues},$
\begin{align}
\label{eq: MAL land}
& \sum\limits_{sr', l,} lusc^{BECCS}(sr, sr', b, l, t) \leq \sum\limits_{ez} LRP\&MAL^{OWR3}(sr, ez) \nonumber \\
& + MAL^{OWR3}(sr)
\end{align}

\noindent $\forall \; sr, \; t, \; b \neq \mbox{wheat} \; or \; \mbox{forest residues},$
\begin{align}
\label{eq: LRP MAL land}
& \sum\limits_{ez, a} lu^{AF}(sr, ez, t, a) + \sum\limits_{sr', l,} lusc^{BECCS}(sr, sr', b, l, t) \leq  MAL^{OWR3}(sr) \nonumber \\
& + \sum\limits_{ez} (LRP^{OWR3}(sr, ez) + LRP\&MAL^{OWR3}(sr, ez))
\end{align}

\noindent $\forall \; sr', \; t, \; b = \mbox{wheat},$
\begin{align}
\label{eq: WH land}
    \sum\limits_{sr',l,} lusc^{BECCS}(sr, sr', b, l, t) \leq LHW(sr)
\end{align}

\noindent where:
\begin{list}{\labelitemi}{\leftmargin=1em}
\setlength\itemsep{-0.5em}
\item parameter $LRP^{owr3}(sr', ez)$ is the maximum available LRP (excluding overlaping LRP with MAL), restricted to low OWR (< 3), in sub-region $sr'$ and ecological zone $ez$;

\item parameter $MAL^{OWR3}(sr)$ is the maximum available MAL (excluding overlapping MAL with LRP), restricted to low OWR (< 3), in sub-region $sr'$ and ecological zone $ez$;

\item parameter $LRP\&MAL^{owr3}(sr', ez)$ is the maximum available LRP and MAL (overlapping), restricted to low OWR (< 3), in sub-region $sr'$ and ecological zone $ez$;

\item parameter $LHW(sr)$ is the maximum available LHW in sub-region $sr$;

\item and variable $lusc^{BECCS}(sr, sr', b, l, t)$ is the land use of BECCS biomass supply chain with biomass \textit{b}, cultivated on land type \textit{l} in sub-region \textit{sr}, and transported to sub-region \textit{sr'} in year \textit{t} (ha);
\end{list}

%
%
%

\subsubsection*{B.2.5 ~~ AR expansion constraints.}
AR deployment is also bounded by a maximal annual expansion rate for newly afforested lands.
Because the optimisation model is solved over a 10 years time-step --- decades 2020--2030, 2030--2040,..., 2090--2100 --- it is averaged over a 10 years time-period.
This is shown below in Eq. \ref{eq: AR deployment}:

\begin{align}
\label{eq: AR deployment}
    \sum\limits_{ez, t = a} lu^{AF}(sr', ez, t, a) \leq ExpR^{AF}(sr') \times \frac{\sum_{k = 1}^{k = 10} k}{10}, \qquad \forall \; sr', \; t
\end{align}

\noindent where: $ExpR^{AF}(sr')$ is the maximum annual expansion rate of AR in sub-region $sr'$.


\smallbreak
Moreove, Eq. \ref{eq: schedul} ensures that once AR is deployed, it is maintained throughout the whole planning horizon.

\begin{align}
\label{eq: schedul}
    lu^{AF}(sr',ez,t,a) = \begin{cases} lu^{AF}(sr',ez,t=a)  & \qquad \forall \; sr',\; ez,\; t \geq a \\
0 & \qquad \forall \; sr',\; ez,\; t < a \end{cases}
\end{align}

\subsubsection*{B.2.6 ~~ BECCS expansion constraints.}

It is assumed that once a BECCS plant is built, it is operated throughout its whole lifetime (30 years). 
Total numbers of BECCS plants $np^{BECCS}(sr',t)$ in sub-region \textit{sr} and in year \textit{t} are tracked by BECCS fleet ages --- the number of first generation plants (10 years) $np1g^{BECCS}(sr',t)$, second generation (20 years) $np2g^{BECCS}(sr',t)$ and third generation (30 years) $np3g^{BECCS}(sr',t)$.
The time-period linking BECCS plants balance constraints are shown below in Eqs. \ref{eq: BECCS fleet 1}--\ref{eq: BECCS fleet 2}:
\\

\noindent $\forall \; sr', \; t:$
\begin{align}
\label{eq: BECCS fleet 1}
np1g^{BECCS}(sr',t) = np^{BECCS}(sr',t) - np2g^{BECCS}(sr',t) \nonumber \\
- np3g^{BECCS}(sr',t)
\end{align}

\begin{align}
np2g^{BECCS}(sr',t) = np1g^{BECCS}(sr',t-1), \qquad t > 1 
\end{align}

\begin{align}
np3g^{BECCS}(sr',t) = np2g^{BECCS}(sr',t-1), \qquad t > 1
\end{align}

\begin{align}
\label{eq: BECCS fleet 2}
np^{BECCS}(sr',t) = np1g^{BECCS}(sr',t), \qquad t = 1 
\end{align}

\noindent where: parameter $Unit^{BECCS}(sr, sr', b, l, t)$ is the number of BECCS plants per tonne of CO$_2$ removed, with biomass \textit{b}, cultivated on land type \textit{l} in sub-region \textit{sr}, and transported to sub-region \textit{sr'} in year \textit{t} (unit/tCO2 removed).

\smallbreak
Eq. \ref{eq: BECCS CO2 removal plants} computes the CO$_2$ removal resulting from all operating BECCS plants:
\\

\noindent $\forall \; sr', \; t:$
\begin{align}
\label{eq: BECCS CO2 removal plants}
np^{BECCS}(sr',t) = \sum\limits_{sr,b,l} arco2^{BECCS}(sr, sr', b, l, t) 
\nonumber \\
\times Unit^{BECCS}(sr, sr', b, l, t)
\end{align}

\smallbreak

Finally, the deployment of new-built BECCS plants of 500 MW/yr capacity $CAP^{BECCS}$ is bounded by a maximal built rate $BR^{BECCS}$ of 2GW/sub-region, averaged over a 10 years time-period. This is shown below in Eq. \ref{eq: BECCS plants max dep}:

\begin{align}
\label{eq: BECCS plants max dep}
np1g^{BECCS}(sr',t) = \frac{BR^{BECCS}}{CAP^{BECCS}} \times \frac{\sum_{k = 1}^{k = 10} k}{10}, \qquad \forall \; sr', \; t
\end{align}

\subsubsection*{B.2.7 ~~ DACCS expansion constraints.}

As for BECCS, once built, DAC plants are operated throughout their whole lifetime (30 years). 
A similar time-period linking DAC plants balance constraints are applied where $np^{DACCS}_{tk,sr',t}$ is the total numbers of DAC plants of technology \textit{tk}, in sub-region \textit{sr} and in year \textit{t}, and $np1g^{DACCS}_{tk,sr',t}$, $np2g^{DACCS}_{tk,sr',t}$ and $np3g^{DACCS}_{tk,sr',t}$ the number respective of 1G, 2G and 3G DAC plants.
This is shown below in Eqs.\ref{eq: DACCS fleet 1}--\ref{eq: DACCS fleet 2}:

\noindent $\forall \; tk, \; sr', \; t:$
\begin{align}
\label{eq: DACCS fleet 1}
np1g^{DACCS}(tk,sr',t) = np^{DACCS}(tk,sr',t) - np2g^{DACCS}(tk,sr',t) \nonumber \\
- np3g^{DACCS}(tk,sr',t)
\end{align}

\begin{align}
np2g^{DACCS}(tk,sr',t) = np1g^{DACCS}(tk,sr',t-1), \qquad t > 1
\end{align}

\begin{align}
np3g^{DACCS}(tk,sr',t) = np2g^{DACCS}(tk,sr',t-1), \qquad t > 1
\end{align}

\begin{align}
\label{eq: DACCS fleet 2}
np^{DACCS}(tk, sr', t) = np1g^{DACCS}(tk, sr', t), \qquad t = 1
\end{align}

\noindent where: parameter $aCCO2^{DACCS}(sr',tk,t)$ is the annual CO2 captured \textit{via} DACCS in sub-region \textit{sr}, with technology \textit{tk}, in year \textit{t} (tCO$_2$ captured/tCO$_2$ removed/yr).

\smallbreak
Eq. \ref{eq: DACCS CO2 removal plants} computes the CO$_2$ removal resulting from all operating DAC plants:
\\

\noindent $\forall \; tk, \; sr', \; t:$
\begin{align}
\label{eq: DACCS CO2 removal plants}
np^{DACCS}(tk,sr',t) \times CAP^{DACCS}(tk) =  \frac{arco2^{DACCS}(sr',tk,t)}{aCCO2^{DACCS}(sr',tk,t)}
\end{align}

\smallbreak

Finally, the deployment of new-built DAC plants of CO$_2$ capture capacity $CAP^{DACCS}$ is bounded by a maximal built rate $BR^{DACCS}$ of 2 MtCO$_2$/sub-region, averaged over 10 years time-period. This is shown in Eq. \ref{eq: DACCS plants max dep}: 

\begin{align}
\label{eq: DACCS plants max dep}
\sum\limits_{tk} (np1g^{DACCS}_{tk, sr', t} CAP^{DACCS}_{tk}) = BR^{DACCS} \frac{\sum_{k = 1}^{k = 10} k}{10} && \forall \; tk, \; sr', \; t 
\end{align}

\subsection*{B.3 ~~ Additional equations}
The scenarios discussed in Section \ref{sec: Switch Policy} are solved in the optimisation  using results from the ISOLATION scenario --- more specifically AR, BECCS and DACCS decisions variables' outputs --- as an additional constraint to AR, BECCS and DACCS co-deployments:
\\

\noindent $\forall \; sr',\; ez,\; t < switch, \;a:$
\begin{align}
lu^{AF}(sr', ez, t, a) = ISOLlu^{AF}(sr', ez, t, a)
\end{align}

\noindent $\forall \; sr, \; sr', \; b, \; l, \; t < switch:$
\begin{align}
arco2^{BECCS}(sr, sr', b, l, t) = ISOLarco2^{BECCS}(sr, sr', b, l, t)
\end{align}

\noindent $\forall \; tk,\; sr',\; t < switch:$
\begin{align}
arco2^{DACCS}(tk, sr', t) = ISOLarco2^{DACCS}(tk, sr', t)
\end{align}

\noindent where:

\begin{list}{\labelitemi}{\leftmargin=1em}
\item parameter $ISOLlu^{AF}_{sr', ez, t, a}$ is the land use of AR in sub-region \textit{sr'} and ecological zone \textit{ez}, in year \textit{t}, and deployed from year \textit{a}, as projected in the ISOLATION scenario (ha), 

\item parameter $arco2^{BECCS}_{sr, sr', b, l, t}$ is the annual CO$_2$ removed \textit{via} BECCS with biomass \textit{b}, cultivated on land type \textit{l} in sub-region \textit{sr}, and transported to sub-region \textit{sr'} in year \textit{t}, as projected in the ISOLATION scenario (tCO$_2$ removed/yr), 

\item parameter $arco2^{DACCS}_{tk, sr', t}$ is the annual CO$_2$ removed \textit{via} DACCS with technology \textit{tk}, in sub-region \textit{sr'} in year \textit{t}, as projected in the ISOLATION scenario (tCO$_2$ removed/yr),

\item $switch$ is the year at which the model 'switches' from an ISOLATION to a COOPERATION policy paradigm.
\end{list}

\section*{Appendix C ~~ Datasets of MONET}

In this study, AR, BECCS and DACCS deployments are constrained by bio-geophysical limits.
Here, we describes in details the methodologies applied to obtain estimates of CO$_2$ storage capacity at the sub-regional level, that limit BECCS and DACCS deployment, and estimates of land with reforestation potential at the sub-regional level, that limit AR deployment.
The methodologies used to obtain estimates of MAL at the sub-regional level, and exclude lands with higher OWR, that limit BECCS deployment, have been introduced previously \cite{Fajardy2018,Fajardy2020b} and will not be repeated here.

\subsection*{C.1 ~~ CO$_2$ storage}

\subsubsection*{C.1.1 ~~ Overview.}

The deployment of CDR options involving geological CO$_2$ storage (\textit{i.e.}, BECCS and DACCS) relies on the presence of suitable geological storage sites and the existence of CO$_2$ transportation methods between the CO$_2$ capture facility and the storage site.
Whilst global aggregated CO$_2$ storage capacity is generally not considered as a barrier for CCS deployment \cite{GlobalCCSInstitute2018,Dooley2013}, existing regional storage assessments are limited \cite{Kearns2017}. In addition to the lack/absence of completeness, regional storage assessments differ in their methodology, making direct comparisons or summation of their evaluated CO$_2$ capacities inaccurate \cite{Kearns2017,Heidug2013}. 
\newline
\indent Particularly, CO$_2$ storage capacity can be estimated and classified into theoretical, effective, practical and source-sink matched, as defined by the “Geologic CO$_2$ storage Resource Capacity Pyramid” classification \cite{Bachu2008}. This concept estimates CO$_2$ storage capacity based on physical-geological and engineering/technical limits, legal, regulatory and economic barriers and the presence of matching CO$_2$ sources.
CO$_2$ storage capacity can also be classified into geological classes: deep saline aquifers, depleted oil and gas reservoirs, and less frequently unmineable coal seams \cite{Bachu2008}. Deep saline aquifers have been generally acknowledged as containing the vast majority of geological CO$_2$ storage capacity \cite{Kearns2017}. 
\linebreak
\indent Globally, when conducted, the capacity and reliability of regional storage assessments vary significantly and may lead to regional storage constraints.

\subsubsection*{C.1.2 ~~ Medium/Base-Case CO$_2$ storage scenario.}

Although we recognise the need for further work in developing a more complete and consistent regionally-disaggregated CO$_2$ storage assessment, our study includes CO$_2$ storage availability and capacity for Brazil, China, India, the EU (27- UK) and the USA, at the sub-regional level, based on the existing literature.
Data is obtained from country, regional and basin-scale CO$_2$ storage assessments. Our study evaluates CO$_2$ storage capacity in deep saline aquifers and depleted oil and gas reservoirs, and includes both on-shore and off-shore storage sites.
As a result of this, base-case (medium), low and high scenarios are estimated.
Table \ref{tab:CO2 Capa} provides data sources for CO$_2$ storage capacity in this study and compares our estimates with the ones of recent literature.

\begin{table*}[h]
\small
  \caption{Geological CO$_2$ storage capacity}
  \begin{tabular*}{\textwidth}{@{\extracolsep{\fill}}lllllllll}
    \hline
    \multirow{4}{*}{Country} & \multicolumn{6}{c}{This study} & \multicolumn{2}{c}{GCCSI \cite{GlobalCCSInstitute2018}} \\
    & \multicolumn{2}{c}{Low} & \multicolumn{2}{c}{Medium} & \multicolumn{2}{c}{High} & & \\
    & (GtCO$_2$) & Data sources & (GtCO$_2$) & Data sources & (GtCO$_2$) & Data sources & (GtCO$_2$) & (\%)\\
    \hline
    Brazil & 0.95 & \cite{Rockett2013} & 0.95 & \cite{Rockett2013} &  641 & \cite{Rockett2013,Ketzer2014} & 2,000 & 56-75\\
    China & 54 & \cite{Zhongyang2009,Pearce2011,NZEC2009,COACH2009,Vincent2010,Vincent2011,Poulsen2010,Jiao2011,Ellett2013,Wang2011,Zhou2020} & 3,106 & \cite{Li2009,Dahowski2009} & 3,106 & \cite{Li2009,Dahowski2009} & 2,000 & 75-100\\
    EU & 180 & \cite{Vangkilde-Pedersen2009a,Vangkilde-Pedersen2009b,Poulsen2014} & 180 & \cite{Vangkilde-Pedersen2009a,Vangkilde-Pedersen2009b,Poulsen2014} & 180 & \cite{Vangkilde-Pedersen2009a,Vangkilde-Pedersen2009b,Poulsen2014} & 300 & 75-100\\
    \textit{of which UK} & \textit{78} & \cite{Gammer} & \textit{78} & \cite{Gammer} & \textit{78} & \cite{Gammer} & \textit{80} & \textit{75-100}\\
    India & 0 & - & 0 & - & 53 & \cite{Holloway2009} & 50 & 56-75 \\
    USA & 2,565 & \cite{USDOE2015} & 8,533 & \cite{USDOE2015} & 21,865 & \cite{USDOE2015} & 8,150 & 75-100 \\
    \hline
  \end{tabular*}
  \label{tab:CO2 Capa}
\end{table*}

Whilst our estimates for India are in accordance with the Global CCS Institute (GCCSI), our Brazilian estimate is 3 times lower. This significant difference is probably due to our conservative choice of the default CO$_2$ storage factor when estimating the CO$_2$ storage of Brazil. A sensitivity analysis on this factor indicates that a less conservative, average value of 3,520 Gt of CO$_2$ could be obtained. Nevertheless, due to the high incertitude concerning CO$_2$ storage capacity, we use the conservative value.

\subsubsection*{C.1.3 ~~ Global methodology.}
In this study, CO$_2$ storage availability and capacity are assessed for Brazil, China, India, the EU (-28) and the USA, at the sub-regional level.
In the absence of quantitative CO$_2$ storage capacity assessments, the methodology described in Wildenborg \textit{et al.} is applied \cite{Wildenborg2004}. This method estimates the CO$_2$ capacity of saline aquifers from the surface area of the basin in which the aquifer is located, as follows in Equation \ref{eq: Co2S_Capa_methodo}:
 
\begin{align}
\label{eq: Co2S_Capa_methodo}
V_{CO_{2}} = A \times ACF \times SF
\end{align}

\noindent where:
\begin{list}{\labelitemi}{\leftmargin=1em}
\setlength\itemsep{-0.5em}
    \item $V_{CO_{2}}$ is the CO$_2$ storage capacity (MtCO$_2$)
    \item $A$ is the basin area (km$^2$)
    \item $ACF$ is the aquifer coverage factor (-)
    \item and $SF$ is the storage factor (MtCO$_2$/km$^2$)
\end{list}

Conservatively, it is assumed that a single aquifer is present in each basin, covers approximately half of the basin area, and is well sealed. Resulting default values of the ACF and the SF are respectively 50\% and 0.2 MtCO$_2$/km$^2$.

\subsubsection*{C.1.4 ~~ Brazil.}

To this date, no country-scale quantitative CO$_2$ storage analysis has been performed for Brazil.
Instead, a qualitative prospection of 31 Brazilian basins, covering an area of approximately 6.4 million km$^2$, was generated through a basin-by-basin analysis, ranking them into three categories: low, medium or high potential for storage \cite{Ketzer2014}.
Additionally, one quantitative analysis on the Campos basin, a depleted oil field, off the south-east coast of Brazil, comprised of 17 off-shore oil fields was performed, resulting in a CO$_2$ storage capacity of 950 MtCO$_2$ \cite{Rockett2013}.
The method described in Wildenborg \textit{et al.} \cite{Wildenborg2004} is thus applied to the 31 Brazilian basins, resulting in a CO$_2$ storage capacity estimate of 640 Gt. 
These basin-scale estimates are then spatially defined in the software ArcGIS 10.6 by using the corresponding map, and further aggregated by Brazilian sub-regions.
\newline
\indent In this study, the high CO$_2$ storage capacity estimate is equal to both the basins’ aquifers and the oil field CO$_2$ storage capacities, and the medium and low estimates are equal to the oil field CO$_2$ storage capacity only. 
It is worth mentioning that when using a value of 2 MtCO$_2$/km$^2$ for the SF (\textit{i.e.,} default value for unconfined aquifers), the CO$_2$ storage capacity of Brazil increases to 6,400 Gt CO$_2$, resulting in a less conservative estimate of 3,520 Gt of CO$_2$.

\subsubsection*{C.1.5 ~~ China.}

In China, the national and quantitative CO$_2$ storage capacity in sedimentary basins \cite{Li2009,Dahowski2009} and in oil \& gas fields \cite{Dahowski2009} was evaluated. 
Overall, 27 saline aquifers, 17 gas fields and 19 oil fields were assessed, resulting in CO$_2$ storage capacities of respectively 3,096 Gt CO$_2$, 5.2 Gt CO$_2$ and 4.6 Gt CO$_2$. 
In this work, each basin’s CO$_2$ storage capacity is spatially allocated to its respective basin in the software ArcGIS 10.6 by using the spatial dataset on sedimentary basins in China from the US Geological Survey (USGS), and further aggregated by Chinese sub-regions. 
The resulting values account for the medium and high estimates in our study.
\newline
\indent Numerous region-, basin-, and formation-scale studies have also been performed. 
In particular, the Near Zero Emissions from Coal (NZEC) project examined CO$_2$ storage projects on the Sangliao and the Subei Basin \cite{Zhongyang2009,Pearce2011,NZEC2009}. 
The CO$_2$ storage capacity of the Songliao basin’s aquifer has been evaluated at 593 Mt CO$_2$, and the effective capacities of the Daquig and Jilin Oilfields’ at 71.5 Mt and 692 Mt respectively. 
\linebreak
\indent Similarly, the Cooperation Action Carbon Capture and Storage China-EU (COACH) project investigated options in the Bohai Basin \cite{COACH2009,Vincent2010,Vincent2011,Poulsen2010}. 
The effective CO$_2$ storage capacity of the Dagang ang the Shengli oilfields have been estimated at 22 and 472 Mt CO$_2$ respectively. 
The largest capacity has been identified in the aquifers of the Huimin sub-basin within the Jiyang Depression, with an upper bound estimate of 22 Gt, and latest interest on the Guanyao Formation within the Huimin sub-basin has resulted in the estimation of 700 Mt of CO$_2$. 
Significant uncertainty around these aquifers’ estimates must be acknowledged due to the limited availability of geological data.
\linebreak
\indent Several studies have been also investigated the Ordos basin CO$_2$ storage potential.
An initial upper estimate of 287 Gt CO$_2$ was suggested for the aquifer CO$_2$ storage capacity by Jiao et al. \cite{Jiao2011}, later re-evaluated around 25 Gt CO$_2$ by Ellet et al. \cite{Ellett2013}. 
In the work of Wang et Carr \cite{Wang2011}, oil \& gas fields CO$_2$ storage capacity were evaluated at 38 Mt and 3.2 Gt respectively. 
In the recent work of Zhou et al \cite{Zhou2020}, up to 60 oil fields were considered, resulting in a CO$_2$ storage capacity of 1,222 Mt.
\linebreak
\indent Overall, these basin-scale estimates are spatially represented in the software ArcGIS 10.6 by estimating their geological formation’s surface areas from corresponding maps, provided by the aforementioned studies. 
They are then aggregated by Chinese sub-regions and account for the low estimates in our study.

\subsubsection*{C.1.6 ~~ EU.}

The UK storage appraisal project (UKSAP) of the Energy Technologies Institute (ETI) provided a detailed UK CO$_2$ storage assessment, in which a total CO$_2$ storage capacity of 78 Gt CO$_2$ was identified \cite{Gammer}.
\newline
\indent At the EU-scale, the EU GeoCapacity Project and later the CO$_2$Stop project developed a CO$_2$ storage potential capacity database \cite{Vangkilde-Pedersen2009a,Vangkilde-Pedersen2009b,Poulsen2014}. 
The GeoCapacity project involved 25 countries and evaluated a total CO$_2$ capacity of 95.7 Gt in deep saline aquifers and 20.2 Gt in Hydrocarbon fields. 
The CO$_2$StoP project involved 27 countries and presented a standardised CO$_2$ storage capacity of 157.8 Gt in deep saline aquifers, and 13.3 Gt in hydrocarbon fields, which are 1.6 times higher and 0.7 times lower than the respective estimates provided by the GeoCapacity project. 
Overall, the CO$_2$StoP identified CO$_2$ storage capacity is 1.5 times higher than the one of the GeoCapacity project.
\linebreak
\indent Regarding CO$_2$ storage in aquifers, it would appear that the CO$_2$ storage capacity of France, Germany and Poland in the GeoCapacity project is equivalently allocated to Poland, only, in the CO$_2$StoP project. 
Similarly, the reported CO$_2$ storage capacities of Denmark and Norway in the GeoCapacity project are also equivalent to the one of Denmark, only, in the CO$_2$StoP project. 
Finally, when subtracting the CO$_2$ capacity of the UK from the total CO$_2$ capacity in the CO$_2$StoP project, the resulting total CO$_2$ capacity is only equal to 0.7 of the one in the GeoCapacity project. 
Regarding CO$_2$ storage in oil \& gas fields, no similar hypotheses can be drawn. 
\linebreak
\indent Overall, value ranges vary significantly between the two projects, both at the country-scale and the global-scale, whether for aquifers, oil \& gas fields, or both. 
Additionally, in the CO$_2$StoP project, important differences can also be observed between the “user-entered”, the “calculations” and the “standardised calculations”. 
As a result of this, our study uses the GeoCapacity data, which accounts for the low, medium and high CO$_2$ storage capacity estimates.

\subsubsection*{C.1.7 ~~ India.}

To this date, only a country-scale qualitative CO$_2$ storage analysis has been performed for India, categorising sedimentary basins into good, fair, and limited \cite{Holloway2009}.
The method described in Wildenborg \textit{et al.} \cite{Wildenborg2004} is thus applied to 9 “good” and 3 “fair” basins, resulting in a high/upper CO$_2$ storage capacity estimate of 63.3 Gt. 
This CO$_2$ storage capacity was further aggregated by Indian sub-regions in the software ArcGIS 10.6.

\subsubsection*{C.1.8 ~~ USA.}

Two national quantitative CO$_2$ storage capacity assessments have been completed.
The USGS provided an estimate of 2,924 (2,135-4,013) Gt of CO$_2$ storage capacity for 36 sedimentary basins in the United States \cite{Warwick2013,USGS2013}.
More recently, the U.S. Department of Energy’s (DOE) National Energy Technology Laboratory (NETL) published the 5th edition of the Carbon Storage Atlas (Atlas V), in which 205 (186-232) Gt and 8,328 (2,379-21,633) Gt of CO$_2$ storage capacity have been identified in oil \& gas fields and aquifers respectively \cite{USDOE2015}. 
Spatial data was also available.
\linebreak
\indent 
In this study, we aggregate the DOE’s NETL spatial dataset into U.S. States in the software ArcGIS 10.6. 
The P5 (min), mean and P95 (max) estimates accounts respectively for the low, medium and high estimates.
 
\subsection*{C.2 ~~ Areas with reforestation opportunities}

The type of land on which afforestation tales places, also referred as cover type, is included within the afforestation value chain framework, thereby impacting afforestation GHG and cost balance, through the respective contributions of direct \& indirect land use change and land cost.

In this study, afforestation deployment is constrained to areas with reforestation (RP) potential, as provided by (Griscom et al.) \cite{Griscom2017}.
RP lands are defined by non-forests lands in areas ecologically appropriate for forests (\textit{i.e.}, areas with a low tree cover (< 25\%) but within the boundaries of a native forest cover type).
Land categorised as grassland or savanna, cropland, and land within boreal ecological zones are excluded from the scope of the dataset respectively to limit negative impacts on biodiversity, to account for food security, and the albedo effect. 
\linebreak
\indent The 740m geographic resolution spatial dataset has been processed in the software ArcGIS 10.6 (ESRI) to obtain the availability of RP land in each sub-region \textit{sr} and ecological zone \textit{gez}.
Because of the low tree cover of RP lands, and the little information on their current use, we assume that LUC and iLUC factors, and cost of land are equal to zero.

\section*{Appendix D ~~ Additional results}

\subsection*{D.1 ~~ Cost-optimal cumulative CO$_2$ removal}

\begin{figure*}[h!]
\centering
  \includegraphics[width=15cm]{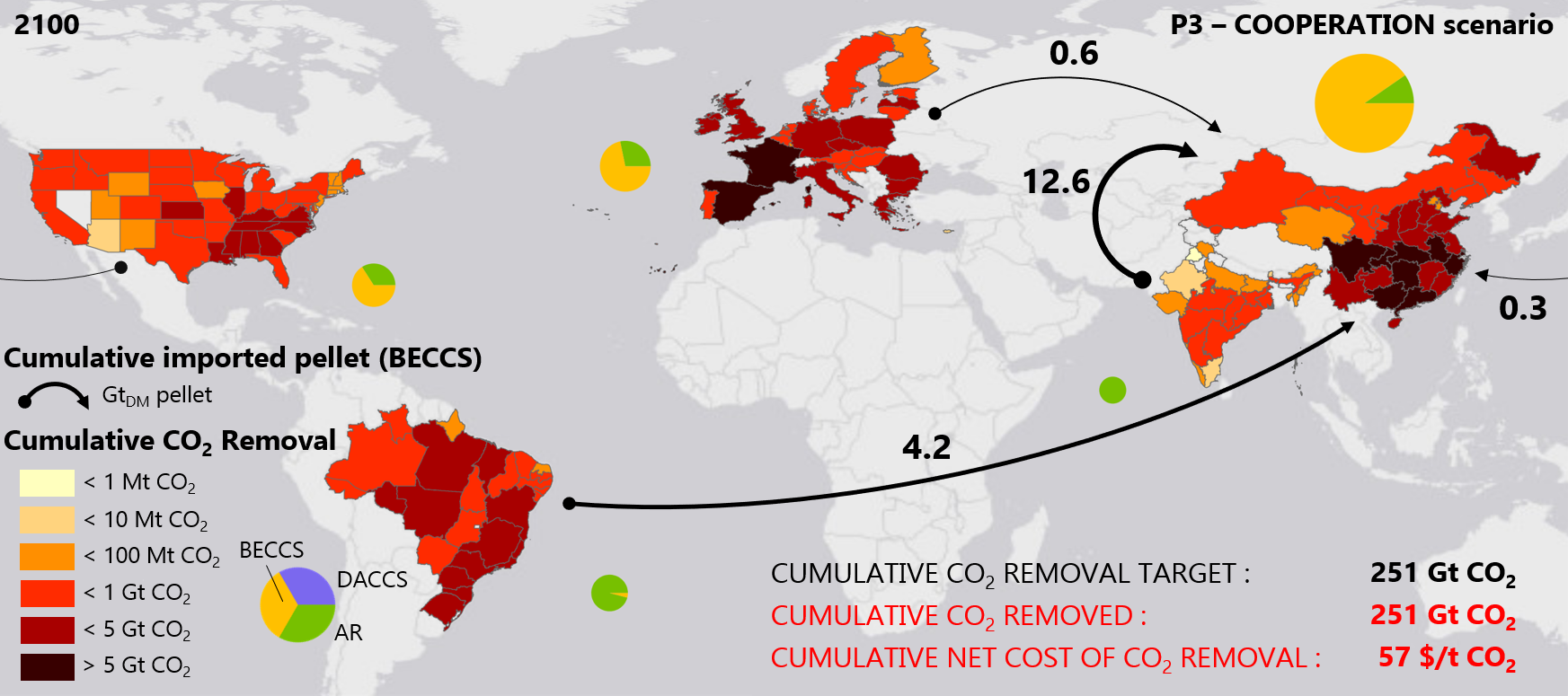}
  \includegraphics[width=15cm]{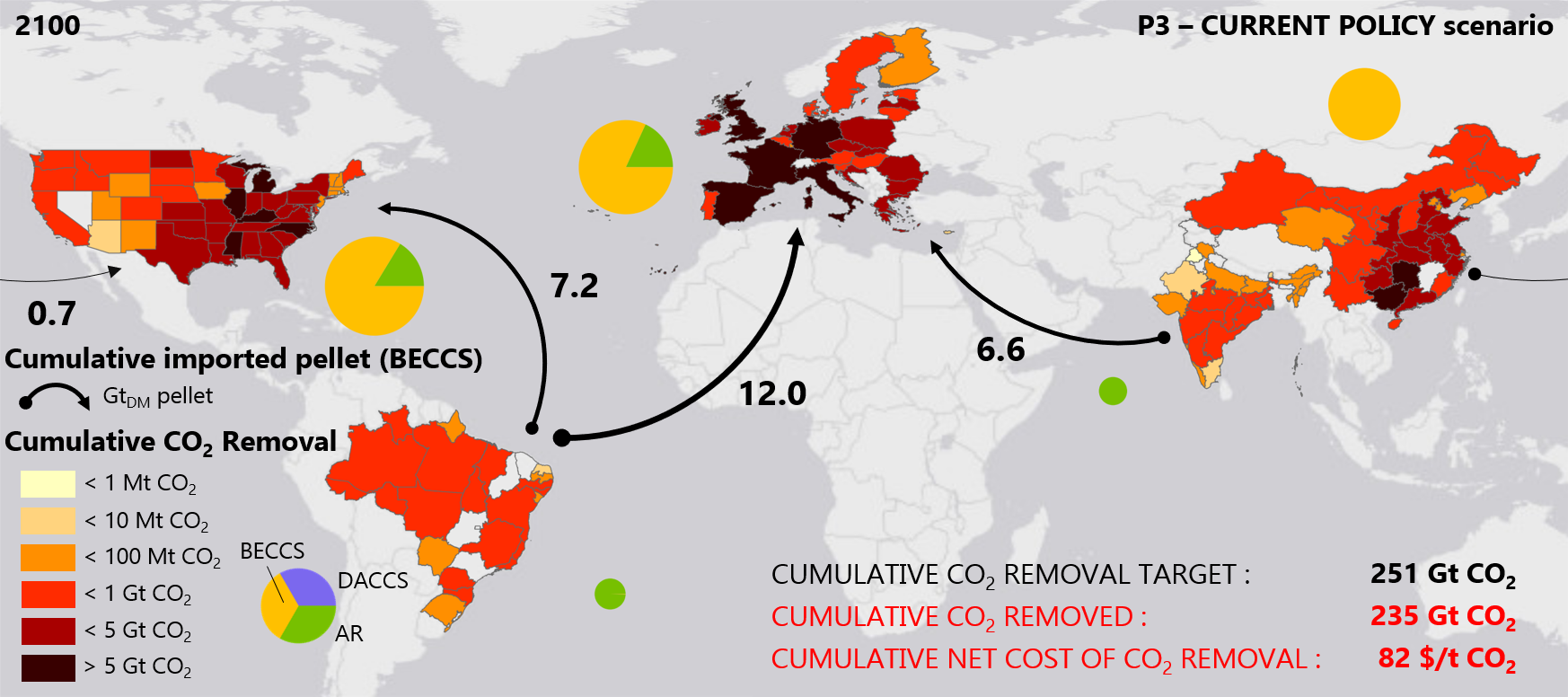}
  \includegraphics[width=15cm]{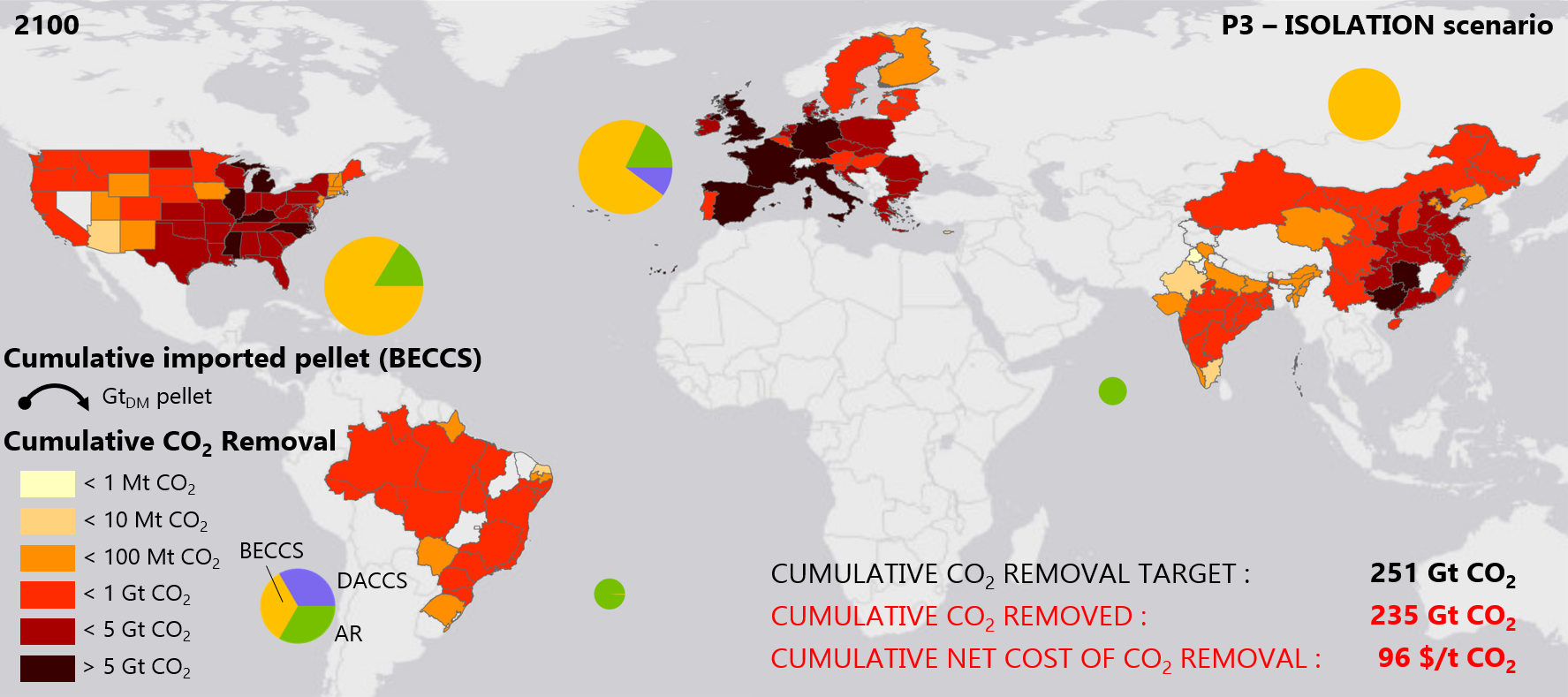}
  \caption{Maps of cost-optimal CDR pathways deployed by 2100, under alternative P3-consistent policy scenarios: COOPERATION, CURRENT POLICY and ISOLATION scenarios.}
  \label{fig: Opt P3 Maps}
\end{figure*}

Fig.\ref{fig: Opt P3 Maps} illustrates the cost-optimal CDR pathways in 2100 under the different P3-consistent policy scenarios discussed in Section \ref{subsec: Opt Scenarios}. 
The amount of CO$_2$ removed cumulatively by 2100 is indicated at the sub-regional scale, and pied charts showing the contribution of each CDR option (\textit{i.e.,} AR, BECCS, and DACCS) are at the regional scale. 
Cumulative imported biomass (pellet) for BECCS by 2100 are also indicated \textit{via} arrows at the regional scale.

\subsection*{D.2 ~~ Cost supply curves}

\begin{figure*}[t]
\centering
  \includegraphics[width=17.1cm]{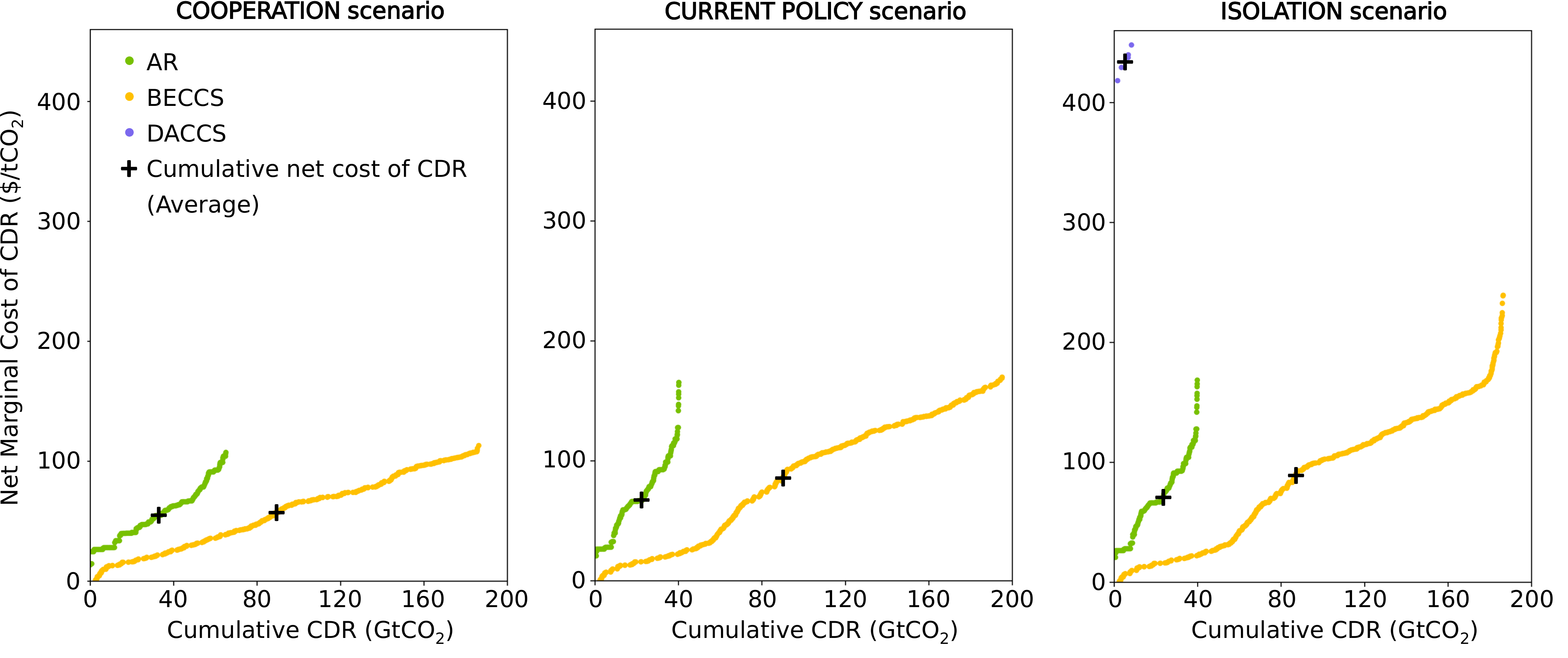}
  \caption{Cost supply curves of each CDR option deployed by 2100, under alternative P3-consistent policy scenarios. BECCS delivers the most cost-efficient CDR -- its cost supply curve is below all the ones of the other CDR options -- and therefore the most CDR, and that in all scenarios.}
  \label{fig: Opt P3 Cost Supply CDR Options}
\end{figure*}

Fig. \ref{fig: Opt P3 Cost Supply CDR Options} shows the cost-efficiency of each CDR option deployed -- how the cost of CDR increases as more CO$_2$ is removed from the atmosphere -- in all P3-consistent policy scenarios.

\section*{Appendix E ~~ Sensitivity analysis}
\label{sec: Sensitivity Analysis}

\begin{figure*}[h!]
\centering
  \includegraphics[width=17.1cm]{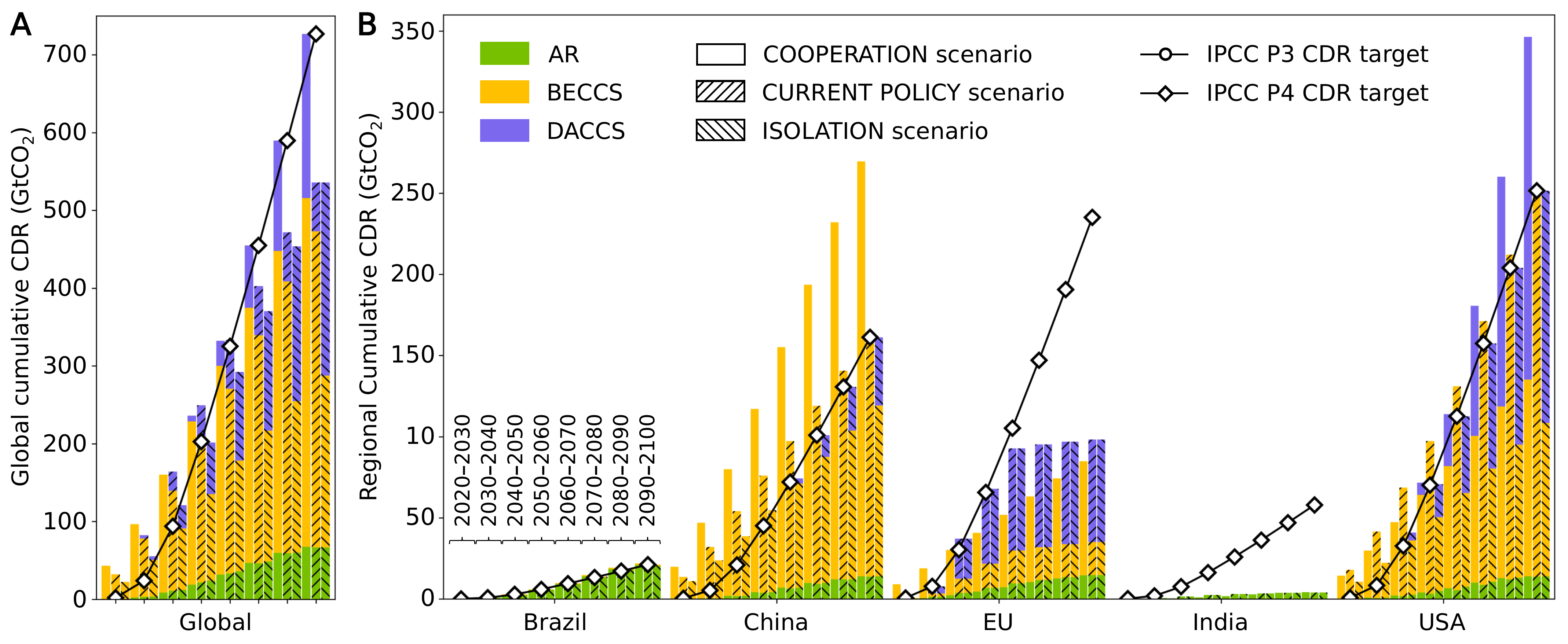}
  \includegraphics[width=17.1cm]{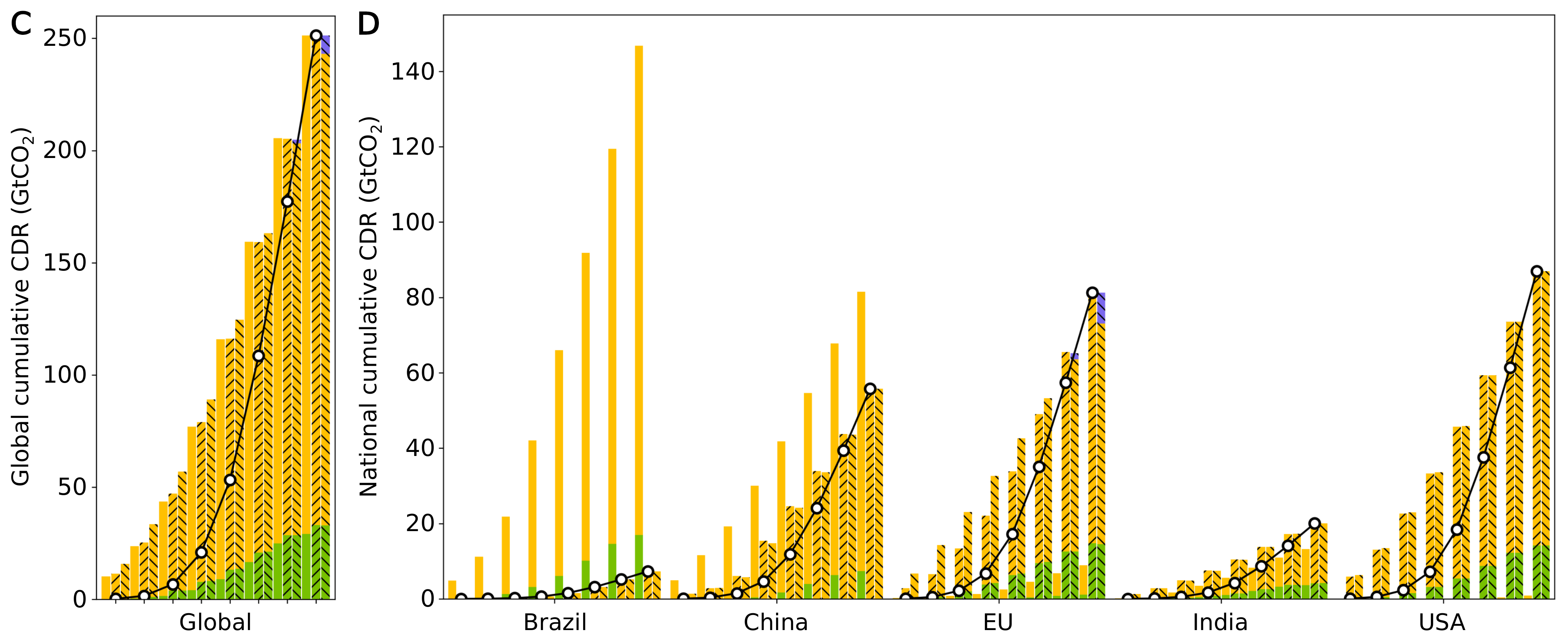}
  \includegraphics[width=17.1cm]{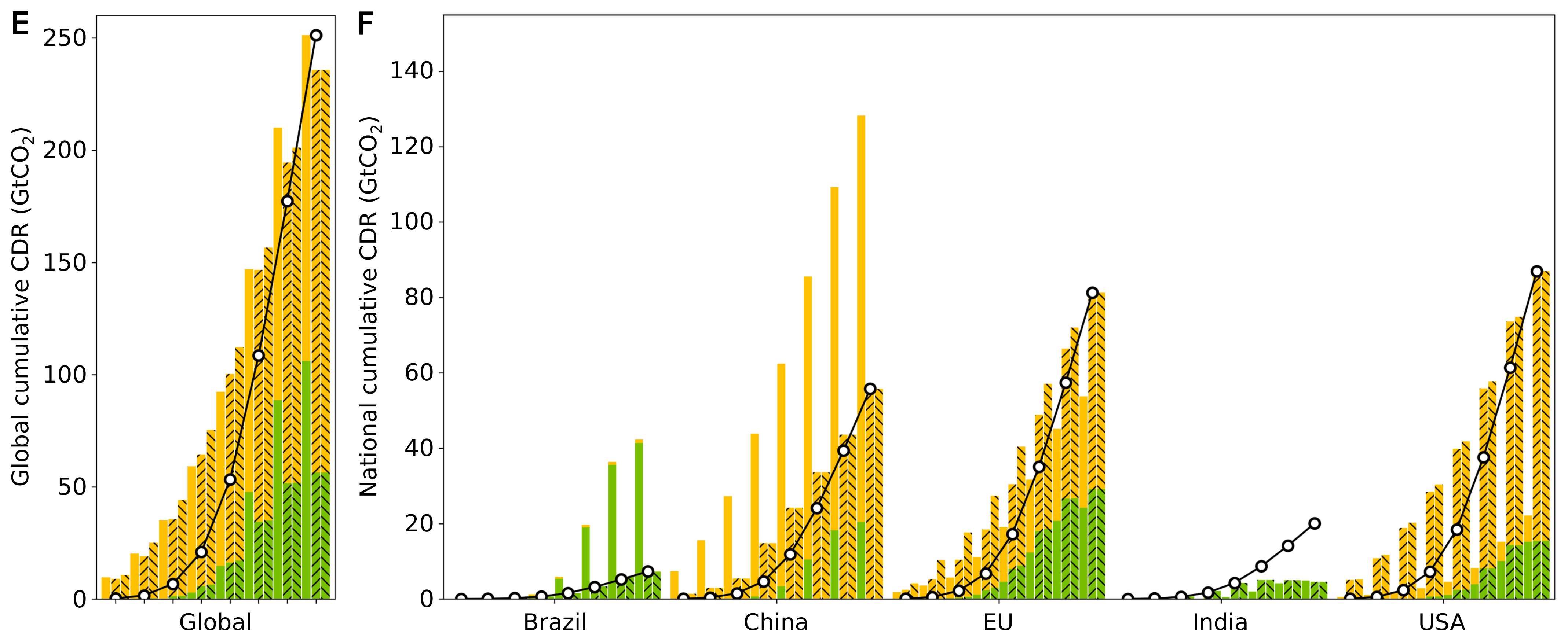}
  \caption{Sensitivity analysis are performed on: increasing CDR targets, globally (A) and nationally (B); increasing CO$_2$ storage availability, globally (C) and nationally (D); and on increasing built (\textit{i.e.,} BECCS and DACCS) and expansion (\textit{i.e.,} AR) rates, globally (E) and nationally (F).}
    \label{fig: Sensitivity Analysis}
\end{figure*}

The scenarios discussed in Section \ref{sec: Opt CDR} are dependant on the level of CDR targets that must be met, globally and nationally. 
They also rely on the availability of CO$_2$ storage, and the maximum deployment rates for AR, BECCS and DACCS.

Here, we investigates and discusses the influence of varying CDR targets, CO$_2$ storage availability and maximum deployment rates on the cost-optimal composition of CDR pathways.

\subsection*{E.1 ~~ Increasing CDR targets}
\label{sec: SA CDR targets}

IAMs have shown that a diversity of climate mitigation pathways were consistent with the Paris Agreement's 1.5$^{\circ}$C objective, differing in socio-economic drivers, demand for energy and/or land, and cumulative GHG emissions, and therefore in temperature overshoots and resulting CDR achieved.
The IPCC classifies these climate mitigation pathways into four illustrative scenarios, based on their levels of CDR achieved by the end of the century.
Specifically, the IPCC P3 scenario aims at being representative of a middle-of-the-road climate mitigation pathway, in which BECCS --- used as a proxy for CDR solutions --- is deployed up to 407 cumulative GtCO$_2$ by 2100.
In this study, the IPCC P3 scenario was used to set out base-case CDR targets between 2020--2100.
Results in Section \ref{sec: Opt CDR} showed that delivering the Paris Agreement' CDR objectives at the IPCC P3 scale is feasible, yet challenging --- less feasibly and more expensively --- without international cooperation policy in  climate mitigation.
Therefore, we perform/run a sensitivity analysis on higher CDR targets by using the IPCC P4 scenario, representative of a fossil-fuel intensive and high energy demand pathway, and characterised by the aggressive deployment of BECCS --- 1,176 cumulative GtCO$_2$ by 2100.
As a reference, the IPCC P3 and P4 annual levels of CDR in 2100 are equivalent to 12.2 and 35.1 times the current level of CO$_2$ emissions --- 33.5 GtCO$_2$ in 2018).

As shown in Fig. \ref{fig: Sensitivity Analysis}-A, higher (P4) CDR targets results into the deployment of DACCS in all scenarios, making up between 12--46\% of the CO$_2$ cumulatively removed from the atmosphere globally by 2100.
The Paris Agreement's CDR objectives are successfully delivered at the P4 scale in the COOPERATION scenario, but the gap by which they are missed in the CURRENT POLICY and ISOLATION scenarios increases to 191 GtCO$_2$ in 2100, equivalent to 26\% of the 2100 global P4 target.

In the COOPERATION scenario, DACCS is deployed in the USA only, up to 201 GtCO$_2$ in 2100 (Fig. \ref{fig: Sensitivity Analysis}-B), and contributes to overtaking the USA's 2100 national CDR target by 38\%.
This is achieved with the DACCS-CE archetypal configuration only, as it is cheaper than the DACCS-CW archetypal one, and increases the (global) cumulative cost of CDR by 3.5 times to \$203/tCO$_2$ by 2100.
With higher CDR targets, DACCS needs therefore to be added to the cost-optimal CDR pathway to deliver the Paris Agreement at scale, whilst taking best advantage of international cooperation policy, which results inevitably into higher CDR costs.

In the CURRENT POLICY scenario, because of the absence of international cooperation in climate mitigation, we find that DACCS is deployed in the EU as early as 2030.
However,  this is insufficient to reach the EU's national CDR targets as the EU's CO$_2$ storage sites are exhausted by 2070.
Although less CO$_2$ is removed by the end of the century than in the previous scenario, both DACCS-CE and DACCS-Cw archetypal configurations are deployed, as the DACCS-Cw archetypal configuration is more efficient to remove CO$_2$ from the atmosphere, once the electricity system has fully transitioned to net-zero. 
However, the DACCS-Cw archetypal configuration is also more expensive than the DACCS-CE one. 
Therefore the cumulative cost of CDR also increases to \$205/tCO$_2$ by 2100, which is lower than in the previous scenario.

The cumulative cost of CDR is the highest in the ISOLATION scenario --- \$285/tCO$_2$ by 2100 --- as international biomass trading is not allowed, and therefore leads to a higher contribution of DACCS to the CDR pathway than in the CURRENT POLICY scenario.
This is observed in the USA and in China, where DACCS displaces BECCS using imported biomass from Brazil and India.

The above emphasises the importance of developing a geopolitical framework for negative emissions trading in order to guarantee the international ability of delivering the Paris Agreement's objectives, even in the pessimistic P4 scenario of the IPCC.
However, as more CO$_2$ will need to be removed from the atmosphere, the financial burden of the Paris Agreement will increase significantly, due to the inevitable deployment of DACCS.

\subsection*{E.2 ~~ Increasing CO$_2$ storage availabilities \& capacities}
\label{sec: SA CO$_2$ storage}

This study highlighted that CDR solutions involving CO$_2$ storage play a key role in delivering the Paris Agreement at scale, with BECCS and DACCS contributing between 74--84\% to the total CO$_2$ removal by 2100, in all scenarios discussed in Section \ref{sec: Opt CDR}.
More specifically, results in Section E.1 showed the importance of well distributed CO$_2$ storage sites with high capacity.
Specifically, nations can be categorised as follows:
No or very limited CO$_2$ storage potential --- < 1 GtCO$_2$ --- such as Brazil and India; limited CO$_2$ storage potential --- $\geq$ 1 GtCO$_2$ and < 1,000 GtCO$_2$ --- such as the EU; and unlimited CO$_2$ storage potential --- $\geq$ 1,000 GtCO$_2$ --- such as China and the USA.
Here, we investigate the effect of a globally better distributed and higher CO$_2$ storage capacity on the cost-optimal deployment of CDR solutions.
The description of the CO$_2$ storage assessment methodology and the resulting regional CO$_2$ storage estimates are provided in Appendix B.


Fig. \ref{fig: Sensitivity Analysis}-C shows that better distributed and higher CO$_2$ storage capacities across the world allow for the successful delivering of the Paris Agreement, under any alternative policy options.
This is because, in the CURRENT POLICY and ISOLATION scenarios, India's national CDR target can be met \textit{via} BECCS, owing to the availability of CO$_2$ storage site.
This also results into a slight decrease of the cumulative net cost of CDR by 4--5\% to \$79/tCO$_2$ and \$92/tCO$_2$, respectively.
Specifically, the cumulative cost of CDR is the lowest in the COOPERATION scenario --- \$31/tCO$_2$ --- owing to the the availability of CO$_2$ storage sites in Brazil, and thus the ability to deploy the least-cost BECCS configurations across the world, and where 58\% of the total CO$_2$ removal by 2100 is therefore achieved. 
This is equivalent to 20 times Brazil's 2100 national CDR target.

This highlights the importance of well-distributed and high CO$_2$ storage capacity at the national level, as this would unlock the delivering of the Paris Agreement under any alternative policy options, and decrease its financial burden, specifically with an international cooperation policy.

\subsection*{E.3 ~~ Increasing deployment rates}
\label{sec: SA BE rates}

The cost-optimal composition of a CDR pathway that is consistent with the Paris Agreement is subject to a set of constraints.
Specifically, the rate at which each CDR solution can be deployed in this study is determined by maximum expansion rates --- an afforestation rate for AR, and built rates for new BECCS and DACCS plants.
There are, however, no historic records for nascent CDR technologies such as BECCS and DACCS, or for AR for the purpose of CDR, which makes maximum expansion limits, now and in the future, highly uncertain.
This study showed that, AR deployment was partially limited by such expansion constraints, at the rates specified as in Section \ref{sec: CDR Expansion}.
Here, we perform a sensitivity analysis on higher deployment rates for all CDR options, in the scenarios discussed in Section \ref{sec: Opt CDR}.
Specifically, AR's maximum regional land expansion rate is multiplied by a factor 10, from 50 kha/yr to 500kha/yr, 
BECCS's maximum regional built rate is increased from 2GW/yr to 5GW/yr --- equivalent to a CO$_2$ capture rate of 42 MtCO$_2$/yr in each region --- and DACCS's maximum regional built rate is increased accordingly to 42 MtCO$_2$/yr as well.

As shown in Fig \ref{fig: Sensitivity Analysis}-E, higher deployment rates lead to higher, yet limited, deployments of AR to the detriment of BECCS and DACCS, and reduce by 7--18\% the cumulative cost of CDR to \$47/tCO$_2$, \$76/tCO$_2$ and \$79/tCO$_2$ in 2100 in the COOPERATION, CURRENT POLICY and ISOLATION scenarios, respectively. 
However, this does not improve the ability to meet the Paris Agreement's 2100 CDR objectives in the absence of international cooperation in climate mitigation policy (in the CURRENT POLICY and ISOLATION scenarios).

In the COOPERATION scenario, in which AR is the most deployed, total CO$_2$ removal \textit{via} AR increases by 32 GtCO$_2$ in 2100, but BECCS contribution to the cost-optimal CDR pathway is still predominant.
This is because, even with higher afforestation rates, AR deployment is bio-geophysically limited by land availability.
In India, 92\% of the land with a potential for reforestation is already allocate to AR as early as 2050. 
In Brazil and in the EU, AR is deployed by 2060 over 89\% and 91\% of the available land, respectively.
This is also because BECCS is still more cost-efficient than AR.
Overall, we find that 98--100\% of the total CO$_2$ removed by 2100 in Brazil and India is achieved \textit{via} AR, but 45--69\% in the EU and the USA, and only 16\% in China.

In the CURRENT POLICY scenario, BECCS is also replaced by AR, but only by 16 GtCO$_2$.
As in the base-case CURRENT POLICY scenario, this is because of the absence of international cooperation policy in climate mitigation.
Even with higher afforestation rates, BECCS is still more cost-efficient and therefore more deployed than AR by 2100.

Finally, in the ISOLATION scenario, higher deployment rates results into the full replacement of DACCS by AR.

The above therefore shows that higher deployment rates can alleviate the Paris Agreement's financial burden by deploying AR instead of BECCS and DACCS.
Without international cooperation in climate change policy, this is however insufficient to deliver the Paris Agreement at scale by 2100.
In a pessimistic 'isolation' paradigm, DACCS's contribution to the CDR pathway is very low, but could nonetheless still compromise the sustainability of the CDR pathway by requiring the deployment of new power capacity as well.

\begin{figure}[t]
\centering
  \includegraphics[width=8.3cm]{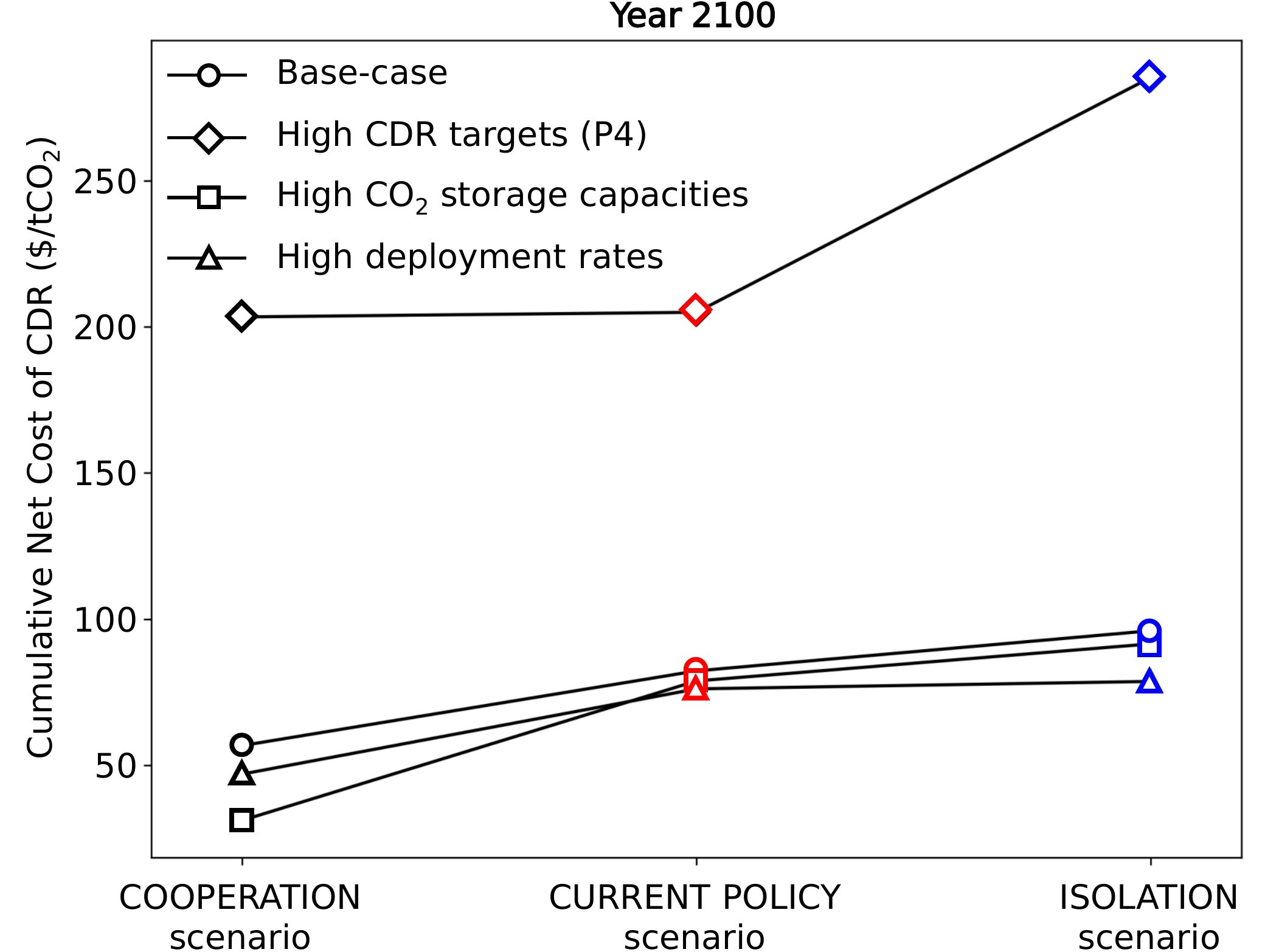} 
  \caption{Cumulative net cost of CDR in 2100, under all policy options --- 'cooperation', current policy and 'isolation' paradigms --- in the base-case and sensitivity analysis scenarios.
  The sensitivity analysis on higher CDR targets shows that high afforestation rates can slightly reduce the financial burden of financial burden of the Paris Agreement's CDR objectives. 
  From the sensitivity analysis on higher deployment rates, we find that high afforestation rates can slightly reduce the financial burden of financial burden of the Paris Agreement's CDR objectives, but from the one on higher CO$_2$ storage capacity, we find that the biggest reduction potential arises from well distributed CO$_2$ storage with high capacity, with international cooperation in climate mitigation policy.}
  \label{fig: SA Summury}
\end{figure}

Fig. \ref{fig: SA Summury} summarises cumulative net costs of CDR under all three alternative policy options, for both the base-case scenarios and the sensitivity analysis.
It emphasises the urgency of mitigating climate change as early as possible, to minimise the need for CDR, and therefore its financial burden (high CDR targets).
It also shows that high afforestation rates can slightly reduce the financial burden of financial burden of the Paris Agreement's CDR objectives (high deployment rates), but that the biggest reduction potential arises from well distributed CO$_2$ storage with high capacity, with international cooperation in climate mitigation policy (high CO$_2$ storage capacity).

\section*{Acknowledgements}
The authors thank thank Imperial College London and the Centre for Environmental Policy for the funding of a PhD Scholarship.



\balance

\renewcommand\refname{References}

\bibliography{main}
\bibliographystyle{ieeetr} 

\end{document}